\documentclass[12pt]{article}

\usepackage[margin=1in]{geometry}

\usepackage[utf8]{inputenc}
\usepackage{setspace}
\usepackage{latexsym}
\usepackage{amssymb,amsmath, bm}
\usepackage{graphicx}
\usepackage{marvosym}
\usepackage{here}  
\usepackage{subcaption} 
\usepackage{float}
\usepackage{caption}
\usepackage{chngcntr} 
\usepackage{multirow}
 
\usepackage[dvipsnames]{xcolor}
\usepackage{makecell}
\usepackage{tikz}
\usetikzlibrary{shapes,arrows}
\usetikzlibrary{positioning}

\usepackage{natbib}
\usepackage{color}
\usepackage[hidelinks, pdftex, bookmarksopen=true, bookmarksnumbered=true,
pdfstartview=FitH, breaklinks=true, urlbordercolor={0 1 0}, citebordercolor={0 0 1}]{hyperref}
\definecolor{darkgreen}{rgb}{0,0.545,0}
\definecolor{darkyellow}{rgb}{0.933,0.604,0}

\usepackage{theorem}
\theoremstyle{break}
\theoremheaderfont{\scshape}
\theorembodyfont{\rm}

\newcommand\btheta{{\boldsymbol \theta}}

\newcommand\boldeta{{\boldsymbol \eta}}
\newcommand\btau{{\boldsymbol \tau}}
\newcommand\bPsi{{\boldsymbol \Psi}}

\usepackage{xr}
\graphicspath{{figs/}} 

\renewcommand{\appendixname}{Supplementary Information }
\begin{document}

\newcommand\spacingset[1]{\renewcommand{\baselinestretch}%
{#1}\small\normalsize}

\newcommand*\circled[1]{\tikz[baseline=(char.base)]{
            \node[shape=circle,draw,inner sep=2pt] (char) {#1};}}

\spacingset{1.1}

\newcommand{\blind}{0}
\newcommand{\submitba}{0}

\newcommand{\titbase}{\bf{A Unified Model of Text and Citations for Topic-Specific Citation Networks
}}

\newcommand{\tit}{\titbase}
\if0\blind

{
\title{
  \tit\thanks{
  \protect We thank Kevin Quinn and Stuart Benjamin for their comments on the draft. We also thank Christopher Lucas, Max Goplerud and the audience of the 39th annual summer meeting of the Society for Political Methodology for their constructive comments.
  }
}

\author{
ByungKoo Kim\thanks{These authors have contributed equally to this work.} \thanks{
    Assistant Professor, KDI School of Public Policy, Sejong, Republic of Korea, Email:~\href{mailto:kimbk@kdischool.ac.kr}{\tt kimbk@kdischool.ac.kr}.
    }
\and
Saki Kuzushima\footnotemark[2] \thanks{
    Postdoctoral~Fellow,~Program~on~U.S.-Japan~Relations,~Weatherhead~Center~for~International~Affairs,~Harvard~University,~Cambridge,~Massachusetts,~USA. Email:~\href{mailto:sakikuzushima@fas.harvard.edu}{\tt sakikuzushima@fas.harvard.edu}.
    }
\and
Yuki Shiraito\thanks{
    Assistant Professor, Department of Political Science, University of Michigan. Center for Political Studies, 4259 Institute for Social Research, 426 Thompson Street, Ann Arbor, MI 48104-2321.  Phone: 734-615-5165, Email: \href{mailto:shiraito@umich.edu}{\tt shiraito@umich.edu}, URL: \href{https://shiraito.github.io}{\tt shiraito.github.io}.
    }
 }

\date{
First draft: July 13, 2022 \\ 
This draft: February 10, 2025
}

}\fi

\if1\blind
{
 \title{\tit}
 \date{}

}\fi

\maketitle

\pdfbookmark[1]{Title Page}{Title Page}

\thispagestyle{empty}
\setcounter{page}{0}

\begin{abstract}
Social scientists analyze citation networks to study how documents influence subsequent work across various domains such as judicial politics and international relations. 
However, conventional approaches that summarize document attributes in citation networks often overlook the diverse semantic contexts in which citations occur. 
This paper develops the paragraph-citation topic model (PCTM), which analyzes citation networks and document texts jointly. 
The PCTM extends conventional topic models by assigning topics to paragraphs of citing documents, allowing citations to share topics with their embedding paragraphs. 
Our empirical analysis of U.S. Supreme Court opinions in the privacy issue domain, which includes cases on reproductive rights, demonstrates that citations within individual documents frequently span multiple substantive areas, and citations to individual documents show considerable topical diversity. 

\end{abstract}

\clearpage

\tableofcontents
\thispagestyle{empty}
\setcounter{page}{0}

\onehalfspacing

\section{Introduction}

Social scientists often use citation network data to study the influence of documents, such as academic articles, books, laws, and court opinions. 
Research in judicial politics has analyzed the citation networks of the United States Supreme Court (SCOTUS) opinions, revealing how some cases exert significant influence on future rulings \citep{clark2012genealogy, fowler2007network}. 
Similarly, in international relations, scholars explore how citations shape power dynamics in areas like trade \citep{pelc2014politics}, human rights \citep{lupu2012precedent}, and jurisdictional conflicts \citep{larsson2017speaking}.

Conventional approaches seek to summarize document attributes within a network, but often overlook the diverse semantic contexts in which citations occur.
Since the semantic content of documents influences citation network structures \citep{bai2018neural, chang2010hrtm, zhang2022dynamic}, accounting for semantic heterogeneity in document networks can reveal information that might otherwise remain hidden.
The measures of precedential importance for various courts of law, for instance, implicitly treat the absence of citations as a reflection of limited precedential value rather than a potential semantic disconnect between documents \citep{fowler2007network, lupu2012precedent, pelc2014politics}. 
However, a high volume of citations to a court case may indicate its status as a landmark case, the popularity of legal topics it addresses, or both.

Recognizing the importance of semantic heterogeneity in document networks, previous studies have used human-coded topics to ensure semantic coherence, restricting their analyses to documents within discrete semantic domains such as criminal justice  \citep{olsen2017finding} or reproductive rights \citep{clark2012genealogy}.
However, human coding often captures broad categories, leaving significant semantic variation within these groups unaddressed. 
Also, researchers may wish to automatically detect semantic heterogeneity at the granularity that fits their research purpose, or the semantic context itself can be of research interest rather than an object to control for. 

This paper develops a Bayesian topic model that systematically integrates citation network and document text. 
Our proposed model, the paragraph-citation topic model (PCTM), extends conventional topic models by assigning a topic to each paragraph of the citing document, allowing citations to share topics with text of the paragraphs that they are in. 
This marks a departure from other topic models for document networks (i.e. Relational Topic Models) by allowing citations in one document to have heterogeneous topics. 
Our empirical analysis demonstrates that citations within individual documents frequently span multiple substantive areas. 
Moreover, our findings reveal considerable topical diversity in citations to individual documents, illustrating how a single opinion can intersect multiple domains of legal discourse (i.e., \textit{Roe v. Wade} engages with various legal issue areas, including civil procedure, constitutional law, healthcare policy, privacy rights, and beyond).

\subsection{Related Models}
\label{sec:relatedmodels}

A growing body of scholarship has developed models for joint analysis of texts and citation networks \citep{chang2010hrtm, liu2009topic, bai2018neural, le2014probabilistic, zhang2020topic}. 
Early LDA-based approaches leverage citations to improve topic estimation, with semantically similar documents more likely to be connected through citations \citep{chang2010hrtm, liu2009topic, nallapati2008joint}. 
More recent advances employ deep learning techniques to represent texts and citations in lower-dimensional latent embedding spaces \citep{bai2018neural, zhang2022dynamic}. 
The PCTM extends this growing literature in three substantive ways.

First, the PCTM assigns topics to paragraphs rather than individual tokens. 
This modeling strategy stems partly from the observation that paragraphs written by trained professionals often represent coherent units of idea, but more importantly, it is the modeling choice that allows researchers to identify the semantic context of each citation by finding a topic (i.e., a distribution of words) within which the citation is embedded.
Existing models, by contrast, do not have direct connections between a citation and words around it.
In \citet{chang2010hrtm} and \citet{liu2009topic}, the generative process of citations is based on the mixture of topics in the entire document, rather than assigning topics to individual citations.
The Pairwise Link-LDA model and the Link-PLSA-LDA model developed by \citet{nallapati2008joint} assign topics to individual citations, but these topics are conditionally independent of the topics assigned to words given the document-level parameters.
The PCTM is unique in that it explicitly takes into account the proximity of citations to words in the same paragraph, allowing for a more nuanced understanding of the semantic context of citations.

Second, the PCTM allows a document to send multiple citations---possibly of different topics---to another document. 
Past research has focused on topic estimation in document networks, treating citations primarily as binary linkages between documents. 
Consequently, the semantic context of individual citations has remained largely unaddressed in existing models.
While we build on previous work by utilizing citations to enhance topic estimation, our approach differs by explicitly modeling the semantic context of each citation.
Specifically, the PCTM assigns topics to each paragraph and its embedded citations, allowing citations within the same document to represent distinct topics.

Finally, the PCTM models paragraph-level citation propensities through a regression framework, offering researchers flexibility in modeling strategic citation dynamics. 
This approach aligns with social science studies that emphasize how social and political processes influence citation patterns and frequency \citep{hansford2006politics,lupu2013strategic,pelc2014politics}. 
In its current form, our model's regression layer incorporates both precedential authority and topic similarity between citing paragraphs and cited documents. 
Researchers can include any variables at the paragraph, document, or paragraph-document dyadic level to model strategic citation behavior.\footnote{In this sense, our model is similar to the Structural Topic Model (STM) by \cite{roberts2013structural} where exogenous covariates shape the topic prevalence of documents through a generalized linear model. One can imagine our model as a variation of STM where the regression layer includes endogenous processes of citation formation.}

\section{The United States Supreme Court Opinions}
\label{sec:applicationBK}

The SCOTUS as the highest judicial authority in the United States holds a pivotal role in social science studies with implications for social norms, public policy, and individual rights. 
At the center of the SCOTUS ruling is the principle of \textit{stare decisis} in which a decision must ``stand by things decided.''
The \textit{stare decisis} establishes that precedents take crucial importance in the SCOTUS as they exert varying levels of influence on future rulings.
In particular, landmark cases such as \textit{Roe v. Wade} are a crucial subject of study in social sciences. 
With the \textit{stare decisis}, a SCOTUS ruling is not just about the case at hand, but also about how to interpret the relevant precedents that together shape the boundary of social norms and behavior. 

Due to the significance of precedents in the SCOTUS, many social science studies have been dedicated to exploring various aspects of precedents.
Many scholars have focused on the political processes involved in the choice and the representation of precedents in the SCOTUS majority opinions \citep{hansford2006politics, bailey2008does,clark2010locating}.
How precedents are treated by future cases and eventually fade away was another focal point of research \citep{black2013citation, broughman2017after}.

Mapping the SCOTUS cases and citations into a network, past studies employed network analysis to measure the structural properties of precedents in the citation network.
\cite{clark2012genealogy}, for instance, fits the latent tree model to the SCOTUS citation network and uncovers the hierarchy of precedents as an estimation of the evolution of legal doctrine.
Another strand of research highlights the positions precedents take in the citation network \citep{fowler2005authority, fowler2007network}.
\cite{fowler2007network} and \cite{fowler2008authority} propose a variation of the eigenvector centrality score to gauge how legally ``central'' a case is for the SCOTUS at a given point in time.

While recognizing the usefulness of the network analysis for the SCOTUS citation network, we find that existing approaches commonly overlook the topic heterogeneity of the citation network.
Network analysis of the SCOTUS citation network treats presence and absence of citations as informative signals.
In \cite{fowler2007network} and \cite{fowler2008authority}, for example, precedents that attract many citations are likely to be structurally central and precedents without many citations are considered to be peripheral.
While the presence of citations can be an informative signal for the importance of a case, the absence of citations may simply be due to topic inconsistency rather than its importance.
That is, we do not expect a case to cite a precedent if the given precedent addresses completely distinct legal topics. 
When the network analytic methods are applied to the universe of cases without special attention to the topic differences between them, one may mistakenly interpret the topic differences as indicators of importance.

Another point we highlight is that the topic space of an opinion is multidimensional. 
For example, \textit{Roe v. Wade} is mostly known for the right to privacy in abortion, but it also addresses other legal topics such as substantive due process, end-of-life decisions, and legislative restraints.
A citation to \textit{Roe v. Wade} can be concerning the right to privacy, but it could also be about other topics such as legislative restraints. 
This suggests that subsetting down to a broad legal category of cases for network analysis, such as seen in \cite{clark2010locating} where authors limit their scope to search and seizure and the freedom of religion opinions, may not be sufficient to capture nuanced legal topics that a case touches upon. 

To address the above key challenge, we propose to incorporate the text of the SCOTUS majority opinions with the citation network. 
In the following sections, we propose a model that incorporates both the text and the citation network of the SCOTUS majority opinions. 
Our model can uncover the topic structure of the majority opinions with topic model while utilizing the network linkage in the citation network.
We apply our model to all privacy opinions in the SCOTUS and demonstrate that the resulting topic-homogenous citation subnetwork can be used for further network analysis.

For the application of our model, we obtain the universe of the SCOTUS majority opinions on the privacy issue area from the Caselaw Access Project\footnote{\url{https://case.law}}.
Subsetting follows the issue area categorization provided by Supreme Court Database \citep{scdb}. 
The privacy issue area is chosen for our application because existing literature on citation networks of the SCOTUS cases often focuses on this issue \citep{fowler2007network, clark2012genealogy}.
It is also an important application given the recent controversial decision that overruled the landmark case on constitutional rights to abortion. 
The Privacy opinions subset consists of 106 documents with 4,669 paragraphs, 5,838 unique words, and 452 citations.
More details of data pre-processing for each subset are available in Supplementary Information, Section~\ref{sec:data}.

\section{The Proposed Model}
\label{sec:model}
Our proposed model is built on a topic model, a popular model to discover latent clusters or topics of documents \citep{blei2003lda, blei2007correlated}. 
A topic model that analyzes documents with citation networks must address the following questions: 
By what process do authors decide to cite another document? 
How does the topic structure enter into citation decisions, and conversely, how do citations help determine the topic structure of citing and cited documents? 

To address these questions, we augment a topic model by latent citation propensity to model authors' decisions to make citations in relation to the topic structure. 
The latent citation propensity is shaped by a regression model that reflects the known factors of strategic citation behavior such as the authority (or popularity) of the cited document \citep{larsson2017speaking, lupu2012precedent, lupu2013strategic, pelc2014politics} as well as the similarity of topics between citing and cited documents.

Additionally, we propose to use paragraphs as the unit for the topic assignment. 
We view citations as the directed reference from a paragraph to another document. 
The advantage of this perspective is that it reflects a more realistic data-generating process. 
A paragraph is often the vehicle of one coherent topic, and citations within that paragraph are likely to refer to documents of very similar, if not the same, topic prevalence. 
For example, an opinion in the SCOTUS typically identifies multiple legal doctrines that apply to a given case and addresses them in different paragraphs. 
Therefore, citations within one paragraph are likely to point to a collection of opinions that address the same legal doctrine. 
In other words, citations in paragraphs of different topics are likely to be references to different legal contexts, even if they are from the same document. 
We believe such characteristics are not limited to legal documents of the SCOTUS, but a general feature of any document network, and they should be reflected in the process of uncovering topic structure. 
Below, we delineate our modeling strategy that addresses the above questions in detail. 

\subsection{Paragraph-citation Topic Model}
First, we introduce the notation.
Let $N$, $G$, $V$, and $K$ be the total number of documents, total number of paragraphs, and total number of unique words, and the number of topics, respectively.
We use $N_{ip}$ to denote the number of words in paragraph $p$ of document $i$.
Our data consist of words, $\textbf{W}$, and citations, $\textbf{D}$.
$\textbf{W}$ is a matrix of size $G \times V$ where each row is $\textbf{w}_{ip}$, a vector of length $V$ that represents the number of times each unique word appears in a paragraph $p$ of document $i$.
$\textbf{D}$ is a matrix of size $G \times N$ where each element, $D_{ipj}$ is a binary variable that indicates the existence of a citation from $p$th paragraph in $i$th document towards $j$th document.
$\textbf{D}^*$ is a matrix of size $G \times N$ and its element, $D_{ipj}^*$, is a latent variable that represents the latent citation propensity of $p$th paragraph in $i$th document to cite $j$th document.
We have another latent variable $\textbf{Z}$, a vector of length $G$ where each element is $z_{ip}$, a scalar that takes a value from $\{1,\ldots, K\}$, and it represents the topic assignment of $p$th paragraph in $i$th document.
We have three main parameters to estimate: $\pmb\eta$, $\pmb\Psi$, and $\pmb\tau$.
$\pmb\eta$ is a matrix of size $N \times K$ where each row is $\pmb\eta_i$, a vector of length $K$ that represents the topic proportion of document $i$, generated from a multivariate normal distribution with mean $\pmb\mu$ and covariance $\pmb\Sigma$.
$\pmb\mu$ is further generated from a normal distribution with mean $\pmb\mu_0$ and covariance $\pmb\Sigma_0$.
$\pmb\Psi$ is a matrix of size $K \times V$ where each row is $\pmb\Psi_k$, a vector of length $V$ that represents the word distribution of topic $k$. 
$\pmb\Psi_k$ is generated from a Dirichlet distribution with parameter $\pmb\beta$.
$\pmb\tau$ is a vector that represents the coefficients of the regression model that shapes the latent citation propensity, generated from a multivariate normal distribution with mean $\pmb\mu_{\tau}$ and covariance $\pmb\Sigma_{\tau}$.

The data-generating process is modeled as follows.

\begin{align}
  \begin{split}
	D_{ipj} &= \begin{cases}
		1 \text{ if } D_{ipj}^* \geq 0 \\
		0 \text { if } D_{ipj}^* < 0
	\end{cases}\\
    D_{ipj}^* &\sim \mathcal{N}(\pmb\tau^T\textbf{x}_{ipj},1)\quad \text{where}\ \textbf{x}_{ipj} = [1, \kappa_{j}^{(i)}, \eta_{j,z_{ip}}]\\
	\textbf{w}_{ip} &\sim \text{Multinomial}(N_{ip},\pmb\Psi_{z_{ip}})  \\
	z_{ip} &\sim \text{Multinomial}(1,\text{softmax}(\pmb\eta_i))  \\
	\pmb\Psi_k &\sim \text{Dirichlet}(\pmb\beta)  \\
	\pmb\eta_i &\sim \mathcal{N}(\pmb\mu,\pmb\Sigma)  \\
	\pmb\mu &\sim \mathcal{N}(\pmb\mu_0, \pmb\Sigma_0)  \\
	\pmb\tau &\sim \mathcal{N}(\pmb\mu_{\tau},\pmb\Sigma_{\tau})
  \end{split}
\end{align}

\noindent where $\mathbf{x}_{ipj}$ is a vector of covariates that shape the latent citation propensity for $p$th paragraph in document $i$ to cite document $j$. $\mathbf{x}_{ipj}$ consists of 3 terms -- the intercept, indegree, and $\eta_{j,z_{ip}}$, and $\pmb\tau = [\tau_0, \tau_1, \tau_2]$ is a vector of coefficients.
The intercept in $\textbf{x}_{ipj}$ is to capture the overall sparsity of the citation network. Since networks in the real world are generally very sparse, we expect the intercept $\tau_0$ to be negative. 
The indegree of a precedent is included to capture the authority. 
This follows existing studies of strategic citation that commonly point to the importance of the authority of a precedent as one of the major attracting factors of citations  \citep{hansford2006politics,lupu2012precedent, lupu2013strategic}. 
This is also consistent with a well-known dynamic in social networks called ``rich-get-richer'' or, more technically, ``preferential attachment'' where popular individuals become more popular \citep{newman2001clustering,wang2008measuring}. 
The indegree term is denoted $\kappa_j^{(i)}$, with superscript $(i)$ to indicate the authority of the $j$th document at the time of $i$'s writing. We expect its coefficient $\tau_1$ to be positive. 
Finally, $\eta_{j,z_{ip}}$ is added to capture the topic similarity between the citing paragraph $ip$ and document $j$. 
Since we expect that citations are more likely to occur between documents of similar topics, we expect its coefficient $\tau_2$ to be positive. 
 
While we currently include 3 document-level covariates in $\textbf{x}$, researchers can add other covariates that fit their research purposes. 
For instance, the political ideology of judges in a precedent and a citing case can be an important factor in citation decisions \citep{lupu2013strategic}. 
Then researchers can include a binary copartisanship indicator in $\textbf{x}_{ipj}$ that takes 1 if the author of opinion $i$ and the author of opinion $j$ are appointed by presidents of the same party and 0 otherwise. 

Given words and citations, $\textbf{W}$ and $\textbf{D}$, our posterior probability is 
{\small
\begin{align}
	p(\pmb\eta,\pmb\Psi,\textbf{Z},\pmb\tau|\textbf{W},\textbf{D}) \propto p(\pmb\mu|\pmb\mu_0,\pmb\Sigma_0)p(\pmb\tau|\pmb\mu_{\tau},\pmb\Sigma_{\tau})p(\pmb\eta|\pmb\mu,\pmb\Sigma)p(\pmb\Psi|\pmb\beta)p(\textbf{Z}|\pmb\eta)p(\textbf{W}|\pmb\Psi,\textbf{Z})p(\textbf{D}|\textbf{D}^*)p(\textbf{D}^*|\pmb\tau,\pmb\eta,\textbf{Z},\textbf{D})
\end{align}
}

\subsection{Bayesian Inference}
\label{sec:inference}
Unfortunately, the inference of the given posterior distribution is hard due to the non-conjugacy between normal prior for $\pmb\eta$ and the logistic transformation function \citep{blei2007correlated}. Variational inference is the most frequently employed tool to address this problem, with the additional advantage of computational speed. However, obtained parameters are for the variational distribution which is an approximation to the target posterior. The quality of the approximation is often not sufficiently explored. Furthermore, the variational inference is an optimization method that outputs point estimates. This requires additional steps to obtain a measure of uncertainty in estimation. Quantifying uncertainty in variational inference is often done through bootstrapping \citep{chen2018use,imai2016fast}. However, obtaining bootstrap samples representative of the pseudo population can be highly challenging for network data since observations are connected \citep{chen2019snowboot,levin2019bootstrapping}. It often requires block sampling which entails computing other network quantities (i.e. geodesic distance in \cite{raftery2012fast}) but these additional processes could defeat the advantage of the computational efficiency of using variational inference. 

To remedy this problem, we follow the recent advances in the inference of Correlated Topic Model (CTM) that adopts partial collapsing \citep{held2006bayesian,chen2013scalable,linderman2015dependent}. We first partially collapse the posterior distribution by integrating out the topic-word probability parameter $\pmb\Psi$. Then we introduce an auxiliary Polya-Gamma variable $\pmb\lambda$ and augment the collapsed posterior. Partial collapsing and data augmentation enables us to use Gibbs sampling which is known to produce samples that converge to the exact posterior. With $\pmb\Psi$ integrated out, our new posterior is proportional to
{\small
\begin{align}
	\int_{\pmb\Psi}p(\pmb\eta,\pmb\Psi,\textbf{Z},\pmb\tau|\textbf{W},\textbf{D}) \propto p(\pmb\mu|\pmb\mu_0,\pmb\Sigma_0)p(\pmb\tau|\pmb\mu_{\tau},\pmb\Sigma_{\tau})p(\pmb\eta|\pmb\mu,\pmb\Sigma)p(\textbf{Z}|\pmb\eta)p(\textbf{W}|\textbf{Z})p(\textbf{D}|\textbf{D}^*)p(\textbf{D}^*|\pmb\tau,\pmb\eta,\textbf{Z},\textbf{D})
\end{align}
}
where $p(\textbf{W}|\textbf{Z})$ results from collapsing $\pmb\Psi$ as follows.
\begin{align}
	p(\textbf{W}|\textbf{Z}) &= \int_{\pmb\Psi} p(\textbf{W},\pmb\Psi|\textbf{Z})d\pmb\Psi \nonumber \\
	&= \int_{\pmb\Psi} p(\textbf{W}|\pmb\Psi,\textbf{Z})p(\pmb\Psi|\textbf{Z})d\pmb\Psi \nonumber \\
	&= \int_{\pmb\Psi} p(\textbf{W}|\pmb\Psi,\textbf{Z})p(\pmb\Psi)d\pmb\Psi
\end{align}
The above takes the form of Dirichlet-multinomial distribution which enters in the conditional posterior distribution of $\textbf{Z}$ below. The conditional posterior distribution of \textbf{Z} for $ip$th paragraph is 
{ \small
\begin{align}   
    p(z_{ip}^k=1|\mathbf{Z}_{-ip},\pmb\eta,\mathbf{W},\mathbf{D}^*) &\propto p(z_{ip}^k=1|\pmb\eta_i)p(\mathbf{W}_{ip}|z_{ip}^k=1,\mathbf{Z}_{-ip},\mathbf{W}_{-ip}) \prod_{j=1}^{i-1}p(D_{ipj}^*|z_{ip}^k=1,\textbf{Z}_{-ip},\pmb\tau,\pmb\eta,\kappa) \nonumber \\
    &\propto \pi_{ipj,k}
\end{align}
}
where 
{ \small
\begin{align}
	\pi_{ipj,k} = \text{exp}\Bigg\{  \eta_{ik} + \text{log}\prod_v \Gamma(\beta_v + c_{k,ip}^v + c_{k,-ip}^v) - \text{log}\Gamma(\sum_v \beta_v + c_{k,ip}^v + c_{k,-ip}^v) \nonumber \\
	- \frac{1}{2}\Big(\tau_2^2\eta_{jk}^2 + 2\big(\tau_0\tau_2 + \tau_1\tau_2\kappa_j^{(i)} - \tau_2 D_{ipj}^* \big)\eta_{jk} \Big) \Bigg\}
\end{align}
}

We use $\textbf{Z}_{-ip}$ and $\textbf{W}_{-ip}$ to denote the set of all topic assignments and words except for the $ip$th paragraph, respectively.
Here, $c_{k,ip}^v$ denotes the total number of times the $v$th word appears in paragraph $ip$ of topic $k$ such that $c_{k,ip}^v = \sum_{l=1}^{n_{ip}}\mathbb{I}(W_{ipl}=v)\mathbb{I}(z_{ip}^k=1)$. Likewise, 
$c_{k,-ip}^v$ is the total number of times the $v$th term appears in paragraphs with $k$th topic except for $ip$. 
The form of the conditional posterior for the $ip$th paragraph-level topic $z_{ip}^k$ offers a convenient interpretation on the \textit{source of information}. The first part $p(z_{ip}^k=1|\pmb\eta_i)$ displays the topic information from document-level topic prevalence. The second part represents topic information from the words in $ip$th paragraph. The third part $\prod_{j=1}^{i-1}p(D_{ipj}^*|z_{ip}^k=1,\textbf{Z}_{-ip},\pmb\tau,\pmb\eta,\kappa)$ is equivalent to the total amount of topic information from citations. 

The conditional posterior distribution of $\pmb\eta$ for $i$th document is jointly defined with the augmenting Polya-Gamma distribution for $\pmb\lambda$. The conditional posterior distribution for $\lambda_{ik}$ is
\begin{align}
    p(\lambda_{ik}|\mathbf{Z},\mathbf{W},\pmb\eta) &\propto PG(N_i,\rho_{ik})
\end{align}
\noindent where $\rho_{ik} = \eta_{ik} - \text{log}(\sum_{l \neq k} e^{\eta_{il}})$.

With $\lambda_{ik}$, we can obtain the conditional posterior of $\pmb\eta$ for $i$th document as follows.
\begin{align}
    p(\eta_{ik}|\eta_{i,-k},\mathbf{Z},\mathbf{W},\mathbf{D},\pmb\tau,\lambda_{ik}) &\propto \mathcal{N}(\eta_{ik}|\tilde{\mu}_{ik},\tilde{\sigma}_k^2)
\end{align}
where
\begin{align}
    \tilde{\sigma}_k^2 &= (\sigma_k^{-2} + \lambda_{ik} + v_{i,kk}^{-1})^{-1} \nonumber \\
    \tilde{\mu}_{ik} &= \tilde{\sigma}_k^2 \big( v_{i,kk}^{-1}m_{ik} + \sigma_k^{-2}\nu_{ik} + t_{ik} - \frac{N_i}{2} + \lambda_{ik}\text{log}(\sum_{l\neq k}e^{\eta_{il}}) \big)
\end{align}
For the definition of $v_{i,kk}$, $m_{ik}$, $\nu_{ik}$, and $t_{ik}$ as well as the detailed derivation, see Supplementary Information, Section~\ref{sec:si_inference}.

The conditional posterior for latent citation propensity parameter $\textbf{D}^*$ is 
{ \footnotesize
\begin{align}
    p(D^*_{ipj}|\pmb\eta,\mathbf{Z},\pmb\tau, \mathbf{D}) &\propto \begin{cases} TN_{[0,\infty)}(\tau_0+\tau_1\kappa_j^{(i)}+\tau_2\eta_{j,z_{ip}},1) & \text{ if } D_{ipj} = 1 \\
    TN_{(-\infty,0]}(\tau_0+\tau_1\kappa_j^{(i)}+\tau_2\eta_{j,z_{ip}},1) & \text{ if } D_{ipj} = 0
    \end{cases}
\end{align}
}
\noindent where $TN_{[a,b)}(\mu,\sigma^2)$ denotes the truncated normal distribution with mean $\mu$ and variance $\sigma^2$ truncated to the interval $[a,b)$.
The conditional posterior for $\pmb\tau$ follows the following distribution. Let $\mathbf{x}_{ipj} = [1,\kappa_j^{(i)},\eta_{j,z_{ip}}]^T$ and $\pmb\tau = [\tau_0,\tau_1,\tau_2]^T$
\begin{align}
    p(\pmb\tau|\pmb\eta,\mathbf{Z},\mathbf{D}^*) &\propto exp\Bigg\{-\frac{1}{2} \sum_{ipj}\Big(D_{ipj}^* - \mathbf{x}_{ipj}^T\pmb\tau \Big)^2 \Bigg\} N(\pmb\mu_{\pmb\tau},\pmb\Sigma_{\pmb\tau})\nonumber \\
    &\propto N(\tilde{\pmb\tau},\tilde{\pmb\Sigma_{\pmb\tau}})
\end{align}
\noindent where $\tilde{\pmb\Sigma_{\pmb\tau}} = \Bigg(\Big(\sum_{ipj}\mathbf{x}_{ipj}\mathbf{x}_{ipj}^T \Big) + \pmb\Sigma_{\tau}^{-1} \Bigg)^{-1}$ 
and $\tilde{\pmb\tau} = \tilde{\pmb\Sigma_{\pmb\tau}}\Bigg(\Big(\sum_{ipj}\mathbf{x}_{ipj}^TD_{ipj}^*\Big) + \pmb\Sigma_{\tau}^{-1}\pmb\mu_{\pmb\tau}\Bigg)$

Using simulation data, we confirm that the proposed Gibbs sampler recovers the true latent topics from random initialization.
Our discussion about the initialization of the Gibbs sampler is presented in Supplementary Information, Section~\ref{sec:init}, and the results of the simulation studies are presented in Supplementary Information, Section~\ref{sec:simulation}.

\section{Empirical Results}
\label{sec:result}

This section presents the results of applying the PCTM to the SCOTUS dataset, focusing on the privacy issue area.\footnote{We also present additional results with a dataset on voting rights issue area in Supplementary Information, Section~\ref{sec:voting}.}
We present three main results.
First, we fit the PCTM and existing alternatives, LDA and RTM, to the SCOTUS opinions on the privacy issue area and discuss the advantages of the PCTM over the existing models.\footnote{The convergence diagnostics of the PCTM and discussions about the parameters not discussed in this section are provided in Supplementary Information, Section~\ref{sec:si_privacy}}
We find that the main advantage of the PCTM is its ability to use paragraph-level topics to extract informative topic-specific subset of the citation network.
Second, we utilize these topic-specific subnetworks to measure the importance of cases within each topic, following the methodological framework of \cite{fowler2007network}.
Our analysis reveals that case importance varies substantially across topic domains.
Third, we conduct the predictive analysis of the topic structure of \textit{Dobbs v. Jackson Women's Health Organization}, the recent case that overruled \textit{Roe v. Wade}, based on words and citations in its paragraphs.
We find that the predicted topics of \textit{Dobbs v. Jackson Women's Health Organization} address abortion in markedly different ways from post-\textit{Roe v. Wade} cases, but in ways reminiscent of pre-\textit{Roe v. Wade} cases. 
Together, these results demonstrate the advantage of the PCTM in uncovering valuable insights from the text and citation data of the SCOTUS opinions. 

\subsection{Topic Composition of SCOTUS Opinions on Privacy}
Table \ref{word_mat_privacy} displays the top 10 most frequent words for each topic estimated in the PCTM. 
The Supreme Court Database assigns 4 issue codes to opinions of the privacy issue area, but we identify 7 distinct topics in the PCTM.\footnote{The four issue codes identified by the Supreme Court Database are privacy, abortion, right to die and Freedom of Information Act. 
To determine the optimal number of topics for our analysis, we implemented an iterative approach, beginning with a 4-topics specification and systematically increasing the number of topics up to 15. We ultimately selected a 7-topics model as it provided the most coherent representation of legally salient themes within the privacy issue area, based on our substantive knowledge of constitutional law and privacy jurisprudence.}  
The labels in the table are provided by the authors.
\begin{table}[t!]
\centering
\begin{tabular}{c| l l l l l l l}
  \hline
Topic & \textcolor{SkyBlue}{Regulation} & \textcolor{Blue}{Procedural} & \textcolor{Aquamarine}{Const.} & \textcolor{Green}{Speech} & \textcolor{ForestGreen}{Damage} & \textcolor{Brown}{Privacy} & \textcolor{Red}{Public}\\ 
Label & \textcolor{SkyBlue}{of} & \textcolor{Blue}{Posture} & \textcolor{Aquamarine}{Rights} & \textcolor{Green}{\&} & \textcolor{ForestGreen}{to} & \textcolor{Brown}{vs} & \textcolor{Red}{Disclosure}\\ 
 & \textcolor{SkyBlue}{Abortion} & \textcolor{Blue}{} & \textcolor{Aquamarine}{to} & \textcolor{Green}{Protest} & \textcolor{ForestGreen}{Privacy} & \textcolor{Brown}{Govnt.} & \textcolor{Red}{of Private}\\ 
 & \textcolor{SkyBlue}{Procedure} & \textcolor{Blue}{} & \textcolor{Aquamarine}{Abortion} & \textcolor{Green}{} & \textcolor{ForestGreen}{} & \textcolor{Brown}{Interest} & \textcolor{Red}{Information}\\ 
  \hline
1 & \textcolor{SkyBlue}{abort} & \textcolor{Blue}{appeal} & \textcolor{Aquamarine}{right} & \textcolor{Green}{clinic} & \textcolor{ForestGreen}{damag} & \textcolor{Brown}{drug} & \textcolor{Red}{inform}\\ 
2   & \textcolor{SkyBlue}{parent} & \textcolor{Blue}{district} & \textcolor{Aquamarine}{abort} & \textcolor{Green}{injunct} & \textcolor{ForestGreen}{act} & \textcolor{Brown}{act} & \textcolor{Red}{agenc}\\ 
3   & \textcolor{SkyBlue}{minor} & \textcolor{Blue}{board} & \textcolor{Aquamarine}{constitu} & \textcolor{Green}{right} & \textcolor{ForestGreen}{actual} & \textcolor{Brown}{test} & \textcolor{Red}{exmpt}\\ 
4   & \textcolor{SkyBlue}{physician} & \textcolor{Blue}{ani} & \textcolor{Aquamarine}{protect} & \textcolor{Green}{public} & \textcolor{ForestGreen}{congress} & \textcolor{Brown}{student} & \textcolor{Red}{disclosur}\\ 
5   & \textcolor{SkyBlue}{perform} & \textcolor{Blue}{order} & \textcolor{Aquamarine}{medic} & \textcolor{Green}{speech} & \textcolor{ForestGreen}{person} & \textcolor{Brown}{school} & \textcolor{Red}{record}\\ 
6   & \textcolor{SkyBlue}{woman} & \textcolor{Blue}{agency} & \textcolor{Aquamarine}{amend} & \textcolor{Green}{petition} & \textcolor{ForestGreen}{privaci} & \textcolor{Brown}{respond} & \textcolor{Red}{public}\\ 
7   & \textcolor{SkyBlue}{medic} & \textcolor{Blue}{document} & \textcolor{Aquamarine}{decis} & \textcolor{Green}{protest} & \textcolor{ForestGreen}{right} & \textcolor{Brown}{use} & \textcolor{Red}{govern}\\ 
8   & \textcolor{SkyBlue}{interest} & \textcolor{Blue}{rule} & \textcolor{Aquamarine}{person} & \textcolor{Green}{zone} & \textcolor{ForestGreen}{ani} & \textcolor{Brown}{ani}  & \textcolor{Red}{act}\\ 
9   & \textcolor{SkyBlue}{health} & \textcolor{Blue}{unit} & \textcolor{Aquamarine}{interest} & \textcolor{Green}{interest} & \textcolor{ForestGreen}{general} & \textcolor{Brown}{district}  & \textcolor{Red}{congress}\\ 
10   & \textcolor{SkyBlue}{consent} & \textcolor{Blue}{act} & \textcolor{Aquamarine}{life} & \textcolor{Green}{person} & \textcolor{ForestGreen}{doe} & \textcolor{Brown}{petition}  & \textcolor{Red}{foia}\\ 
   \hline
\end{tabular}
\caption{Top 10 words of highest probability for each topic from the PCTM.}
\label{word_mat_privacy}
\end{table}

The first and the third topics both address abortion as the substantive case in point but differ in the context in how abortion is addressed. 
Paragraphs of the first topic illuminate abortion as a woman's right and discuss the conditions in which the decision can be restricted or unrestricted, such as a woman's health, being a minor, or being ill-informed by her physician, etc. The third topic addresses it in a broader context of a person's right to life and death (e.g., is the right to birth control limited to married couples). The second topic addresses the processes involving lower and higher courts, which we believe to be a byproduct of having paragraphs as the unit for topic assignments. Almost all majority opinions in the SCOTUS have at least one paragraph discussing how the case was appealed from the lower court to higher courts. Since the set of vocabulary and citations in those paragraphs are generally distinct from other paragraphs, the PCTM tends to assign a topic for this category. Paragraphs of the fourth topic mostly concern public protests and speeches surrounding (anti-) abortion decisions in courts. The fifth topic addresses what constitutes damage to privacy under the Privacy Act of 1974. The sixth and seventh topics both concern the public disclosure of private information. The sixth topic, which we label as \texttt{Privacy vs Government Interest}, mainly addresses access to private information, such as the history of drug abuse that might disrupt the operations of government agencies. The seventh topic, on the other hand, concerns whether the way private information is recorded constitutes a violation of Privacy Act of 1974.

\begin{figure}[t!]
     \centering
     \begin{subfigure}[b]{0.32\textwidth}
         \centering
         \includegraphics[width=\textwidth, trim={2cm, 3cm, 2cm, 2cm}, clip]{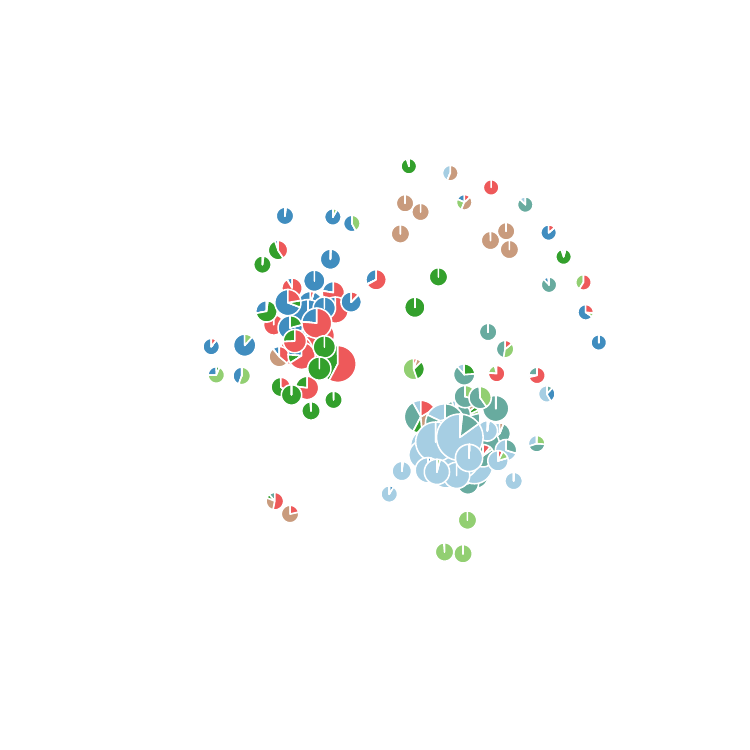}
         \caption{LDA}
         \label{LDA_net}
     \end{subfigure}
     \hfill
     \begin{subfigure}[b]{0.32\textwidth}
         \centering
         \includegraphics[width=\textwidth, trim={2cm, 3cm, 2cm, 2cm}, clip]{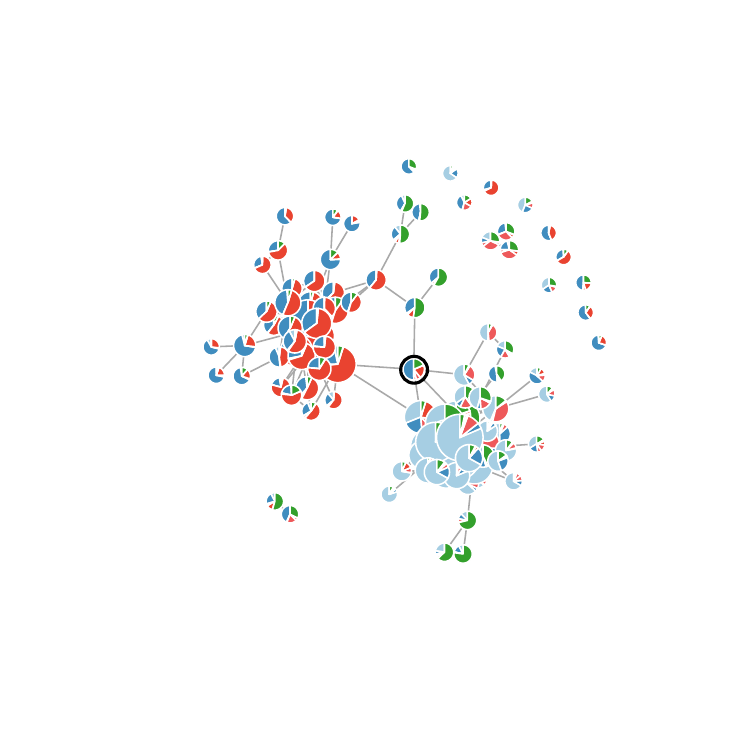}
         \caption{RTM}
	 \label{RTM_net}
     \end{subfigure}
     \hfill
     \begin{subfigure}[b]{0.32\textwidth}
         \centering
         \includegraphics[width=\textwidth, trim={2cm, 3cm, 2cm, 2cm}, clip]{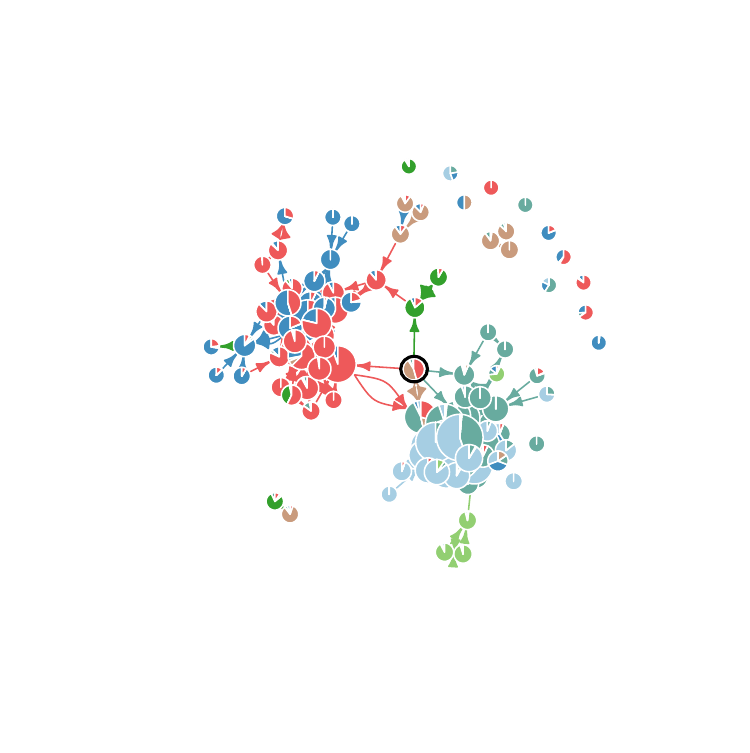}
         \caption{PCTM}
	 \label{PCTM_net}
     \end{subfigure}
        \caption{The result of three topic models, LDA, RTM, and PCTM from (a) to (c), on the US Supreme Court opinions of the privacy issue area.
        A node represents an opinion, and an edge represents a citation between opinions.
        The color composition of a node follows the topic proportion of words (LDA, RTM) or paragraphs (PCTM) in the given opinion.
        The color of an edge is based on the estimated topic of the paragraph where the citation is made.
        Note that the topic spaces of the three models are not exactly the same.
        Same colors are assigned to topics that share the top 5 most frequent words between the three models.
        (a) LDA estimates topic structure of documents without reference to the citation network.
        (b) RTM takes into account the linkage between documents for the estimation of topics, but assumes that edges are undirected and remains agnostic about the topics of citations.
        (c) PCTM recognizes the directions of edges and estimates the topic structure of both documents and citations.
        PCTM offers a semantic context over how documents are connected by identifying the topic of the paragraph in which a citation is made.}
        \label{network_figures}
\end{figure}
Next, we compare the results of LDA, RTM, and PCTM on the privacy issue area of the SCOTUS opinions.
Figure~\ref{network_figures} displays the results of LDA, RTM, and the PCTM on the entire SCOTUS opinions on the privacy issue area. 
LDA assigns topics based on words without reference to how documents are connected.
RTM incorporates the networked structure of documents but assumes that connections between documents are undirected and binary.
Moreover, RTM remains agnostic to the semantic context of citations since it does not consider their location within documents, which is reflected in the uniformly gray edges shown in Figure~\ref{RTM_net}.

By contrast, in Figure~\ref{PCTM_net}, the PCTM assigns topics to paragraphs, which allows citations within the same document to have different topics. 
For example, focus on the case, \textit{NASA vs. Nelson}, represented by the node at the center of the network highlighted by a black circle in Figure~\ref{RTM_net} and Figure~\ref{PCTM_net}.
In Figure~\ref{PCTM_net}, it has six out-going edges colored differently according to the PCTM, which implies that the citations are made in the paragraphs addressing different topics.
By contrast, the same case in Figure~\ref{RTM_net} has six edges colored gray, which means that RTM does not differentiate the topics of the citations. 
This showcases the advantage in the PCTM can provide a richer insight into the topic structure of the citations by identifying the topic of the paragraph in which a citation is made. 

\begin{table}[t!]
     \begin{center}
     \begin{tabular}{| p{5cm} | p{5cm} | }
     \hline
     \textcolor{Brown}{Privacy vs Govnt. Interest} & \textcolor{Red}{Public Disclosure of Private Information} \\ \hline \hline
     \includegraphics[width=0.3\textwidth, height=12mm]{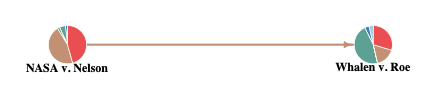}
     &
     \includegraphics[width=0.3\textwidth, height=12mm]{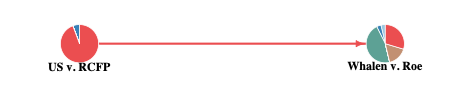}
      \\ \hline
\tiny{
\textcolor{Brown}{With these interests in view, we conclude that the challenged portions of both SF-85 and Form 42 consist of reasonable, employment-related inquiries that further the Government’s interests in managing its internal operations. See Engquist, 553 U. S., at 598-599; \textbf{Whalen v. Roe, 429 U. S.}, at 597-598. As to SF-85, the only part of the form challenged here is its request for information about “any treatment or counseling received” for illegal-drug use within the previous year. ... The Government has good reason to ask employees about their recent illegal-drug use. Like any employer, the Government is entitled to have its projects staffed by reliable, law-abiding persons who will “ ‘efficiently and effectively’” discharge their duties.}
}
      &
\tiny{
\textcolor{Red}{... Here, the former interest, ``in avoiding disclosure of personal matters,'' is implicated. Because events summarized in a rap sheet have been previously disclosed to the public, respondents contend that Medico’s privacy interest in avoiding disclosure of a federal compilation of these events approaches zero. We reject respondents’ cramped notion of personal privacy ... We have also recognized the privacy interest in keeping personal facts away from the public eye. In \textbf{Whalen v. Roe, 429 U. S. 589 (1977)}, we held that ``the State of New York may record, in a centralized computer file, the names and addresses of all persons who have obtained, pursuant to a doctor’s prescription, certain drugs for which there is both a lawful and an unlawful market.'' Id., at 591. In holding only that the Federal Constitution does not prohibit such a compilation, we recognized that such a centralized computer file posed a ``threat to privacy'':}
}
      \\ \hline
      \end{tabular}
      \caption{Paragraphs containing the same citations but assigned with different topics, \texttt{Privacy vs Government Interest} and \texttt{Public Disclosure of Information}.
       The top row displays a pair of opinions and a citation between the two color-coded by topics, and the left node is the citing opinion and the right node is the cited opinion.
       The second row for each topic contains the text of the paragraph where the citation is made in the two citing opinions in the first row.}
      \label{para_table2}
      \end{center}
\end{table}

To highlight the advantage of the PCTM in finding heterogeneous semantic context around citations, we provide example paragraphs containing citations to the same case but with different topics in Table~\ref{para_table2}. 
Since Supreme Court cases typically address multiple legal domains, subsequent citations to these cases often engage with distinct aspects of their jurisprudence.
For instance, \textit{NASA v. Nelson} and \textit{US v. RCFP} in Table \ref{para_table2} both cite \textit{Whalen v. Roe}, but in distinct substantive contexts. 
For \textit{NASA v. Nelson}, the focus was on whether the employer (NASA) should have access to private information (history of drug abuse) of its employees whereas for \textit{US v. RCFP}, citing \textit{Whalen v. Roe} was mainly about the form of record-keeping of private information (in rap sheet in \textit{US v. RCFP} and in computer files in \textit{Whalen v. Roe}) and the consequent public disclosure of that information. 
This demonstrates the semantic heterogeneity of citations even when they refer to the same document, the nuance that the PCTM can capture with paragraph-level topic assignment.

\begin{figure}[t!]
     \centering
     \begin{subfigure}[b]{0.45\textwidth}
         \centering
         \includegraphics[width=\textwidth]{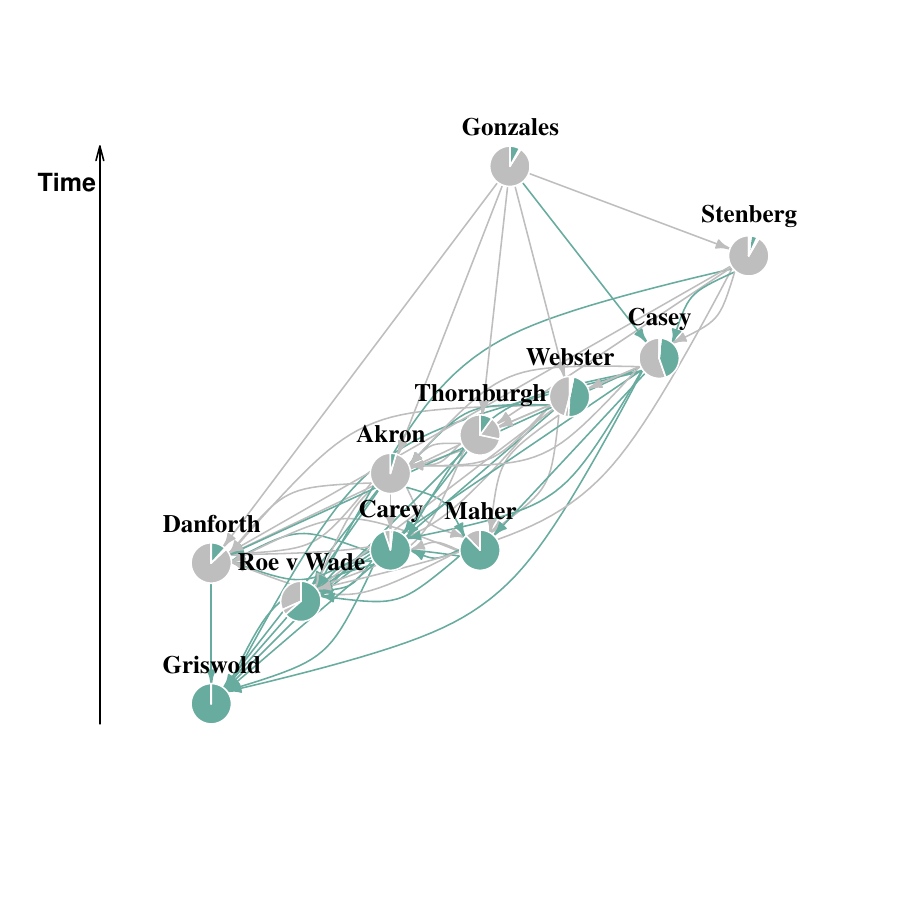}
         \caption{Constitutional Rights to Abortion}
         \label{reprod_3}
     \end{subfigure}
     \hfill
     \begin{subfigure}[b]{0.45\textwidth}
         \centering
         \includegraphics[width=\textwidth]{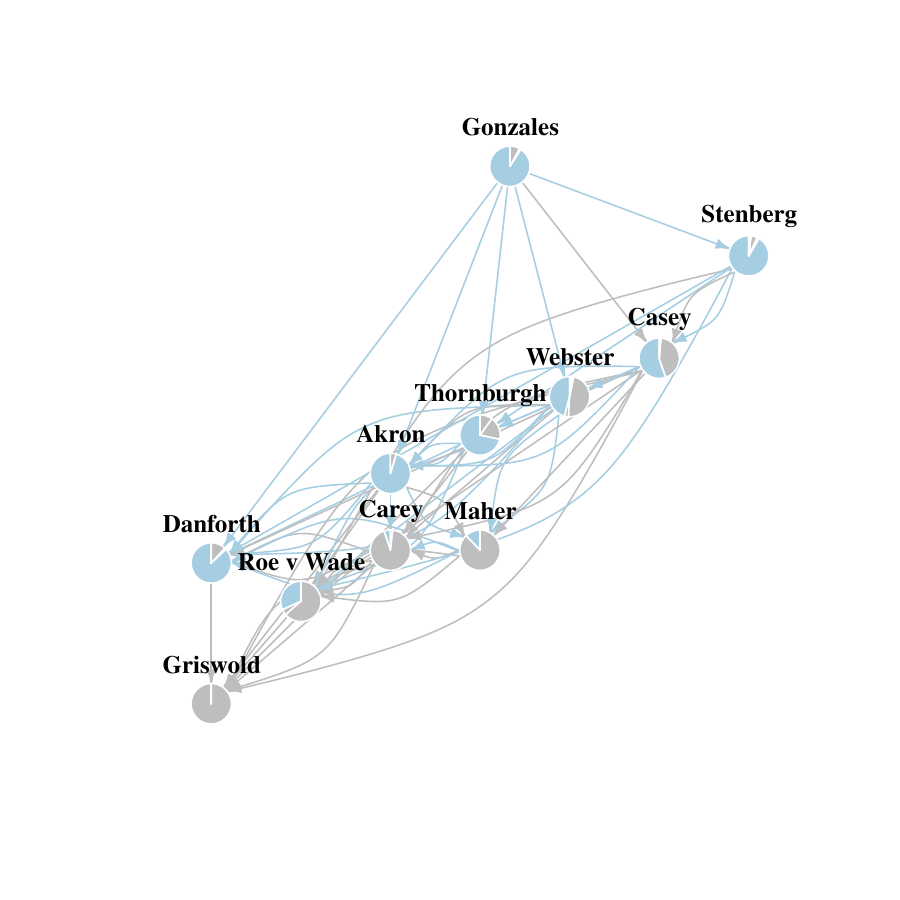}
         \caption{Regulation of Abortion Procedures}
	 \label{reprod_1}
     \end{subfigure}
     \caption{The citation network of 11 selected opinions on reproductive rights. The opinions are part of the SCOTUS subset on the privacy issue area. The left panel highlights the paragraphs and citations of \texttt{Constitutional Rights to Abortion} topic. The right panel colors the paragraphs and citations of \texttt{Regulation of Abortion Procedures} topic. The y-axis represents chronological order such that opinions placed lower indicate older in time and opinions placed in the upper part of the figure are more recent documents.}
        \label{reprod}
\end{figure}

The PCTM also allows us to visualize the evolution of topics over time by extracting topic-specific subnetworks.
We show how the topics on abortion (\texttt{Regulation of Abortion Procedure} and \texttt{Constitutional Rights to Abortion}) have changed over time.
To emphasize this aspect, we extract from our citation network 11 selected opinions on Reproductive rights in Figure \ref{reprod}.\footnote{The 11 opinions on reproductive rights are selected based on Figure 4 of \cite{clark2012genealogy}.}

Figure \ref{reprod} displays the topic structure of the 11 selected opinions on reproductive rights. We observe that the topic structure of the subnetwork is governed mostly by two topics -- \texttt{Regulation of Abortion Procedures} or \texttt{Constitutional Rights to Abortion}. 
Earlier opinions predominantly focus on the \texttt{Constitutional Rights to Abortion} topic, establishing the constitutional foundations through cases like \textit{Griswold v. Connecticut} (1965), which centered on privacy rights and reproductive autonomy. 
Later cases shifted toward the \texttt{Regulation of Abortion Procedures}, addressing specific implementation questions such as viability standards and the undue burden test. 
This evolution is exemplified in \textit{Planned Parenthood v. Casey} (1992), which both reaffirmed constitutional protections and established new regulatory frameworks, stating that ``The ability of women to participate equally in the economic and social life of the Nation has been facilitated by their ability to control their reproductive lives.''

While the discussion so far has focused on the substantive implications the PCTM can provide, we also provide discussion about the advantage of the PCTM in predicting new words and citations compared to existing models in Supplementary Information, Section~\ref{sec:predict_prob}.

\subsection{Document-importance in Topic-specific Citation Networks}
\begin{figure}[t!]
     \centering
     \begin{subfigure}[b]{0.3\textwidth}
         \centering
         \includegraphics[width=\textwidth]{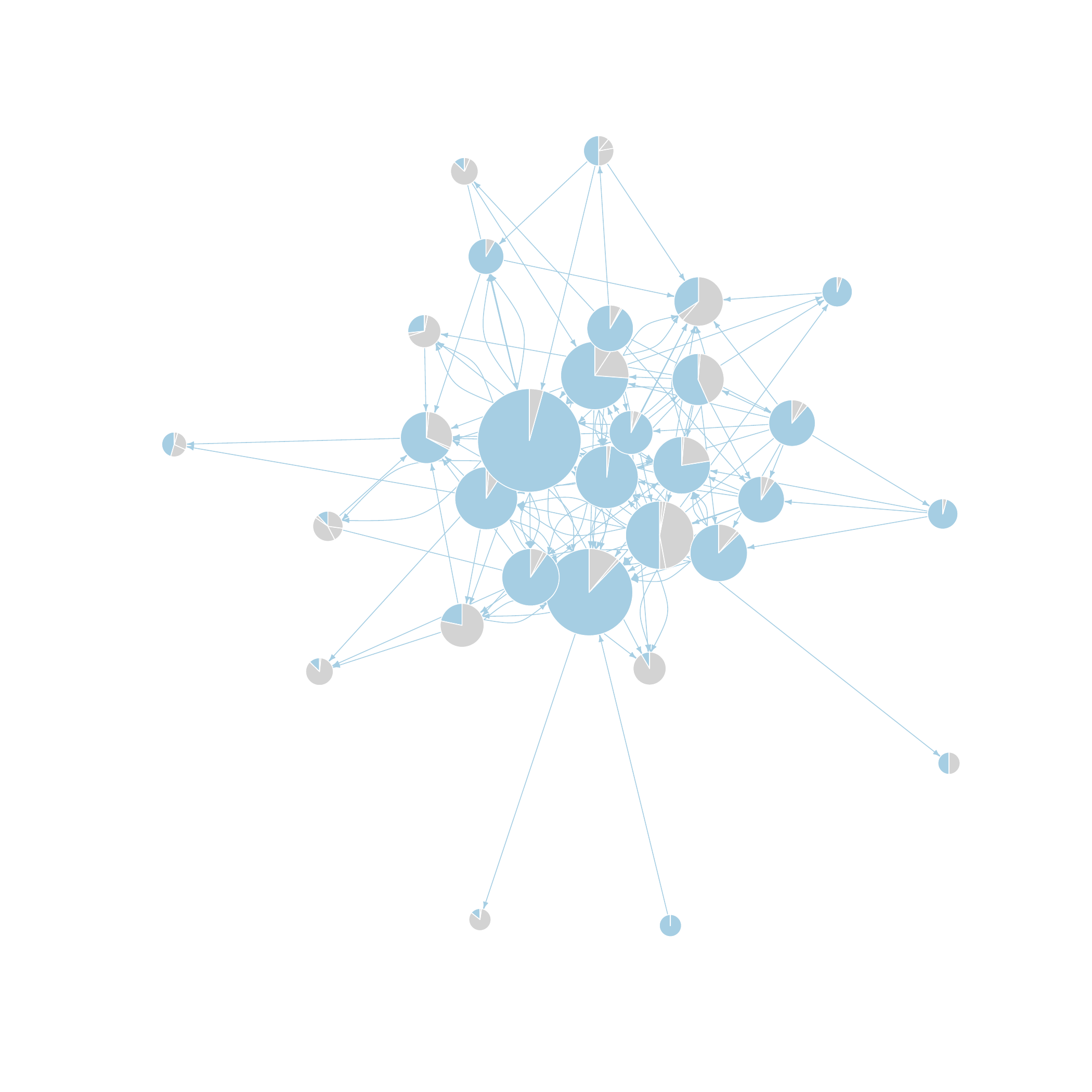}
         \caption{Regulation of Abortion}
         \label{privacy_topic1}
     \end{subfigure}
     \hfill
     \begin{subfigure}[b]{0.3\textwidth}
         \centering
         \includegraphics[width=\textwidth]{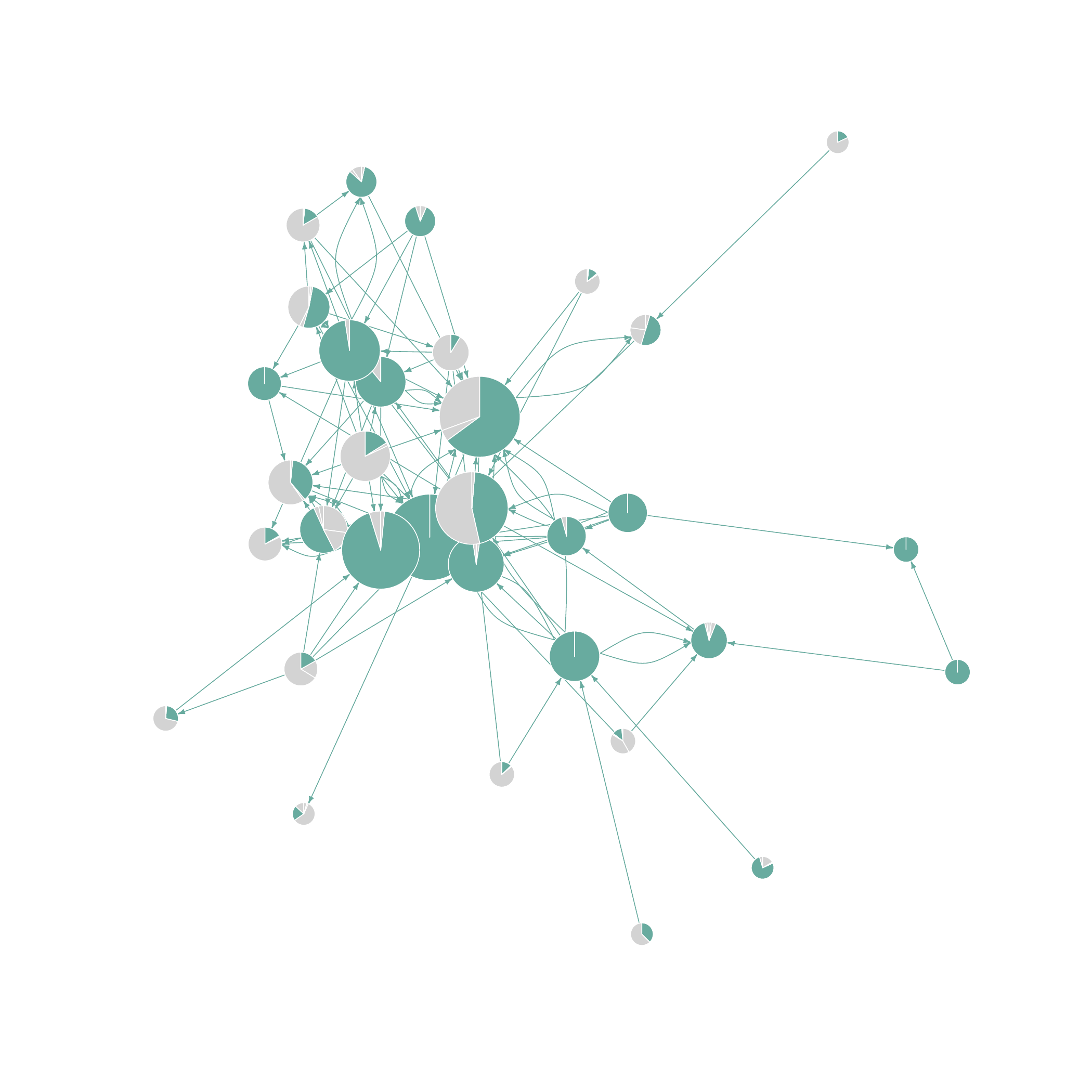}
         \caption{Const. Right Abortion}
	 \label{privacy_topic2}
     \end{subfigure}
     \hfill
     \begin{subfigure}[b]{0.3\textwidth}
         \centering
         \includegraphics[width=\textwidth]{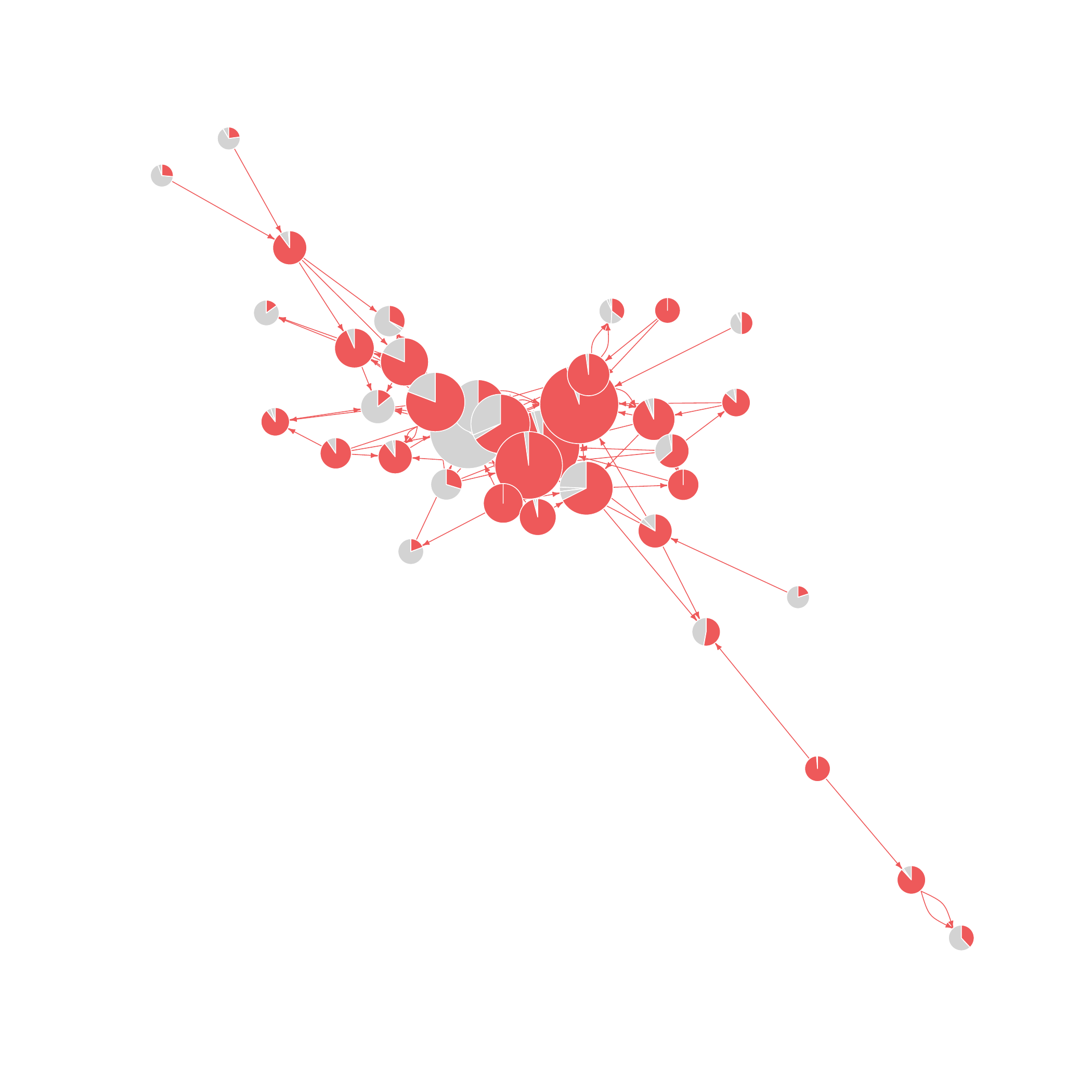}
         \caption{Disclosure \\ of Private Info}
	 \label{privacy_topic3}
     \end{subfigure}
        \caption{Subnetworks specific to each topic. The subnetworks are created by extracting opinions that either send or receive citations of the given topic. The topic-specific subnetworks can be useful in revealing whether and the extent to which topological features of the network varies by topic. For each subnetwork, paragraphs of other topics are all colored in gray for better visualization.}
        \label{subnetwork_privacy_figures}
\end{figure}

The PCTM's ability to assign topics to citations enables extraction of topic-specific subnetworks. 
We construct these subnetworks by including opinions that either send or receive citations of topic $k$. Figure~\ref{subnetwork_privacy_figures} displays the resulting subnetworks for three topics: \texttt{Regulation of Abortion Procedures}, \texttt{Constitutional Rights to Abortion}, and \texttt{Public Disclosure of Private Information}.

Topic-specific subnetworks represent citation patterns within distinct semantic domains, enabling the application of established network analysis methods to semantically coherent subsets of citations. These methods include the ``family tree of law'' approach developed by \cite{clark2012genealogy} and the importance score proposed in \cite{fowler2007network}. Here, we focus on Fowler et al.'s importance scores, which measure an opinion's precedential significance and predict its likelihood of future citations. 
Recognizing semantic differences, however, is critical when computing importance scores because the absence of a citation to a precedent could have two different meanings: that the given precedent does not carry much legal weight or that the given precedent addresses a completely distinct legal issue. 
To demonstrate the significance of semantic context in citation analysis, we compare importance scores computed on the complete network with those derived from topic-specific subnetworks.

The importance score has two parts based on their citation directions. The outward relevance score is based on the number of citations an opinion makes, evaluating the opinion's weight in referencing pertinent legal questions. An opinion with high outward relevance score cites many other opinions that are also deemed important and legally relevant. The inward relevance score is based on the number of citations an opinion receives from other opinions, gauging the extent to which it serves as the integral part of the law as a precedent. An opinion with high inward relevance score is cited by many other important and influential opinions. Since these scores are computed using eigenvectors, they are invariant to scales. In this light, \cite{fowler2007network} suggests using ranks of inward and outward relevance scores as the measure of importance for opinions as precedents. 

\begin{table}[t!]
\centering
\scriptsize
\begin{tabular}{llll}
 & Top 1 Inward-relevant & Top 2 Inward-relevant & Top 3 Inward-relevant \\ 
  \hline
  \hline
  \textit{All Topics} & Planned Parenthood v. Danforth & Roe v. Wade & Griswold v. Connecticut  \\   
  \hline
  \textit{Reg. Abortion} & Planned Parenthood v. Danforth & Colautti v. Franklin & Bellotti v. Baird \\ 
  \textit{Proc. Posture} & Renegotiation Board v. Bannercraft & Hickman v. Taylor & EPA v. Mink \\ 
  \textit{Const. Abortion} & Griswold v. Connecticut & Roe v. Wade & Eisenstadt v. Baird \\ 
  \textit{Speech \& Protest} & Schenck v. Pro-choice Network & Madsen v. Women's Health Center & Roe v. Wade \\ 
  \textit{Damage to Privacy} & Doe v. Chao & US ex rel. Touhy v. Ragen & US v. Reynolds \\ 
  \textit{Privacy v. Govnt.} & Vernonia v. Wayne & Chandler v. Miller & Whalen v. Roe \\ 
  \textit{Pub. Disclosure} & EPA v. Mink & Air Force v. Rose & NLRB v. Sears \\ 
   \hline
\end{tabular}
\caption{Top 3 most inward-relevant cases by topics. The inward relevance scores are computed following \cite{fowler2007network}.}
\label{privacy_inward}
\end{table}

In Table \ref{privacy_inward}, none of the topic-specific top 3 inward-relevant cases exactly match those that are from the entire citation network of the privacy cases. The top 3 inward-relevant for all topics (row 1) seem to be drawing information from two topics -- \texttt{Regulation of Abortion} and \texttt{Constitutional Rights to Abortion}. 
If one is interested in \texttt{Speech \& Protest}, for example, \textit{Schenck v. Pro-choice Network} is the most inward-relevant. \textit{Schenck v. Pro-choice Network} is an influential case that draws the line between public safety and free speech. In \textit{Schenck v. Pro-choice Network}, the SCOTUS concluded that the fifteen feet buffer zone between anti-abortion protestors and abortion clinics was constitutional, but deemed unconstitutional fifteen feet buffer zone between protestors and people seeking entrance to clinics. 
For \texttt{Public Disclosure of Information} topic, \textit{EPA v. Mink} is the most inward-relevant. The case addresses the disclosure of secret documents prepared for a scheduled underground nuclear test, gauging the balance between the Freedom of Information Act (1966) and national security matters. 
Both examples show that one can draw very different conclusions on which case is most inward-relevant, depending on the legal context and area.

\begin{table}[t!]
\centering
\scriptsize
\begin{tabular}{llll}
 & Top 1 Outward-relevant & Top 2 Outward-relevant  & Top 3 Outward-relevant  \\ 
  \hline
  \hline
  \textit{All Topics} & Hodgson v. Minnesota & Akron v. Akron Center & Webster v. Reproductive Health\\  
  \hline
  \textit{Reg. Abortion} & Akron v. Akron Center & Hodgson v. Minnesota & Webster v. Reproductive Health \\ 
  \textit{Proc. Posture}& NLRB v. Sears  & US v. Weber & DOI v. KWUPA \\ 
  \textit{Const. Abortion}  & Carey v. Population Services Int. & Planned Parenthood v. Casey & Hodgson v. Minnesota \\ 
  \textit{Speech \& Protest}  & Hill v. Colorado & Schenck v. Pro-choice Network & Roe v. Wade \\ 
  \textit{Damage to Privacy} & Federal Aviation Admin. v. Cooper & NASA v. Nelson & US ex rel. Touhy v. Ragen \\ 
  \textit{Privacy v. Govnt.}  & Board of Education v. Earls & Chandler v. Miller & Whalen v. Roe \\ 
  \textit{Pub. Disclosure} & DOJ v. Reporters Comm. & FBI v. Abramson & DOJ v. Tax Analysts \\ 
   \hline
\end{tabular}
\caption{Top 3 most outward-relevant cases by topics. The outward relevance scores are computed following \cite{fowler2007network}.}
\label{privacy_outward}
\end{table}

Table~\ref{privacy_outward} shows that while the top three outward-relevant cases in the complete citation network primarily reflect rankings from the \texttt{Regulation of Abortion} and \texttt{Constitutional Rights to Abortion} topics, different patterns emerge when examining specific topics. For instance, in the \texttt{Public Disclosure of Information} topic, \textit{Department of Justice v. Reporters Committee for the Freedom of Press} is the most outward-relevant case. The given case addresses whether the FBI should disclose criminal records to media outlets in the interest of public knowledge and safety. Together with Table~\ref{privacy_inward}, Table~\ref{privacy_outward} shows that legal context can be heterogeneous within the privacy issue area, and such semantic heterogeneity can lead to varying conclusions on the precedential importance of cases.

\subsection{Topic Prediction for a New Abortion Case}

This section presents additional results on a new controversial case regarding abortion. 
On June 24 2022, the Supreme Court made a landmark decision on abortion that invoked a nationwide controversy.
In the case, \textit{Dobbs v. Jackson Women's Health Organization}, the SCOTUS held that abortion is not a part of constitutional rights, and it conferred individual states the right to ban abortion. 
This case overturned both \textit{Roe v. Wade} and \textit{Planned Parenthood v. Casey}, the landmark precedents that have served as the legal basis for the constitutional rights to abortion.  
While qualitative reading of \textit{Dobbs v. Jackson Women's Health Organization} suggests that this case is a clear deviation from the recent trends in abortion rulings in many ways, it is difficult to demonstrate the deviations in a quantitative way. 

Using the PCTM, we examine how the topic structure of \textit{Dobbs v. Jackson Women's Health Organization} differs from the recent rulings on abortion in our corpus.
To do so, we computed the predicted probability of topics of the paragraphs in \textit{Dobbs v. Jackson Women's Health Organization} using the model fitted on our abortion corpus.
We first train the PCTM on the abortion corpus used in the above analysis and then computed the posterior predictive distribution of topics. 
The exact formula to obtain the posterior predictive probability is in Supplementary Information, Section~\ref{sec:predictive}.

To validate that the meaning of the topics is consistent in the new case, \textit{Dobbs v. Jackson Women's Health Organization}, we provide a qualitative analysis of the estimated topics by focusing on the paragraphs that cite the same precedent.
Table~\ref{dobbspara} presents two paragraphs that cite the same precedent, \textit{Planned Parenthood v. Casey} (505 U.S., 878), but with different estimated topics.
The left paragraph has the estimated topic \texttt{Constitutional Rights to Abortion} while the right paragraph has the topic \texttt{Regulation of Abortion Procedure}.
The left paragraph is an introductory paragraph of the judges' criticism of Casey's argument that abortion is a part of the liberty protected by the Fourteenth Amendment.
This is clearly related to whether abortion is a part of constitutional rights or not. 
By contrast, the right paragraph criticizes the ``undue burden'' test that Casey decides.
The undue burden test offers criteria about what kind of state regulations on abortion is prohibited.
Therefore, we can infer that this paragraph discusses a more specific issue about how states regulate abortions. 
By reading these paragraphs, we can verify that the interpretation of the topics in this new case match our interpretations of the topics in the abortion corpus.

\begin{table}[t!]
  \centering
  \begin{tabular}{| m{0.45\textwidth} | m{0.45\textwidth} | }
  \hline
  \textcolor{Aquamarine}{Constitutional Rights to Abortion} & \textcolor{SkyBlue}{Regulation of Abortion Procedures}\\
  \hline
  \hline
  We turn to Casey's bold assertion that the abortion right is an aspect of the ``liberty'' protected by the Due Process Clause of the Fourteenth Amendment. \textbf{505 U.S., at 846}
  &
  The Casey plurality tried to put meaning into the ``undue burden'' test by setting out three subsidiary rules [...] The first rule is that ``a provision of law is invalid, if its purpose or effect is to place a substantial obstacle in the path of a woman seeking an abortion before the fetus attains viability.'' \textbf{505 U.S., at 878}\\
  \hline
\end{tabular}
  \caption{Comparison of Paragraphs in \textit{Dobbs v. Jackson} with Different Estimated Topics on Abortion.\\
  Both paragraphs cite the same precedent, \textit{Planned Parenthood v. Casey} (505 U.S., 878), but with different estimated topics. 
  }
  \label{dobbspara}
\end{table}

\begin{figure}[t!]
  \includegraphics[width=\textwidth]{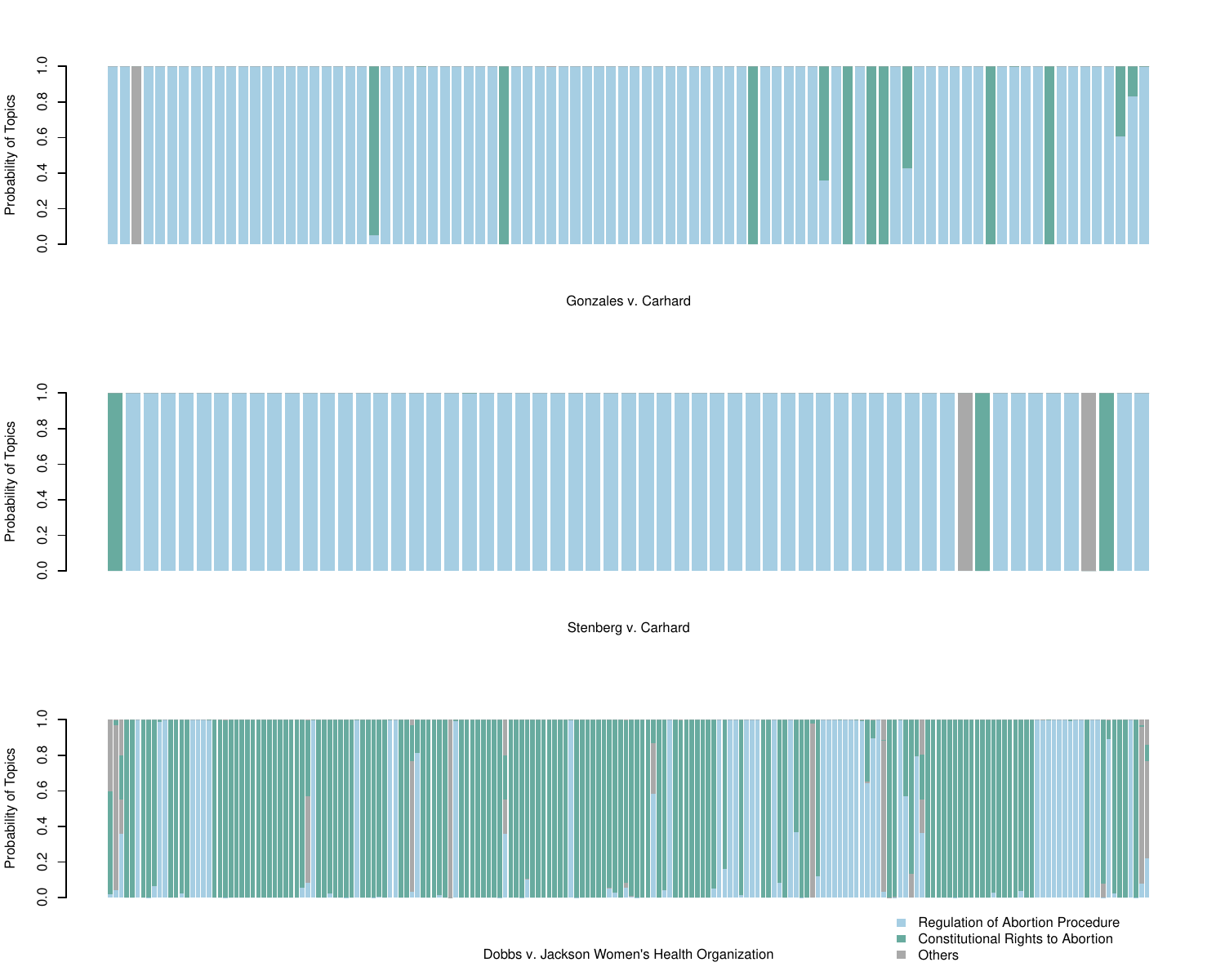}
  \caption{Predicted Probability of Topics for the Paragraphs of Dobbs v. Jackson.\\
    Each vertical bar represents a paragraph. 
    Each paragraph is colored according to the predicted probability of topics.
    We focus on two topics related to abortion: \texttt{Constitutional Rights to Abortion} and \texttt{Regulation of Abortion Procedure}.
    The case are \textit{Gonzales v. Cargard}, \textit{Stenberg v. Carhard}, and \textit{Dobbs v. Jackson Women's Health Organization}, from top to bottom.
    \textit{Dobbs v. Jackson Women's Health Organization} case have more paragraphs with \texttt{Constitutional rights to abortion} topic rather than \texttt{Regulation of abortion procedure} topic while the two recent precedents in our corpus, \textit{Gonzales v. Carhard} and \textit{Stenberg v. Carhard}, are the opposite. 
    This shows that \textit{Dobbs v. Jackson Women's Health Organization} goes against the recent trend in the abortion cases in our corpus, where the stronger emphasis is placed on how abortion can be regulated by the states instead of whether abortion is a part of the constitutional rights, as shown in \textit{Gonzales v. Carhard} and \textit{Stenberg v. Carhard}.
  }
  \label{abortion_topic_bar} 
\end{figure}

How do the topic structure in \textit{Dobbs v. Jackson Women's Health Organization} differ from the recent landmark cases in our corpus? 
For comparison, we also computed the predicted probability of the topics for the two recent precedents about abortion in our corpus: \textit{Gonzales v. Carhard} and \textit{Stenberg v. Carhard}, two recent landmark cases in abortion in our corpus.
Figure~\ref{abortion_topic_bar} shows the predicted probability of topics for each paragraph for the three cases on abortion, \textit{Gonzales v. Carhard}, \textit{Stenberg v. Carhard}, and \textit{Dobbs v. Jackson Women's Health Organization}, from top to bottom.
Each vertical bar represents a paragraph, and each bar is colored according to the predicted probability of topics.
Since we want to focus on the difference in the legal discourse regarding abortion, we focus our analysis on the two topics relevant to abortion: \texttt{Constitutional Rights to Abortion} or \texttt{Regulation of Abortion Procedure}. 
While more than 90\% of the paragraphs of both Gonzales and Stenberg are assigned with \texttt{Regulation of Abortion Procedure} topic, only 28\% of the paragraphs in \textit{Dobbs v. Jackson} are assigned with the \texttt{Regulation of Abortion Procedure} topic and 67\% of the paragraphs are assigned with \texttt{Constitutional Rights to Abortion}. 
This accurately reflects the fact that \textit{Dobbs v. Jackson Women's Health Organization} is distinct from the current trend in the abortion rulings in our corpus. 

\section{Concluding Remarks}
\label{sec:conclusion}

Social scientists often use citation networks to study how documents influence following documents in various domains, such as political science, international relations, and legal studies.
However, conventional approaches to analyzing citation networks often overlook the semantic context in which citations occur.
While existing studies use document-level labels to find the context of citations, this approach assumes that all citations within a document are made under the same context, which may lead to misunderstanding of how citations reflect the influence of documents.
To address this challenge, this paper proposes a novel joint model of text and citations, the paragraph-citation topic model.
The key innovation of the PCTM is to assign topics to paragraphs, which allows citations in different paragraphs to be associated with different topics.
After deriving a collapsed Gibbs sampler for inference, we applied the PCTM to the SCOTUS opinions on privacy issues to highlight the diversity of topics of citations within each document.
Also, the model uncovered informative subnetworks of the judicial opinions that shared citations with the same topic.

The applications of the PCTM need not be limited to citation networks of legal documents. 
The model will help address a number of important research questions in the analysis of document networks. 
For example, a researcher can use the latent citation propensity in the PCTM to understand the role of authors' gender in citation making in academic journals. 
Since academic articles address diverse scholarly subjects, capturing semantic contexts in the analysis of citation formation is critical, and can be properly addressed in our model.
Moreover, as the PCTM estimates topic-specific subnetworks of citations using information from both text and networks in a unified framework, it can be used together with established measures of networks, such as legal importance scores in \cite{fowler2007network}, to produce better academic insights.

\clearpage 

\singlespacing
\pdfbookmark[1]{References}{References}
\bibliography{nettext}

\onehalfspacing

\appendix
\renewcommand{\appendixname}{Supplementary Information }
\counterwithin{figure}{section}
\counterwithin{table}{section}

\section{Constructing SCOTUS Paragraph-document Citation Network}
\label{sec:data}

We construct a new dataset of the SCOTUS opinions that combines text and citation networks. 
The original data is obtained from the Caselaw Access Project, which allows public access to all official and published opinions at all levels of the US courts \citep{caselaw}.
The data contains the full text of majority and minority opinions in addition to their metadata, such as decision dates, reporter names, volumes in the reporter, and page numbers. 
We decided to focus on the text of majority opinions and discard minority opinions since minority opinions rarely receive recognition as legal precedents.
In total, the population data contains 24,000 cases with 749,888 paragraphs with the year ranging from 1834 to 2013. 

The document networks of the SCOTUS consist of two forms of datasets: text and citation networks. 
With respect to the text, we construct a ``paragraph''-feature matrix based on the population corpus. 
A paragraph feature matrix is similar to a common document-feature matrix, where a $(i,j)$ element of the matrix corresponds to the number of times a unique feature $j$ appears in a document $i$.
The only difference is that a paragraph-feature matrix uses paragraphs instead of documents as a unit. 
This is because our proposed model uses paragraphs as a unit of analysis. 
After tokenizing the corpus, we removed punctuations, symbols, special characters, numbers, and common English stopwords.\footnote{We used the set of English stopwords provided in \texttt{quanteda} package in \textsf{R} \citep{quanteda}.}
In addition to the common list of stopwords, we also removed legal terms that are common across the documents in our data such as ``court", ``state", ``law'' and, ``trial''.
After removing too frequent words and too rare words, the population paragraph-feature matrix contains 32,644 unique features.

The other component is a citation network.
While previous studies have constructed citation networks of the SCOTUS cases \citep{fowler2007network, clark2012genealogy}, their unit of analysis is at the document level while ours is at the paragraph level.
In other words, we want to form an adjacency matrix of $G\times N$ where $G$ is the number of paragraphs and $N$ is the number of documents, and the $(ip, j)$ element of the matrix is 1 if paragraph $p$ of document $i$ cites document $j$, and 0 otherwise. 
Since such data is not readily available, we constructed our own citation network of the SCOTUS cases by extracting citations from the text via regular expression matching. 
One of the challenges of this approach is that a citation is recorded by multiple reporters and appears in the paragraph as many times as the number of reporters.
To avoid complication, we focused on the citations to the official reporter, \textit{the United States Reports}, because this is the recommended and the most dominant citation method.
A citation to a case in the United States Reports typically has a relatively consistent format and thus is easier to be extracted through regular expression matching. 
For instance, a citation to \textit{Roe v. Wade} is typically written as \textit{Row v. Wade, 410 U.S. 113 (1973)}. 
Since we focus on the SCOTUS cases only, citations to and from outside of the corpus (e.g. citations to and from the Courts of Appeals and State courts) were discarded. 
This results in 191,173 citations in total.  

In this paper, we focus on a subset of this dataset for our applications. 
For our application, we decided to focus on cases on the Privacy issue area, which includes decisions about abortion and public disclosure of private information.
We chose this as our primary application data since existing literature on citation networks of the SCOTUS cases often focuses on this issue \citep{fowler2007network, clark2012genealogy}.
It is also an important application given the recent controversial decision that overruled the landmark case on constitutional rights to abortion. 
After we subset the data, we performed more preprocessing based on the term frequency within the subset.
More details of data pre-processing for each subset are available in the Supplementary Information document, Section A. 
This subset on the Privacy issue area consists of 106 documents with 4,669 paragraphs, 5,838 unique words, and 452 citations.

Results of topic models can be highly sensitive to how data is preprocessed \citep{denny2018text}. In addition to the simple preprocessing steps we introduced in Section 2, we removed words that appear very commonly across documents. The list of these words are ``Statue",``Supp",``Ann",``Rev",``Stat",``Judgment",``Reverse",``Follow",``Certiorari" and ``Opinion". While words such as ``Follow" or ``Reverse" could convey certain contexts, in legal opinions they are typically used to define how the drafted opinion stands in relation to precedents, and we believe they do not contain useful information with respect to topic discovery. In addition, words such as ``Supp" or ``Ann" are short words for Supplementary and Annex, which are specific collection of legal documents and thus removed for a better detection of topics. 

Since common terms can vary by different subsets, we made additional preprocessing for each subset we used for application of our model. For each subset, we removed terms that appear too frequently as well as terms that appear too infrequently. Terms too common across documents for Privacy subset include ``agent", ``month",``level" and ``unfair" and for Voting Rights subset the removed words include``Vote", ``Voter",``Elect" and ``Candid". For both subsets, terms that were too uncommon turned out to be simple typos or names of people or institutions such as ``Rawlinson".  The above process removed about 40\% of the terms. 

\clearpage

\section{Model inference: collapsed Gibbs sampler}
\label{sec:si_inference}
This section describes the details of the collapsed Gibbs sampler for the proposed model. 
Our model is as follows.

\begin{align}
  \begin{split}
	D_{ipj} &= \begin{cases}
		1 \text{ if } D_{ipj}^* \geq 0 \\
		0 \text { if } D_{ipj}^* < 0
	\end{cases}\\
    D_{ipj}^* &\sim \mathcal{N}(\pmb\tau^T\textbf{x}_{ipj},1)\quad \text{where}\ \textbf{x}_{ipj} = [1, \kappa_{j}^{(i)}, \eta_{j,z_{ip}}]\\
	\textbf{w}_{ip} &\sim \text{Multinomial}(N_{ip},\pmb\Psi_{z_{ip}})  \\
	z_{ip} &\sim \text{Multinomial}(1,\text{softmax}(\pmb\eta_i))  \\
	\pmb\Psi_k &\sim \text{Dirichlet}(\pmb\beta)  \\
	\pmb\eta_i &\sim \mathcal{N}(\pmb\mu,\pmb\Sigma)  \\
	\pmb\mu &\sim \mathcal{N}(\pmb\mu_0, \pmb\Sigma_0)  \\
	\pmb\tau &\sim \mathcal{N}(\pmb\mu_{\tau},\pmb\Sigma_{\tau})
  \end{split}
\end{align}

The full posterior is denoted as follows.
{\small
\begin{align}
	p(\pmb\eta,\pmb\Psi,\textbf{Z},\pmb\tau|\textbf{W},\textbf{D}) \propto p(\pmb\mu|\pmb\mu_0,\pmb\Sigma_0)p(\pmb\tau|\pmb\mu_{\tau},\pmb\Sigma_{\tau})p(\pmb\eta|\pmb\mu,\pmb\Sigma)p(\pmb\Psi|\pmb\beta)p(\textbf{Z}|\pmb\eta)p(\textbf{W}|\pmb\Psi,\textbf{Z})p(\textbf{D}|\textbf{D}^*)p(\textbf{D}^*|\pmb\tau,\pmb\eta,\textbf{Z},\textbf{D})
\end{align}
}

Unfortunately, the inference of the given posterior distribution is hard due to the non-conjugacy between normal prior for $\pmb\eta$ and the logistic transformation function \citep{blei2007correlated}. Variational inference is the most frequently employed tool to address this problem, with an additional advantage of computational speed. However, obtained parameters are for the variational distribution which is an approximation to the target posterior. Moreover, the quality of the approximation is often not sufficiently explored (Add citations here). 

To remedy this problem, we follow the recent advances in the inference of CTM models \citep{held2006bayesian,chen2013scalable,linderman2015dependent}. We first partially collapse the posterior distribution by integrating out $\pmb\Psi$. Then we introduce an auxiliary Polya-Gamma variable $\pmb\lambda$ and augment the collapsed posterior. Partial collapsing and data augmentation enables us to use Gibbs sampling which is known to produce samples that converge to the exact posterior. 

With $\pmb\Psi$ integrated out, our new posterior is proportional to
{\small
\begin{align}
	\int_{\pmb\Psi}p(\pmb\eta,\pmb\Psi,\textbf{Z},\pmb\tau|\textbf{W},\textbf{D}) \propto p(\pmb\mu|\pmb\mu_0,\pmb\Sigma_0)p(\pmb\tau|\pmb\mu_{\tau},\pmb\Sigma_{\tau})p(\pmb\eta|\pmb\mu,\pmb\Sigma)p(\textbf{Z}|\pmb\eta)p(\textbf{W}|\textbf{Z})p(\textbf{D}|\textbf{D}^*)p(\textbf{D}^*|\pmb\tau,\pmb\eta,\textbf{Z},\textbf{D})
\end{align}
}
\subsection{Derivation of the conditional distribution for \textbf{Z}}
For $ip$th paragraph, the conditional distribution of $z_{ip}$ is
{\small
\begin{align}   
    p(z_{ip}^k=1|\mathbf{Z}_{-ip},\pmb\eta,\mathbf{W},\mathbf{D}^*) &\propto p(z_{ip}^k=1|\pmb\eta_i)p(\mathbf{W}_{ip}|z_{ip}^k=1,\mathbf{Z}_{-ip},\mathbf{W}_{-ip}) \prod_{j=1}^{i-1}p(D_{ipj}^*|z_{ip}^k=1,\textbf{Z}_{-ip},\pmb\tau,\pmb\eta,\kappa)
\end{align}
}
The first term is $\frac{\text{e}^{\eta_{ik}}}{\sum_{l}\text{e}^{\eta_{il}}}$ which is proportional to $\text{e}^{\eta_{ik}}$.

The form of second term warrants further elaboration. Integrating out $\pmb\Psi$ as
\begin{align}
	p(\textbf{W}|\textbf{Z}) &= \int_{\pmb\Psi} p(\textbf{W},\pmb\Psi|\textbf{Z})d\pmb\Psi \nonumber \\
	&= \int_{\pmb\Psi} p(\textbf{W}|\pmb\Psi,\textbf{Z})p(\pmb\Psi|\textbf{Z})d\pmb\Psi \nonumber \\
	&= \int_{\pmb\Psi} p(\textbf{W}|\pmb\Psi,\textbf{Z})p(\pmb\Psi)d\pmb\Psi
\end{align}
for $ip$th paragraph with $k$th topic yields the following.
\begin{align}
    p(\mathbf{W}_{ip}|z_{ip}^k=1,\mathbf{Z}_{-ip},\mathbf{W}_{-ip}) &\propto \int_{\pmb{\Psi}_k}\Psi_{k1}^{\beta_1-1}\Psi_{k2}^{\beta_2-1} ... \Psi_{kV}^{\beta_V-1} \prod_v \Psi_{kv}^{\sum_{l=1}^{n_{ip}} \mathbb{I}(W_{ipl}=v)} \nonumber \\
    & \times \prod_v \prod_{(i^{'},p^{'}) \neq (i,p)} \Psi_{kv}^{\sum_{l=1}^{n_{i^{'}p^{'}}} \mathbb{I}(W_{i^{'}p^{'}l}=v)\mathbb{I}(z_{i^{'}p^{'}}^k=1)} d\pmb{\Psi}_k
\end{align}

Here, $N_{ip}$ denotes the total number of words in $ip$th paragraph, and $n_{ip}$ denotes the total number of unique words in $ip$th paragraph. Let $C_k^v = \sum_{i=1}^N\sum_{p=1}^{N_{ip}}\sum_{l=1}^{n_{ip}}\mathbb{I}(W_{ipl}=v)\mathbb{I}(z_{ip}^k=1)$, and $c_{k,ip}^v = \sum_{l=1}^{n_{ip}}\mathbb{I}(W_{ipl}=v)\mathbb{I}(z_{ip}^k=1)$ then the above can be simplified as
\begin{align}
    p(\mathbf{W}_{ip}|z_{ip}^k=1,\mathbf{Z}_{-ip},\mathbf{W}_{-ip}) &\propto \int_{\pmb{\Psi}_k}\Psi_{k1}^{\beta_1+c_{k,ip}^1+ c_{k,-ip}^1-1}\Psi_{k2}^{\beta_2+c_{k,ip}^2+ c_{k,-ip}^2-1} ... \Psi_{kV}^{\beta_V+c_{k,ip}^V+ c_{k,-ip}^V-1} d\pmb{\Psi}_k \nonumber \\
    &= \frac{\prod_v \Gamma(\beta_v + c_{k,ip}^v + c_{k,-ip}^v)}{\Gamma(\sum_v \beta_v + c_{k,ip}^v + c_{k,-ip}^v)}
\end{align}

Imagine a paragraph of 3 words $\mathbf{W}_{ip} = \{1,1,3\}$, two of the first word and one of the third word. Then
\begin{align}
    p(\mathbf{W}_{ip}|z_{ip}^k=1,\mathbf{Z}_{-ip},\mathbf{W}_{-ip}) &\propto \frac{\prod_v \Gamma(\beta_v + c_{k,ip}^v + c_{k,-ip}^v)}{\Gamma(\sum_v \beta_v + c_{k,ip}^v + c_{k,-ip}^v)}
\end{align}

The numerator is 
\begin{align}
    \Gamma(\beta_1 + 2 + c_{k,-ip}^1)\Gamma(\beta_3 + 1 + c_{k,-ip}^3) \times \prod_{v \neq (1,3)}\Gamma(\beta_v + c_{k,-ip}^v) \nonumber \\
    = (\beta_1 + 1 + c_{k,-ip}^1)(\beta_1 + c_{k,-ip}^1)(\beta_3 + c_{k,-ip}^3) \times \prod_v \Gamma(\beta_v + c_{k,-ip}^v)
\end{align}

In the same sense, the denominator is
{\small
\begin{align}
    \Gamma(3 + \sum_v \beta_v + c_{k,-ip}^v) = (2+\sum_v \beta_v + c_{k,-ip}^v)(1+\sum_v \beta_v + c_{k,-ip}^v)(\sum_v \beta_v + c_{k,-ip}^v)\Gamma(\sum_v \beta_v + c_{k,-ip}^v)
\end{align}
}

Rearrange the above and we have
\begin{align}
    \frac{(\beta_1 + 1 + c_{k,-ip}^1)(\beta_1 + c_{k,-ip}^1)(\beta_3 + c_{k,-ip}^3)}{(2+\sum_v \beta_v + c_{k,-ip}^v)(1+\sum_v \beta_v + c_{k,-ip}^v)(\sum_v \beta_v + c_{k,-ip}^v)} \times \frac{\prod_v \Gamma(\beta_v + c_{k,-ip}^v)}{\Gamma(\sum_v \beta_v + c_{k,-ip}^v)}
\end{align}

The second term does not depend on $z_{ip}^k$. Then for $\textbf{W}_{ip}=\{1,1,3\}$, we have 
\begin{align}
    p(\mathbf{W}_{ip}|z_{ip}^k=1,\mathbf{Z}_{-ip},\mathbf{W}_{-ip}) \propto \frac{(\beta_1 + 1 + c_{k,-ip}^1)(\beta_1 + c_{k,-ip}^1)(\beta_3 + c_{k,-ip}^3)}{(2+\sum_v \beta_v + c_{k,-ip}^v)(1+\sum_v \beta_v + c_{k,-ip}^v)(\sum_v \beta_v + c_{k,-ip}^v)}
\end{align}

If a paragraph consists of only one word such that $W_{ip}=l$, the above changes to 
\begin{align}
    p(\mathbf{W}_{ip}|z_{ip}^k=1,\mathbf{Z}_{-ip},\mathbf{W}_{-ip}) \propto \frac{\beta_l + c_{k,-ip}^l}{\sum_v \beta_v + c_{k,-ip}^v}
\end{align}
which matches with the form for the equivalent part in collapsed Gibbs for LDA \citep{porteous2008fast,xiao2010efficient,asuncion2012smoothing}.

The third term $p(D_{ipj}^*|z_{ip}^k=1,\textbf{Z}_{-ip},\pmb\tau,\pmb\eta,\pmb\kappa) = \text{exp}\{ -\frac{1}{2} \big(D_{ipj}^* - (\tau_0 + \tau_1\kappa_j^{(i)} + \tau_2\eta_{j,z_{ip}}) \big)^2 \}$ is proportional to 
\begin{align}
\text{exp}\Bigg\{-\frac{1}{2}\Big(\tau_2^2\eta_{jk}^2 + 2\big(\tau_0\tau_2 + \tau_1\tau_2\kappa_j^{(i)} - \tau_2 D_{ipj}^* \big)\eta_{jk} \Big) \Bigg\}
\end{align}

\subsection{Derivation of the conditional distribution for $\pmb\eta$}
\begin{align}
    p(\pmb{\eta}|\textbf{Z},\textbf{W},\textbf{D}) &= \prod_{i=1}^N \Big( \prod_{p=1}^{N_i} p(z_{ip}|\pmb\eta_i) \Big) \mathcal{N}(\pmb{\eta}_i|\pmb\mu,\pmb\Sigma) \prod_{p=1}^{N_i}\prod_{j=1}^{i-1}p(D^*_{ipj}|\kappa,\pmb\eta_i,\mathbf{Z})\nonumber \\
    &= \prod_{i=1}^N \Big(\prod_{p=1}^{N_i}\frac{e^{\eta_{i,z_{ip}}}}{\sum_{j=1}^K e^{\eta_{ij}}} \Big) \mathcal{N}(\pmb{\eta}_i|\pmb\mu,\pmb\Sigma) \prod_{p=1}^{N_i}\prod_{j=1}^{i-1}p(D^*_{ipj}|\kappa,\pmb\eta_i,\mathbf{Z})
\end{align}

Following \cite{held2006bayesian}, the likelihood for $\eta_{ik}$ conditioned on $\eta_{i,-k}$ is
\begin{align}
    \ell(\eta_{ik}|\eta_{i,-k}) &= \prod_{p=1}^{N_i} \Big(\frac{e^{\rho_{ik}}}{1+e^{\rho_{ik}}} \Big)^{z_{ip,k}} \Big(\frac{1}{1+e^{\rho_{ik}}} \Big)^{1-z_{ip,k}} \nonumber \\
    &= \frac{(e^{\rho_{ik}})^{t_{ik}}}{(1+e^{\rho_{ik}})^{N_i}}
\end{align}
where $\rho_{ik} = \eta_{ik} - \text{log}(\sum_{l \neq k} e^{\eta_{il}})$ and $t_{ik} = \sum_{p=1}^{N_i}\mathbb{I}(z_{ip}=k)$.

Then
\begin{align}
    p(\eta_{ik}|\eta_{i,-k},\mathbf{Z},\mathbf{W},\mathbf{D},\pmb\tau) \propto \ell(\eta_{ik}|\eta_{i,-k})\mathcal{N}(\eta_{ik}|\nu_{ik},\sigma_k^2)p(D^*|\pmb\eta,\pmb\tau,\mathbf{Z})
\end{align}

where
\begin{align}
    \nu_{ik} &= \mu_k - \Lambda_{kk}^{-1}\pmb\Lambda_{k,-k}(\pmb\eta_{i,-k}-\pmb\mu_{i,-k}) \nonumber \\
    \sigma_k^2 &= \pmb\Lambda_{kk}^{-1} \nonumber \\
    \pmb\Lambda &= \pmb\Sigma^{-1}
\end{align}

The third term can be rewritten with respect to $\pmb\eta$ as 
\begin{align}
    p(\mathbf{D}^*|\pmb\eta,\pmb\tau,\mathbf{Z}) &=  \prod_i\prod_p\prod_{j=1}^{i-1}\text{exp}\Bigg\{-\frac{1}{2}\big(D_{ipj}^* - (\tau_0 + \tau_1\kappa_j^{(i)} + \tau_2\eta_{j,z_{ip}})\big)^2 \Bigg\} \nonumber \\
    &\propto \prod_i\prod_p\prod_{j=1}^{i-1}\text{exp}\Bigg\{-\frac{1}{2(1/\tau_2^2)}\Big(\eta_{j,z_{ip}}^2 - 2\frac{D_{ipj}^* - \tau_0 - \tau_1\kappa_j^{(i)}}{\tau_2}\eta_{j,z_{ip}} \Big) \Bigg\}  \nonumber \\
    &\propto \prod_i\prod_p\prod_{j=1}^{i-1} \mathcal{N}(\eta_{j,z_{ip}}|\mu_{ipj}^*,\frac{1}{\tau_2^2}) \nonumber \\
    &= \prod_i\prod_p\prod_{j=1}^{i-1}\prod_k \mathcal{N}(\eta_{jk}|\mu_{ipj}^*,\frac{1}{\tau_2^2})^{\mathbb{I}(z_{ip}=k)}
\end{align}
where $\mu_{ipj}^* = \frac{D_{ipj}^* - \tau_0 - \tau_1\kappa_j^{(i)}}{\tau_2}$.
We notice that the above can be rewritten as a product of univariate normal distributions such that
\begin{align}
    & \prod_k\prod_{s=i+1}^N\prod_{p=1}^{N_s}\mathcal{N}(\eta_{ik}|\mu_{spi}^*,{\sigma^2}^*)^{\mathbb{I}(z_{sp}=k)} \nonumber \\
    &\equiv  \prod_{k=1}^K \mathcal{N}(\eta_{ik} | m_{ik},V_{i,kk})
\end{align}

$\mathbf{V}_i$ is a diagnoal matrix with the $k$th diagonal entry of the inverse of $\mathbf{V}_i$ (or $\mathbf{V}_i^{-1})$ as
\begin{align}
    V_{i,kk}^{-1} &= \frac{1}{{\sigma^2}^*} \sum_{s=i+1}^N\sum_{p=1}^{N_s}\mathbb{I}(z_{sp} = k) \nonumber \\
    &= \tau_2^2 \sum_{s=i+1}^N\sum_{p=1}^{N_s}\mathbb{I}(z_{sp} = k)
\end{align}

The $k$th entry of $\mathbf{m}_i$ then is
\begin{align}
    m_{ik} &= \frac{\tau_2^2\sum_{s=i+1}^N\sum_{p=1}^{N_s}\mu_{spi}^*\mathbb{I}(z_{sp}=k)}{V_{i,kk}^{-1}} \nonumber \\
    &= \frac{\sum_{s}\sum_{p}\mu_{spi}^*\mathbb{I}(z_{sp}=k)}{\sum_s\sum_{p}\mathbb{I}(z_{sp}=k)}
\end{align}

Then the $\eta$ conditional is
\begin{align}
    p(\eta_{ik}|\eta_{i,-k},\mathbf{Z},\mathbf{W},\mathbf{D},\pmb\tau) \propto \ell(\eta_{ik}|\eta_{i,-k})\mathcal{N}(\eta_{ik}|\nu_{ik},\sigma_k^2)\mathcal{N}(\eta_{ik}|m_{ik},V_{i,kk})
\end{align}

We now introduce Polya-Gamma augmentation such that 
{\footnotesize
\begin{align}
    p(\eta_{ik}|\eta_{i,-k},\mathbf{Z},\mathbf{W},\mathbf{D},\pmb\tau,\lambda_{ik}) &\propto \text{exp}\{(t_{ik}-\frac{N_i}{2})\rho_{ik} - \frac{\lambda_{ik}}{2}\rho_{ik}^2 \}\mathcal{N}(\eta_{ik}|\nu_{ik},\sigma_k^2)\mathcal{N}(\eta_{ik}|m_{ik},V_{i,kk}) \nonumber \\
    &\propto \mathcal{N}(\eta_{ik}|\frac{t_{ik}-N_i/2}{\lambda_{ik}}+\text{log}(\sum_{l\neq k}e^{\eta_{il}}),1/\lambda_{ik})\mathcal{N}(\eta_{ik}|\nu_{ik},\sigma_k^2)\mathcal{N}(\eta_{ik}|m_{ik},V_{i,kk})
\end{align}
}
\normalsize
Summing all of the above, the conditional distribution of $\eta_{ik}$ is 
\begin{align}
    p(\eta_{ik}|\eta_{i,-k},\mathbf{Z},\mathbf{W},\mathbf{D},\pmb\tau,\lambda_{ik}) &\propto \mathcal{N}(\eta_{ik}|\tilde{\mu}_{ik},\tilde{\sigma}_k^2)
\end{align}
where
\begin{align}
    \tilde{\sigma}_k^2 &= (\sigma_k^{-2} + \lambda_{ik} + v_{i,kk}^{-1})^{-1} \nonumber \\
    \tilde{\mu}_{ik} &= \tilde{\sigma}_k^2 \big( v_{i,kk}^{-1}m_{ik} + \sigma_k^{-2}\nu_{ik} + t_{ik} - \frac{N_i}{2} + \lambda_{ik}\text{log}(\sum_{l\neq k}e^{\eta_{il}}) \big)
\end{align}

\subsection*{Derivation of conditional distribution for $\pmb\lambda$}
The Gibbs sampling for the augmentation variable $\pmb\lambda$ is obtained by collecting terms that include $\pmb\lambda_i$ in the joint of $\pmb{z}_i$ and $\pmb\eta_i$.
\begin{align}
    p(\lambda_{ik}|\mathbf{Z},\mathbf{W},\pmb\eta) &\propto PG(N_i,\rho_{ik})
\end{align}

\subsection{Derivation of conditional distribution for $\textbf{D}^*$}
{\footnotesize
\begin{align}
    p(D^*_{ipj}|\pmb\eta,\mathbf{Z},\pmb\tau, \mathbf{D}) &\propto \begin{cases} TN_{(0,\infty)}(\tau_0+\tau_1\kappa_j^{(i)}+\tau_2\eta_{j,z_{ip}},1) & \text{ if } D_{ipj} = 1 \\
    TN_{(-\infty,0]}(\tau_0+\tau_1\kappa_j^{(i)}+\tau_2\eta_{j,z_{ip}},1) & \text{ if } D_{ipj} = 0
    \end{cases}
\end{align}
}

\subsection{Derivation of conditional distribution for $\pmb\tau$}

Let $\mathbf{x}_{ipj} = [1,\kappa_j^{(i)},\eta_{j,z_{ip}}]^T$ and $\pmb\tau = [\tau_0,\tau_1,\tau_2]^T$
\begin{align}
    p(\pmb\tau|\pmb\eta,\mathbf{Z},\mathbf{D}^*) &\propto exp\Bigg\{-\frac{1}{2} \sum_{ipj}\Big(D_{ipj}^* - \mathbf{x}_{ipj}^T\pmb\tau \Big)^2 \Bigg\} N(\pmb\mu_{\pmb\tau},\Sigma_{\tau})\nonumber \\
    &\propto N(\tilde{\pmb\tau},\tilde{\Sigma_{\tau}})
\end{align}

where $\tilde{\Sigma_{\tau}} = \Bigg(\Big(\sum_{ipj}\mathbf{x}_{ipj}\mathbf{x}_{ipj}^T \Big) + \Sigma_{\tau}^{-1} \Bigg)^{-1}$ 
and $\tilde{\tau} = \tilde{\Sigma_{\tau}}\Bigg(\Big(\sum_{ipj}\mathbf{x}_{ipj}^TD_{ipj}^*\Big) + \Sigma_{\tau}^{-1}\pmb\mu_{\pmb\tau}\Bigg)$

\subsection{Recovering $\pmb\Psi$}
\label{subsec:Psi}
We estimate the integrated out parameter $\pmb\Psi$ from our posterior samples as follows.
\begin{align}
	\hat{\Psi}_{kv} =  \frac{\sum_{i}\sum_{p} \big(\beta_v + \mathbb{I}(z_{ip}^k=1)\text{W}_{ip,v} \big)}{\sum_{i}\sum_{p}\sum_l \big(\beta_l + \mathbb{I}(z_{ip}^k=1)\text{W}_{ip,l}\big)}
\end{align}

\clearpage

\section{Initialization strategy for collapsed Gibbs sampler}
\label{sec:init}

Similar to other topic models, the PCTM contains a number of parameters for an estimation which increases the concern for multi-modality of the parameter space. Bad initial values can negatively impact the convergence of mcmc chains to the posterior distribution. Initial values distant from the global mode of the parameter space results in slow convergence. Also, for models with high dimensional parameter space, such as LDA or PCTM, bad initial values increase the possibility of the mcmc chain being stuck at local modes that offer suboptimal interpretations at best. To address these concerns, we propose to fit LDA with variational EM to obtain reasonable initial values for $\pmb\eta$, then use them to generate reasonable initial values for other parameters ($\textbf{Z},\pmb\lambda,\textbf{D}^*,\pmb\tau$). 

We first fit LDA with variational EM on document-level document-feature matrix to obtain $\hat{\pmb\theta}$. For $i$th document, 
\begin{align}
    z_{ip}^{(0)} &\sim \text{Categorical}(\hat{\pmb\theta}_i) \quad \forall p=1,2,...,N_i \nonumber \\
    \pmb\eta_{i}^{(0)} &= \text{log}(\hat{\pmb\theta}_i/\hat{\theta}_{iK})
\end{align}

Set $\tilde{\tau}_0$, or the sparsity parameter, using the observed density of the citation matrix and randomly draw the other two parameters as 
\begin{align}
    \tilde{\tau}_0 &= \frac{1}{2}\text{log}(\text{density(\textbf{D})}) \nonumber \\
    \tilde{\tau}_1,\tilde{\tau}_2 &\sim \text{unif}(0,1) 
\end{align}

Sample $\textbf{D}^*$ using the above parameters
\begin{align}
    {D_{ipj}^*}^{(0)} &\sim TN_{(-\infty,0)}(\tilde{\tau}_0 + \tilde{\tau}_1\kappa_j^{(i)} + \tilde{\tau}_2 \eta_{j,z_{ip}^{(0)}}^{(0)},1) \quad \text{ if } D_{ipj} = 0 \nonumber \\
    {D_{ipj}^*}^{(0)} &\sim TN_{[0,\infty)}(\tilde{\tau}_0 + \tilde{\tau}_1\kappa_j^{(i)} + \tilde{\tau}_2 \eta_{j,z_{ip}^{(0)}}^{(0)},1) \quad \text{ if } D_{ipj} = 1
\end{align}

Then set $\pmb\tau^{(0)}$ again using MLE
\begin{align}
    \pmb\tau^{(0)} = (\sum_{ipj}\textbf{x}_{ipj}^{(0)}{\textbf{x}_{ipj}^{(0)}}^T)^{-1} (\sum_{ipj}{\textbf{x}_{ipj}^{(0)}}^T{D_{ipj}^*}^{(0)})
\end{align}
where $\textbf{x}_{ipj}^{(0)} = \{1, \kappa_{j}^{(i)}, \eta_{j,z_{ip}^{(0)}}^{(0)} \}$

Finally, set the values of $\pmb\lambda^{(0)}$ by
\begin{align}
    \lambda_i^{(0)} \sim \text{PG}(N_i,\pmb\eta_i^{(0)})
\end{align}

\clearpage

\section{Simulation Results}
\label{sec:simulation}

\subsection{MCMC Plots of Key Parameters}
\begin{figure}[!ht]
     \centering
         \includegraphics[width=\textwidth]{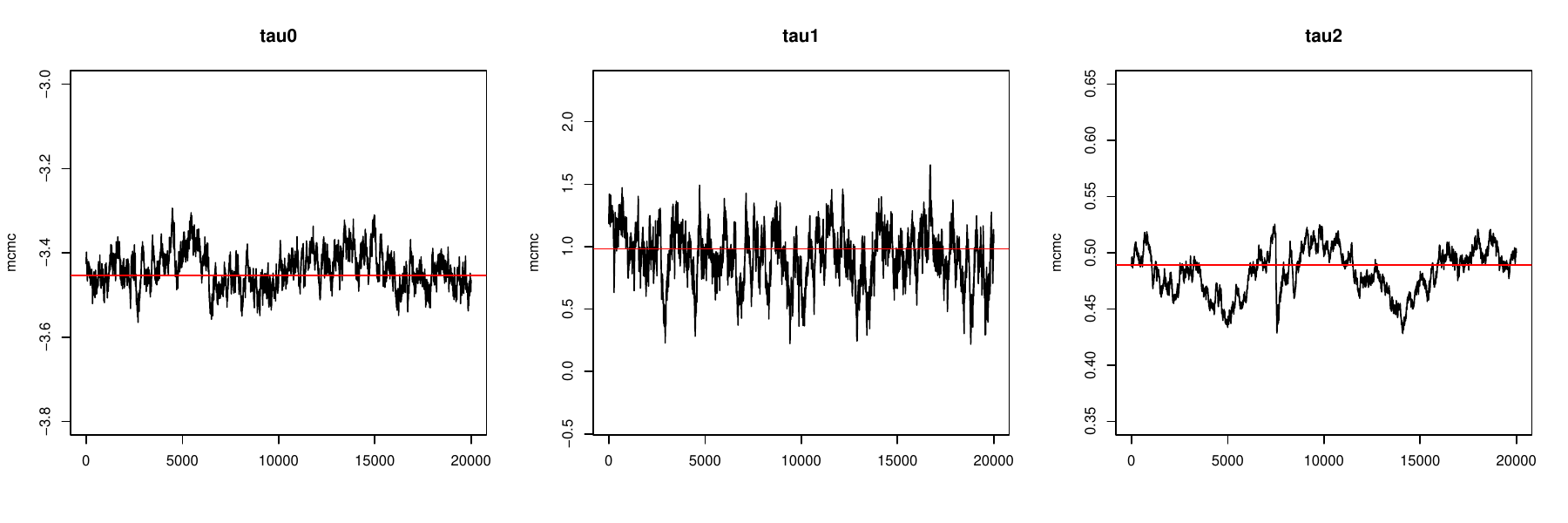}
         \caption{MCMC convergence of $\pmb\tau$ posterior samples in simulation. Horizontal red line indicates the true values of $\pmb\tau$.}
	 \label{sim_tau_mcmc}
\end{figure}

\begin{figure}[!ht]
     \centering
         \includegraphics[width=\textwidth]{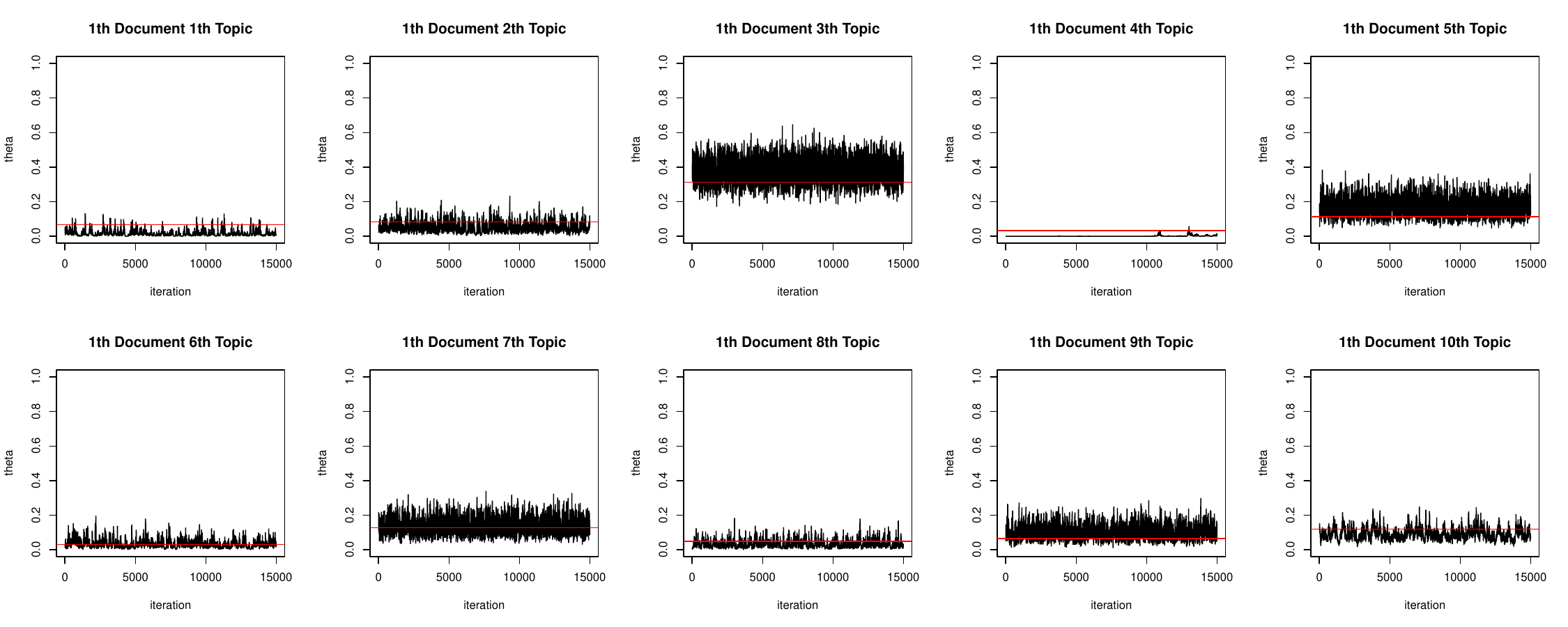}
         \caption{MCMC convergence of $\pmb\theta$ parameters for the first document. $\pmb\theta$ values are obtained by transforming the posterior samples of $\pmb\eta$ of the corresponding document. Horizontal red line indicates the true values of $\pmb\theta$ for the first document for each topic. We do not display the MCMC convergence for other documents, but all documents show similar level of convergence to the true value of $\pmb\theta$.}
	 \label{sim_theta_mcmc}
\end{figure}

\subsection{Recovery of the True Latent Variables}
We generate 100 simulation datasets with similar sizes as our application datasets. Specifically, we set the simulation datasets to have about equal number of documents, paragraphs, unique words and words.\footnote{106 documents, an average of 44 paragraphs per document, 5838 unique words, and an average of 51 words per paragraph.} Citations are generated based on the hyperparameters we input, and we set them so that the number of citations will be similar to those in our application data. This exercise gives us some evidence on the validity of our results on the application datasets.

We show that the PCTM can recover the true parameters from random initialization using our Gibbs sampler. 
We fit the PCTM on one of the simulation datasets while the initial parameters of the paragraph topic, $\mathbf{Z}$, and the distribution of topics, $\pmb\eta$, are randomly initialized.  
Then, we compare the estimated paragraph topics and the distribution of topics with the true values of those parameters.

\begin{figure}[ht!]
     \centering
     \begin{subfigure}[b]{0.4\textwidth}
         \centering
         \includegraphics[width=\textwidth]{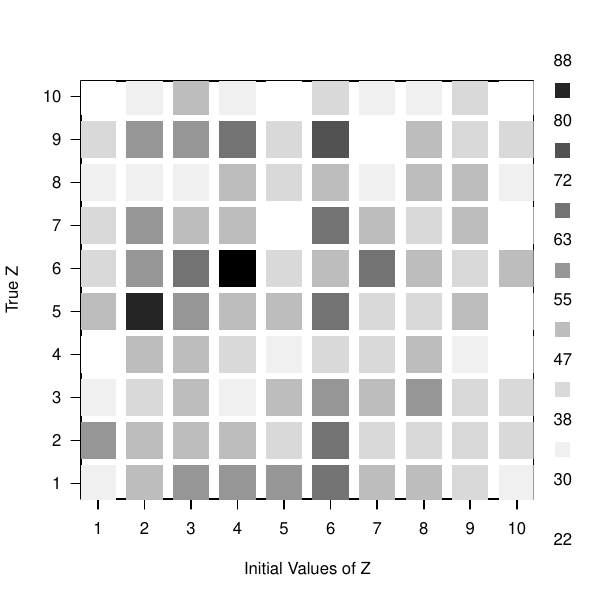}
         \label{Z_mode_init}
     \end{subfigure}
     \hfill
     \begin{subfigure}[b]{0.4\textwidth}
         \centering
         \includegraphics[width=\textwidth]{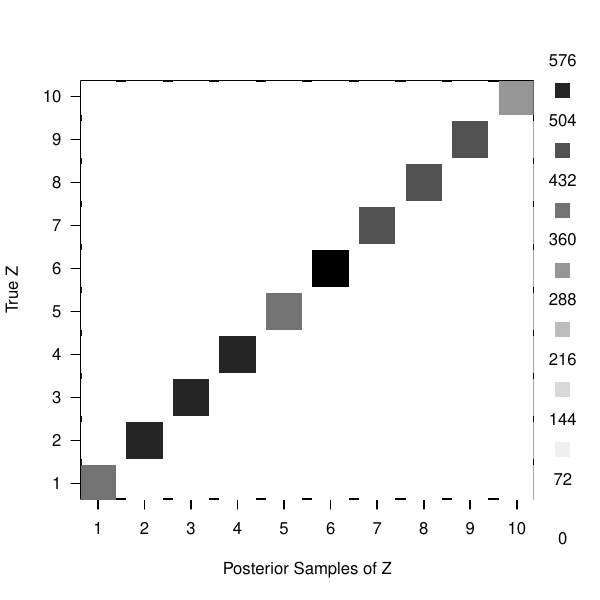}
	 \label{Z_mode_est}
     \end{subfigure}
        \caption{The comparison of the estimated and the true topics of paragraphs.
        On the right panel, the ($k, l$) cell shows the number of paragraphs whose estimated topic is $l$ while the true topic is $k$.
        We estimate topics using the paragraph topic parameter, \textbf{Z}, using the last draw from our Gibbs sampler.
        The cells with darker colors indicate a higher number of paragraphs.
        The concentration on the diagonal elements means that the topics are estimated correctly.
        As a comparison, the left panel plots randomly initialized paragraph topics against true paragraph topics.  
        They show that the PCTM can recover the true topics even when the topics are randomly provided at the initialization of our Gibbs sampler.}
        \label{simulation_Z}
\end{figure}

Figure \ref{simulation_Z} plots the posterior samples of paragraph topics against the true paragraph topics. Numbers on the x-axis and y-axis denote topic labels. The darkness of cell colors is proportional to the number of paragraphs in those cells. The cell in the second row and the third column, for example, denotes the number of paragraphs that are assigned topic 2 in posterior samples when the true topic is 3. Darker colors on the diagonal lines suggest that the model recovers true topics correctly, which we see on the right panel of Figure \ref{simulation_Z}. In comparison, the left panel of Figure \ref{simulation_Z} illustrates that the Gibbs sampler was initiated with randomly generated values of paragraph topics. 

\begin{figure}[ht!]
     \centering
     \begin{subfigure}[b]{0.4\textwidth}
         \centering
         \includegraphics[width=\textwidth]{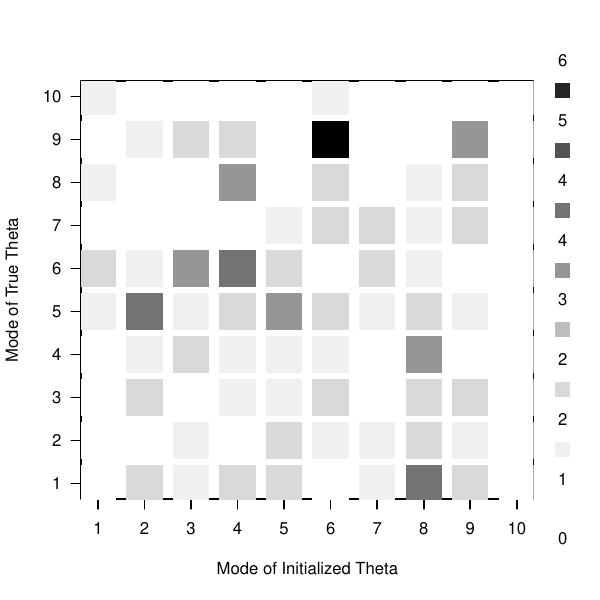}
         \label{theta_mode_init}
     \end{subfigure}
     \hfill
     \begin{subfigure}[b]{0.4\textwidth}
         \centering
         \includegraphics[width=\textwidth]{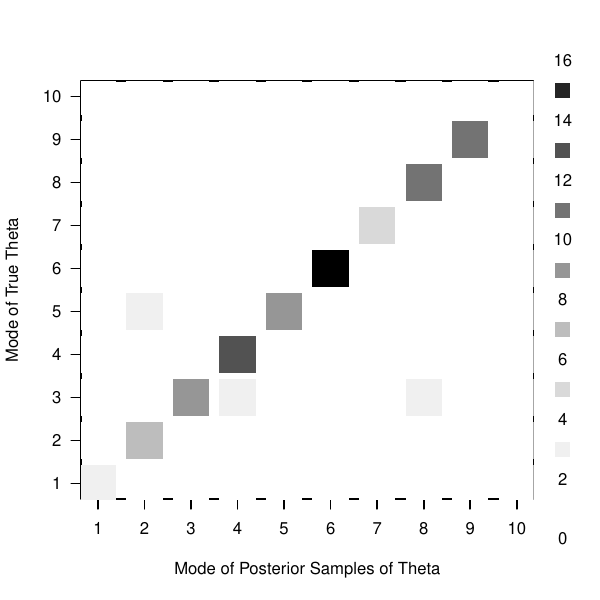}
	 \label{theta_mode_est}
     \end{subfigure}
        \caption{The comparison of the estimated and the true topic distribution of documents.
        On the right panel, the ($k, l$) cell shows the number of documents whose mode of the estimated topic distribution, $\btheta$, across $K$ topics is $l$
        while the mode of the true topic distribution is $k$.
        We obtain $\btheta$ by applying the softmax transformation on each draw of $\pmb\eta$ in our Gibbs sampler, and then obtain the estimated $\btheta$ by their posterior mean. 
        The cells with darker colors mean a higher number of documents are in the cell.
        The concentration on the diagonal elements means that the modes of the topic distributions are estimated correctly.
        As a comparison, the left panel plots the mode of randomly initialized $\btheta$ against true mode of $\btheta$.  
        It shows that the PCTM can recover the true mode of the topic distribution even when the topics are randomly provided at the initialization of our Gibbs sampler.}
        \label{simulation_theta_modes}
\end{figure}

We conduct a similar exercise with the document-level topic mixture $\pmb\eta$. To make the comparison more rooted in conventional topic models, we convert $\pmb\eta$ to $\pmb\theta$ using softmax in this exercise. In Figure \ref{simulation_theta_modes}, we plot the mode of posterior samples of $\pmb\theta$ against the mode of the true topic mixture. The darker colors indicate a higher number of documents in the corresponding cell. Similar to Figure \ref{simulation_Z}, we observe evenly spread colors on the left panel as opposed to the concentrated dark colors on the diagonal entries on the right panel. This shows that the PCTM recovers 

These two results verify that the PCTM can recover true topics from random initialization when applied to simulation data.
This adds to the credibility of the topic estimations in our application since our simulation data resembles our application data. 

\clearpage

\section{Results on the SCOTUS cases on Voting Rights}
\label{sec:voting}

The SCOTUS documents and citations on voting rights proliferated exponentially since the enactment of Voting Rights Act (VRA) in 1965. A number of sections in VRA were challenged over the course of modern American political history, and the majority of those challenges made their way to the Supreme Court. The Supreme Court database assigns 3 issue codes for opinions related to voting.\footnote{The three issue codes on voting are voting, Voting Rights Act of 1965, Ballot Access.} After examining a subset of documents with these issue codes, we decided to set the number of topics to 4 for PCTM.

Table \ref{word_mat_voting} presents the 10 words that appear most frequently for each topic. 
\begin{table}[ht!]
\centering
\begin{tabular}{c| l l l l}
  \hline
Topic & \textcolor{SkyBlue}{Voter} & \textcolor{Blue}{Ballot} & \textcolor{Aquamarine}{Preclearance} & \textcolor{Green}{Voter}  \\ 
Label & \textcolor{SkyBlue}{Eligibility} & \textcolor{Blue}{Access} & \textcolor{Aquamarine}{Requirement} & \textcolor{Green}{Dilution} \\ 
  \hline
1 & \textcolor{SkyBlue}{counti} & \textcolor{Blue}{ballot} & \textcolor{Aquamarine}{chang} & \textcolor{Green}{plan}  \\ 
2   & \textcolor{SkyBlue}{resid} & \textcolor{Blue}{primari} & \textcolor{Aquamarine}{attorney} & \textcolor{Green}{minor} \\ 
3   & \textcolor{SkyBlue}{appel} & \textcolor{Blue}{polit} & \textcolor{Aquamarine}{preclear} & \textcolor{Green}{black} \\ 
4   & \textcolor{SkyBlue}{school} & \textcolor{Blue}{offic} & \textcolor{Aquamarine}{counti} & \textcolor{Green}{major} \\ 
5   & \textcolor{SkyBlue}{properti} & \textcolor{Blue}{counti} & \textcolor{Aquamarine}{practic} & \textcolor{Green}{polit} \\ 
6   & \textcolor{SkyBlue}{citi} & \textcolor{Blue}{file} & \textcolor{Aquamarine}{procedur} & \textcolor{Green}{popul} \\ 
7   & \textcolor{SkyBlue}{tax} & \textcolor{Blue}{interest} & \textcolor{Aquamarine}{cover} & \textcolor{Green}{racial} \\ 
8   & \textcolor{SkyBlue}{board} & \textcolor{Blue}{independ} & \textcolor{Aquamarine}{plan} & \textcolor{Green}{member} \\ 
9   & \textcolor{SkyBlue}{citizen} & \textcolor{Blue}{nomin} & \textcolor{Aquamarine}{section} & \textcolor{Green}{dilut} \\ 
10   & \textcolor{SkyBlue}{test} & \textcolor{Blue}{burden} & \textcolor{Aquamarine}{object} & \textcolor{Green}{white} \\ 
   \hline
\end{tabular}
\caption{Top 10 words of highest probability for each topic from PCTM.}
\label{word_mat_voting}
\end{table}
The first topic \texttt{Voter Eligibility} includes paragraphs that address conditions under which a voter is eligible to register for certain elections. For example, \texttt{Allen et al. v. State Board of Elections et al.} (1969) contains a paragraph of the first topic that discusses whether a 31-year-old man was eligible to cast his vote in a local school district election based on his tax records and property ownership in the neighborhood. The second topic \texttt{Ballot Access} concerns the issue of candidates' access to ballots. A paragraph of this topic in \texttt{Carrington v. Rash et al.} (1965) states that ``... the Texas system creates barriers to candidate access to the primary ballot, thereby tending to limit the field of candidates from which voters might choose.'' Preclearance requirement in Voting Rights Act of 1965 section 5. is the primary issue in the third topic. \texttt{Cipriano v. City of Houma et al.} (1969) contains a paragraph of this topic that stipulates ``... and unless and until the court enters such judgment no person shall be denied the right to vote for failure to comply with such qualification, prerequisite, standard, practice, or procedure: Provided, That such qualification, prerequisite, standard, practice, or procedure may be enforced without such proceeding if the' qualification, prerequisite, standard, practice, or procedure has been submitted by the chief legal officer or other appropriate official ...'' The fourth topic, on the other hand, addresses Voting Rights Act of 1965, section 2 that prohibits voting practices that leads to dilution of voting strength of minority groups. For example, \texttt{Mcdonald et al. v. Board of Election Commissioners of Chicago et al.} (1969) contains multiple paragraphs of this topic one of which states that ``... the Court upheld a constitutional challenge by Negroes and Mexican-Americans to parts of a legislative reapportionment plan adopted by the State of Texas ... .''

\begin{figure}[ht!]
     \centering
         \includegraphics[width=\textwidth]{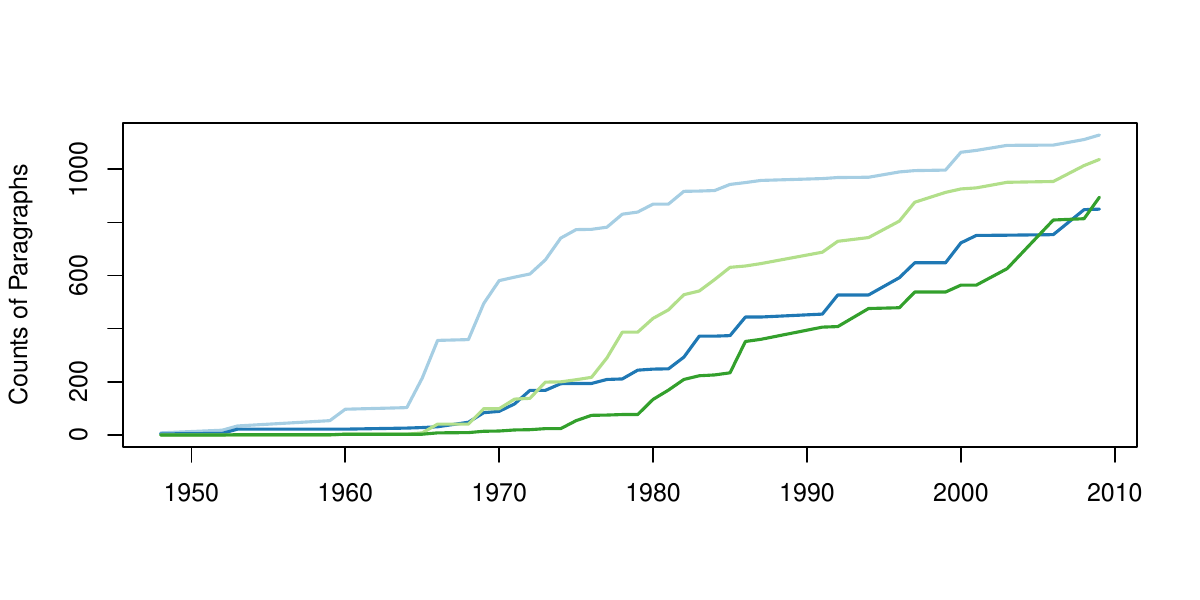}
         \caption{Cumulative number of topics in Voting Rights subset over time.}
	 \label{voting_topic_growth}
\end{figure}

The 4 topics that PCTM identified have varying presence in American political history over time. Figure \ref{voting_topic_growth} shows the cumulative count of paragraphs of each topic. 
The growth of \texttt{Voter Eligibility} topic (in light blue) is most evident until the 1980s and the topics on \texttt{Preclearance Requirement} (in light green) or \texttt{Voter Dilution} (in dark green) become more prevalent in relatively recent periods. This is consistent with \cite{ansolabehere2008end} that describes that discourses on malapportionment was more common in earlier periods, and the topics on equal representation and access to vote, especially with respect to race and minority groups, are becoming more prominent issues in modern American politics.

\begin{figure}[t!]
     \centering
     \begin{subfigure}[b]{0.23\textwidth}
         \centering
         \includegraphics[width=\textwidth]{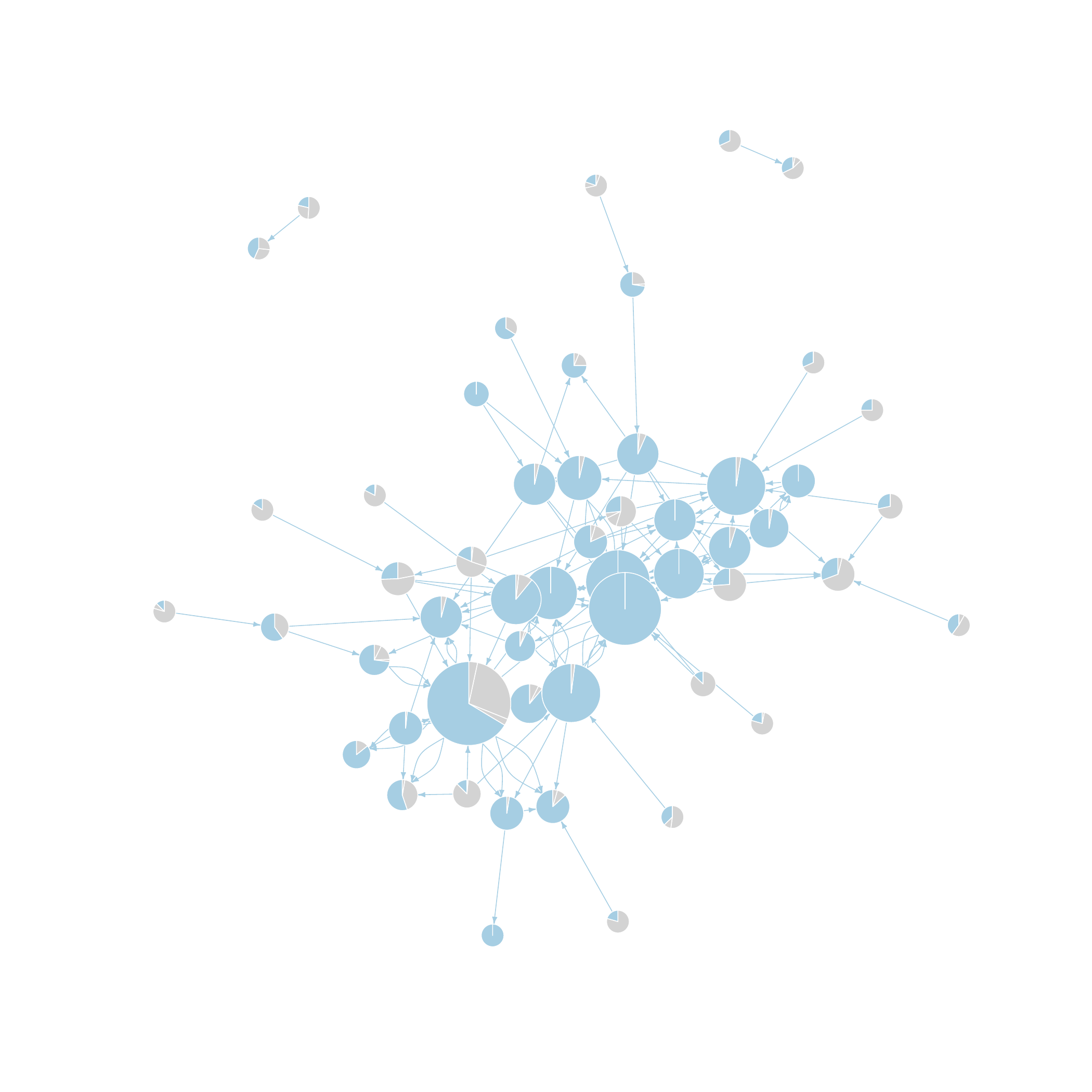}
         \caption{Voter Eligibility}
         \label{voting_topic1}
     \end{subfigure}
     \hfill
     \begin{subfigure}[b]{0.23\textwidth}
         \centering
         \includegraphics[width=\textwidth]{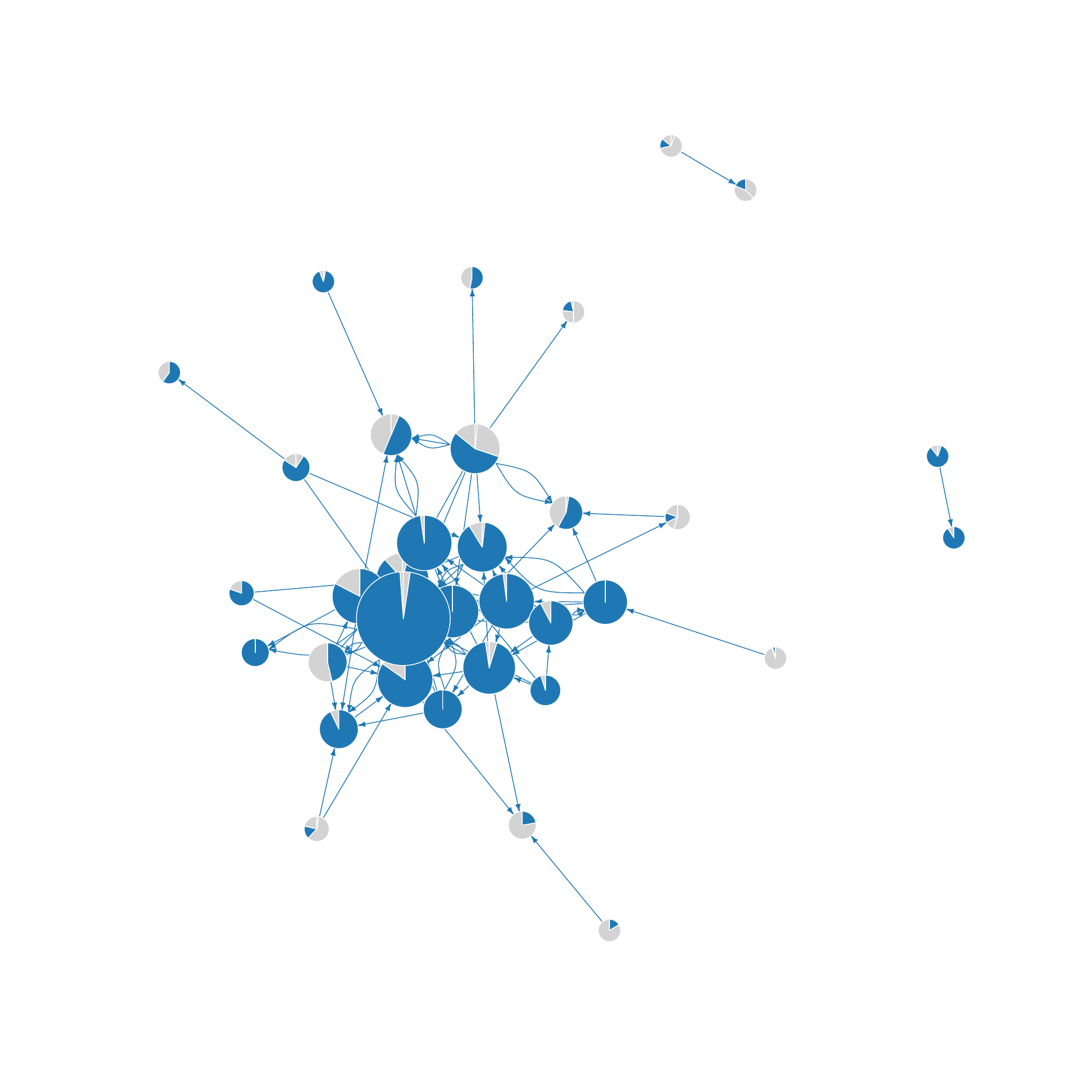}
         \caption{Ballot Access}
	 \label{voting_topic2}
     \end{subfigure}
     \hfill
     \begin{subfigure}[b]{0.23\textwidth}
         \centering
         \includegraphics[width=\textwidth]{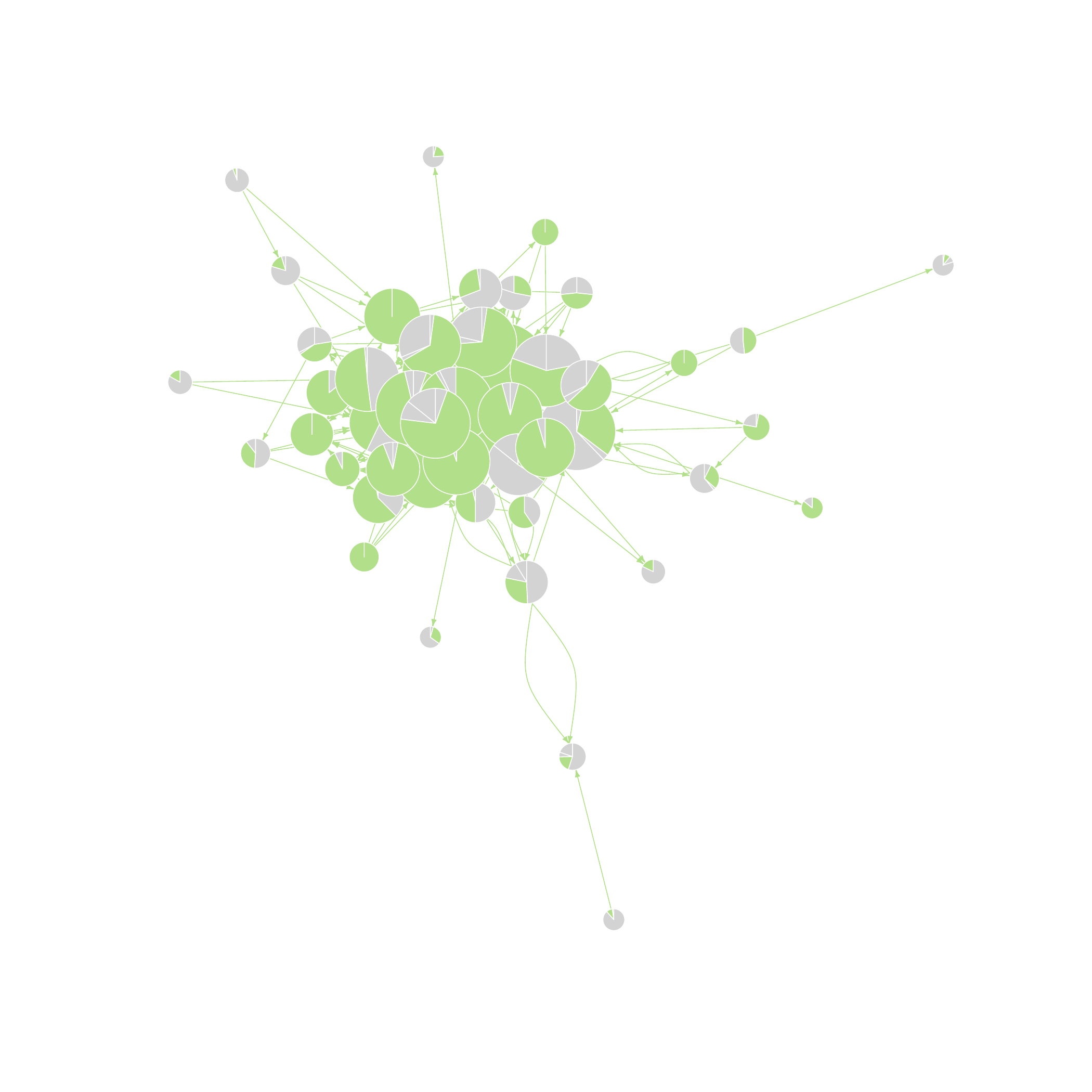}
         \caption{Preclearance \\ Requirement}
	 \label{voting_topic3}
     \end{subfigure}
     \hfill
     \begin{subfigure}[b]{0.23\textwidth}
         \centering
         \includegraphics[width=\textwidth]{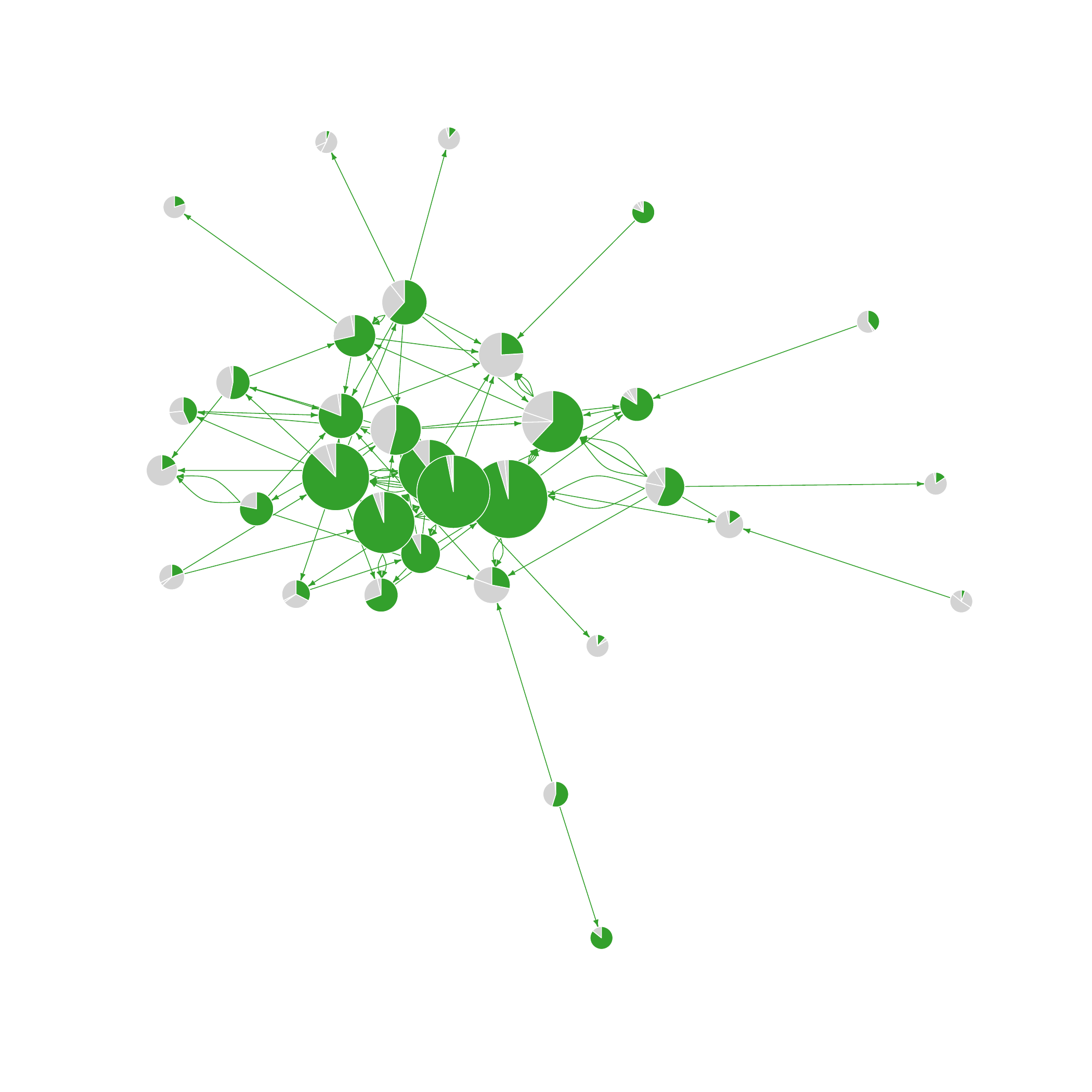}
         \caption{Voter Dilution}
	 \label{voting_topic4}
     \end{subfigure}
        \caption{The subnetwork specific to each topic. The subnetworks are created by extracting opinions that either send or receive citations of the given topic. The topic-specific subnetworks can be useful in revealing whether and the extent to which topological features of the network varies by topic. For each subnetwork, paragraphs of other topics are all colored in gray for better visualization.}
        \label{subnetwork_voting_figures}
\end{figure}

Figure \ref{subnetwork_voting_figures} shows groups of cases that make citations of the given topic. The location of cases on each network is based on their connection patterns such that cases that cite other cases jointly are placed closer to each other. The majority of cases in the third and the fourth panel are located very close to each other, indicating that those cases heavily cite each other. On the other hand, the citation subnetwork in the first panel (\texttt{Voter Eligibility}) is more spread out in comparison. This reflects the fact that opinions on \texttt{Preclearance Requirement} and \texttt{Voter Dilution} have proliferated in a shorter period of time, closely building up on past cases of the same topic whereas opinions on \texttt{Voter Eligibility} have expanded more independently and incrementally over a longer period of time.

The coefficients in the latent citation propensity for Voting subset also have expected signs, with posterior samples of $\tau_1$ and $\tau_2$ both staying above 0. That is, for the citation decisions of opinions for Voting, the authority as well as the topic similarity of precedents have positive impacts. Moreover, the distribution of all $\pmb\tau$ entries stays very similar between the Privacy and the Voting subset, indicating that the citation dynamics do not vary much between different issue areas within the SCOTUS 

\begin{figure}[ht!]
     \centering
     \begin{subfigure}[b]{0.45\textwidth}
         \centering
         \includegraphics[width=\textwidth]{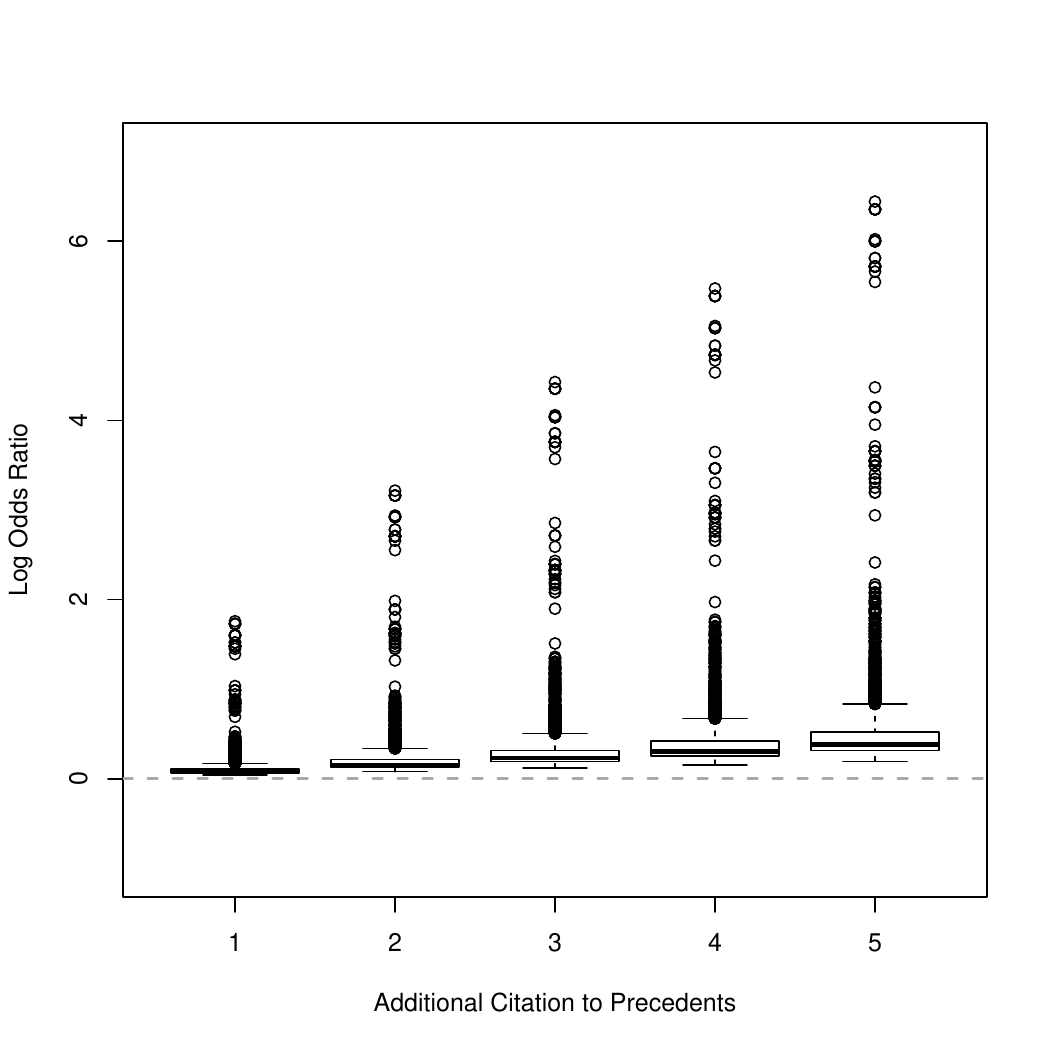}
         \caption{Change in Log Odds Ratio by Additional Citations to Precedents}
         \label{reg_interpret_indegree_voting}
     \end{subfigure}
     \hfill
     \begin{subfigure}[b]{0.45\textwidth}
         \centering
         \includegraphics[width=\textwidth]{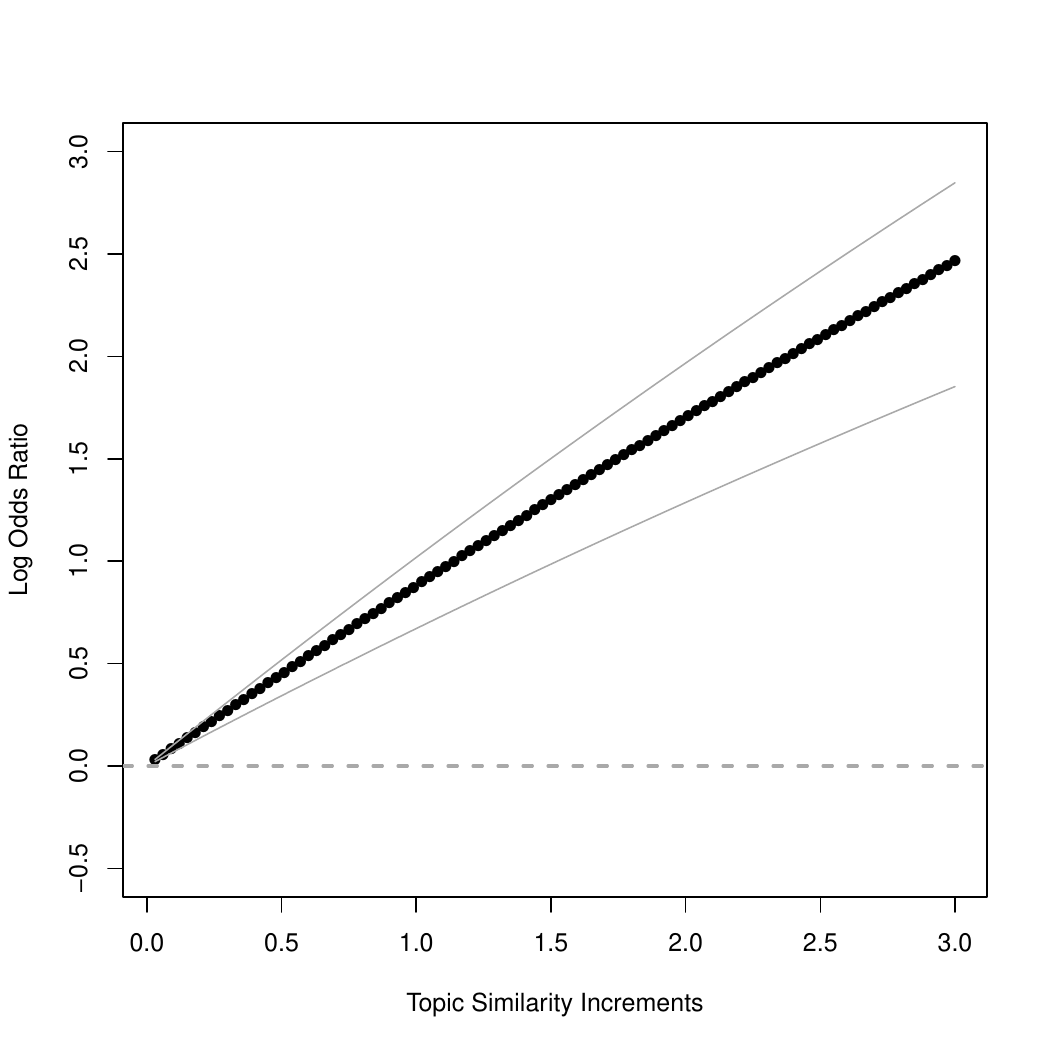}
         \caption{Change in Log Odds Ratio by Increases in $\eta_{j,z_{ip}}$}
         \label{reg_interpret_topic_voting}
     \end{subfigure}
     \caption{Changes in the log odds ratio of citation between a paragraph and a precedent as we increment the authority and the topic similarity of the given precedent. Same exercise used in Figure \ref{reg_interpret_topic_privacy} is employed to create this figure.}
        \label{reg_interpret_voting}
\end{figure}

Similar to the exercise to create Figure \ref{reg_interpret_privacy}, 10,000 randomly drawn pairs of paragraphs and precedents for the Voting subset were used to generate Figure \ref{reg_interpret_voting}. The left panel of Figure \ref{reg_interpret_voting} presents the improvements in the log odds ratio as we increment the authority of the given precedent. For example, if the given precedent had 3 more citations, the odds of the given paragraph citing the given precedent increases by about 25\%. The right panel shows changes in log odds ratio as the topic similarity between the given precedent and the given paragraph increases.

\clearpage
\section{More Results on the SCOTUS cases on Privacy} 
\label{sec:si_privacy}

\subsection{Influence of the In-degree and Topic Similarity on the Probability of Citation}
\begin{figure}[ht!]
     \centering
     \begin{subfigure}[b]{0.45\textwidth}
         \centering
         \includegraphics[width=\textwidth]{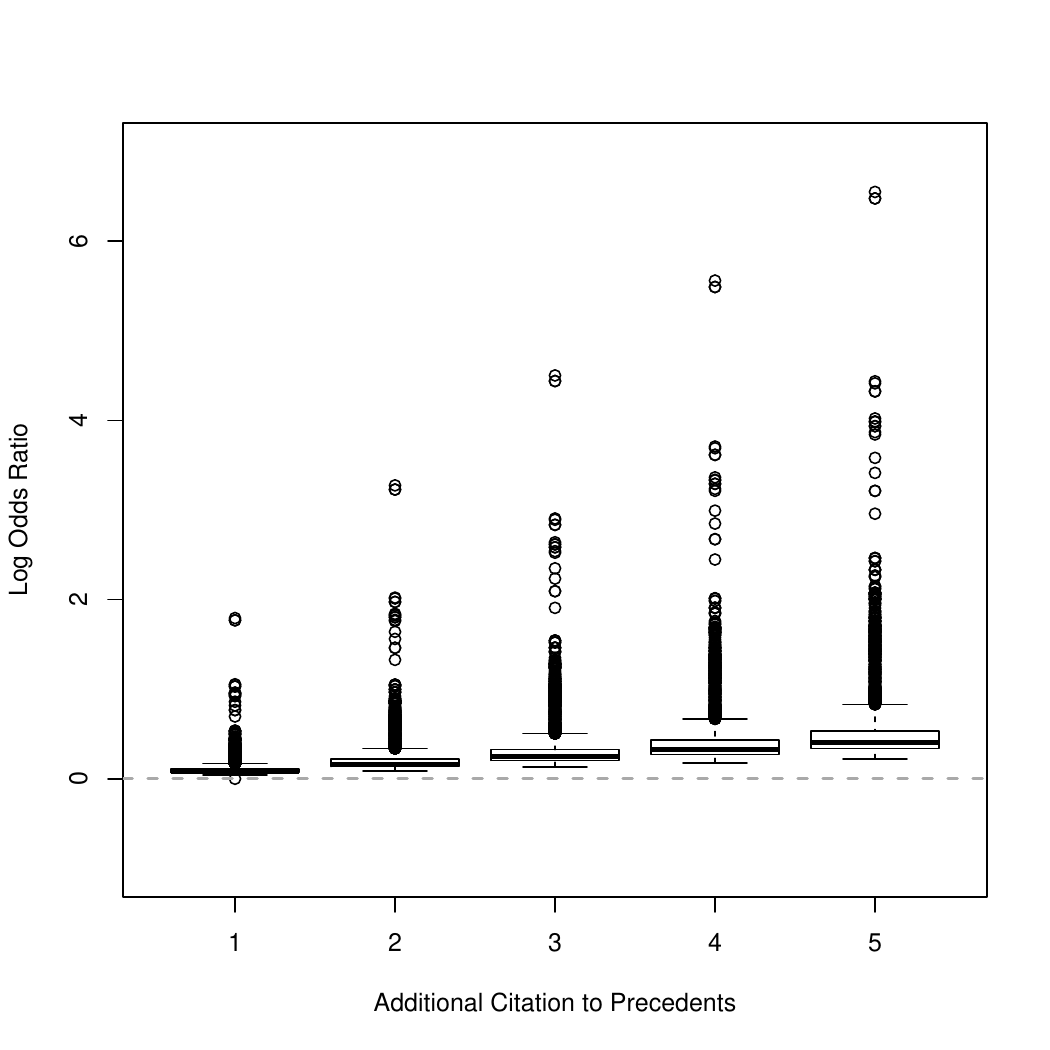}
         \caption{Change in Log Odds Ratio by Additional Citations to Precedents}
         \label{reg_interpret_indegree_privacy}
     \end{subfigure}
     \hfill
     \begin{subfigure}[b]{0.45\textwidth}
         \centering
         \includegraphics[width=\textwidth]{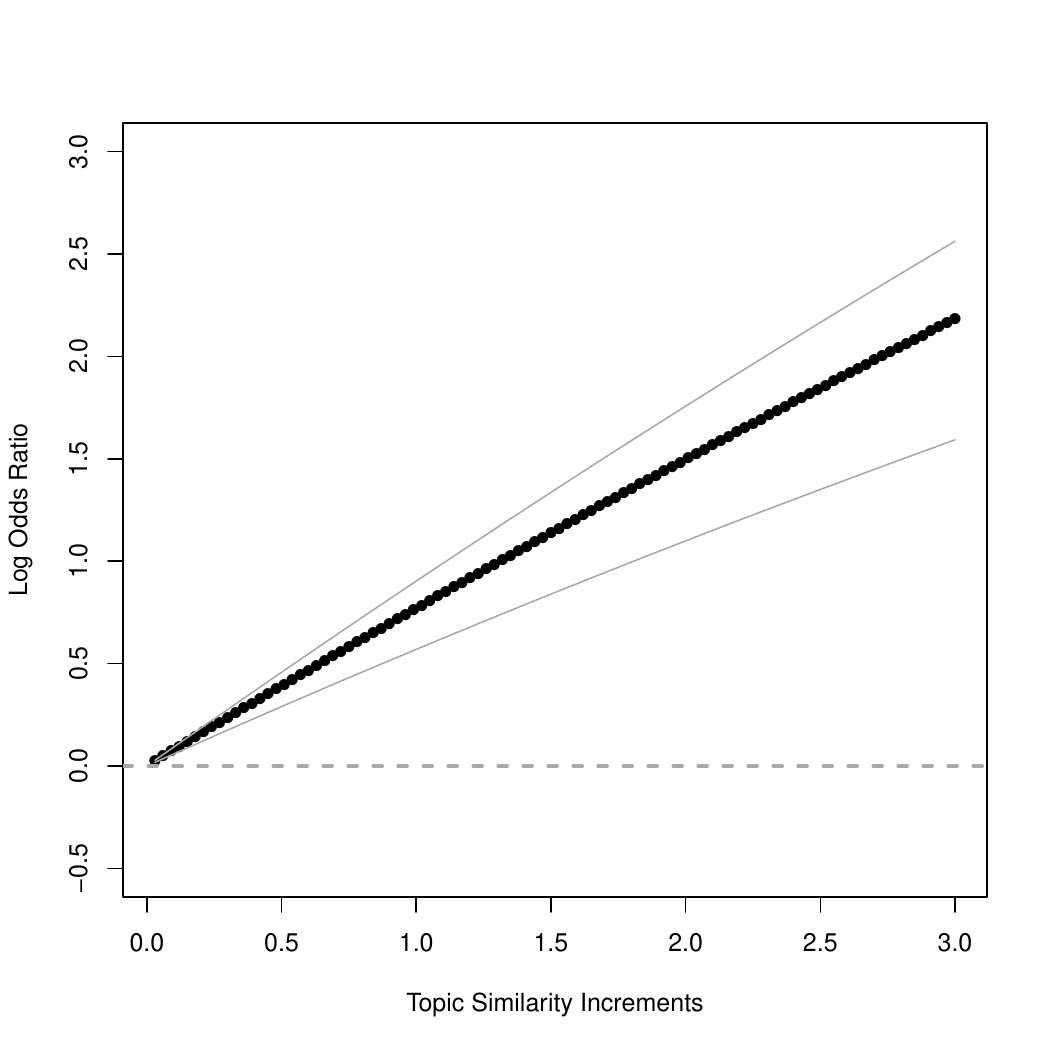}
         \caption{Change in Log Odds Ratio by Increases in $\eta_{j,z_{ip}}$}
         \label{reg_interpret_topic_privacy}
     \end{subfigure}
     \caption{Changes in the log odds ratio of citation between a paragraph and a precedent as we increment the authority and the topic similarity of the given precedent. 10,000 random pairs of paragraphs and precedents were drawn from the data to create this figure. The left panel displays the distribution of improvements in log odds ratio if the given precedent had given additional citations. Each point is one of the 10,000 randomly drawn paragraph-precedent pairs. The right panel shows the improvements in log odds ratio if the given precedent were more topically similar to the given paragraph. The black points represent the average improvements in log odds ratio, and gray lines indicate the 2.5\% and 97.5\% quantile of log odds improvements respectively.}
     \label{reg_interpret_privacy}
\end{figure}

The $\pmb\tau$ coefficients in the latent citation propensity have expected signs. The average value of posterior samples for $\tau_1$ is 0.7 and the 95\% credible interval does not include 0, which suggests that the authority of documents has a positive impact on citation likelihood given topics. Similarly, posterior samples for $\tau_2$ stays above 0, suggesting that topic similarity between precedents and the citing paragraphs has a positive impact on citation decisions.

In Figure \ref{reg_interpret_privacy} we offer one way to interpret coefficients $\pmb\tau$ in latent citation propensity.\footnote{For more detailed information on the posterior samples of $\pmb\tau$, see Supplementary Information E.}
Since the latent citation propensity follows the structure of probit regression, one can employ the conventional approach to interpreting the coefficients where we calculate improvements in predicted probability as we increment one predictor while fixing other predictors at their means. This approach, however, presents two potential challenges. First, citation networks are usually sparse. Under our modeling framework, the sparse feature of citation networks is more emphasized as paragraphs are the unit that makes citations. The citation network for the Privacy subset contains only 452 citations when the fully connected network would have 243,685 citations. Partly due to such sparsity, improvements in predicted probability can be highly marginal. Second, the authority of a precedent, or the indegree, is known to follow the power-law distribution which is highly skewed to the right \citep{eom2011characterizing}. When a distribution is highly skewed, the mean is less likely to be the representative value of the distribution.

To address the above two challenges, we examine improvements in log odds ratio rather than predicted probability. Additionally, when incrementing one predictor we follow \cite{hanmer2013behind} and use observed values of other predictors rather than their means. To create Figure \ref{reg_interpret_privacy} we randomly sampled 10,000 paragraph-precedent pairs from the subset data and computed the extent of improvements in log odds ratio as we increased the authority and topic similarity of the given precedent. The left panel presents the improvements in log odds ratio when the authority of the given precedent is incremented. For example, if the given precedent had 3 more citations, the odds of the given paragraph citing the given precedent increases by about 20\%. Similarly, the right panel displays improvements in log odds ratio as the given precedent becomes more topically similar ($\eta_{j,z_{ip}}$) to the given paragraph.

\subsection{MCMC Convergence Diagnostics}
\begin{figure}[t!]
     \centering
         \includegraphics[width=\textwidth]{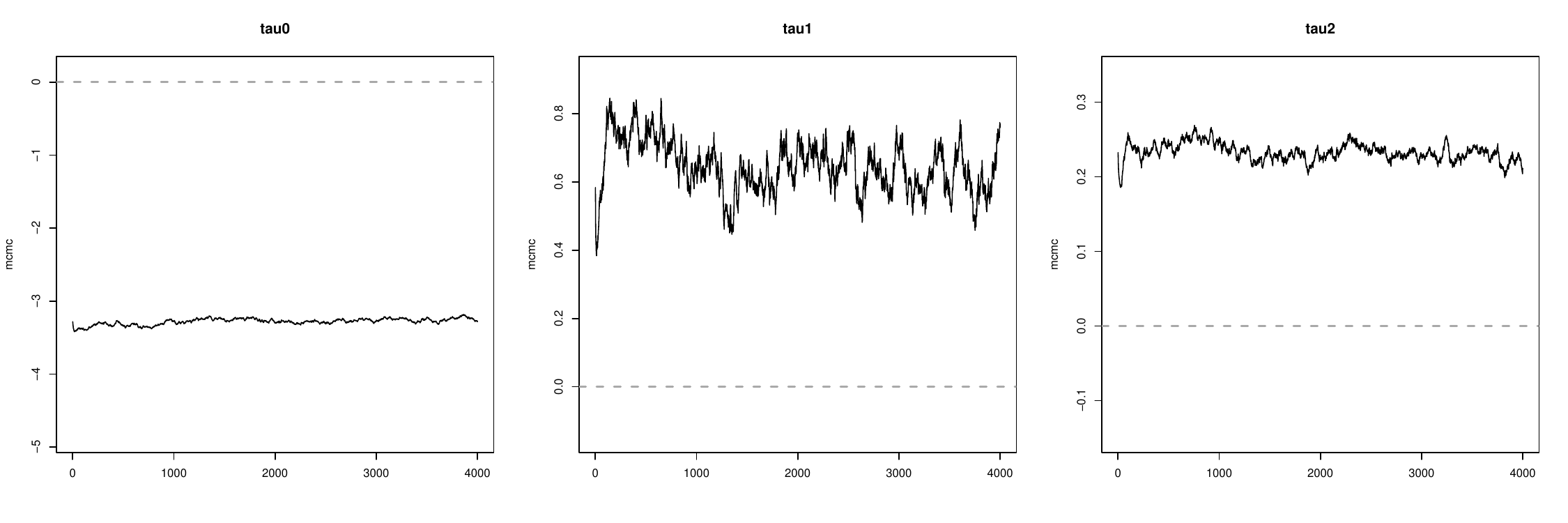}
         \caption{MCMC convergence of $\pmb\tau$ posterior samples for the SCOTUS application on Privacy issue area. Horizontal red line indicates the true values of $\pmb\tau$.}
	 \label{privacy_tau_mcmc}
\end{figure}

\begin{figure}[t!]
     \centering
         \includegraphics[width=\textwidth]{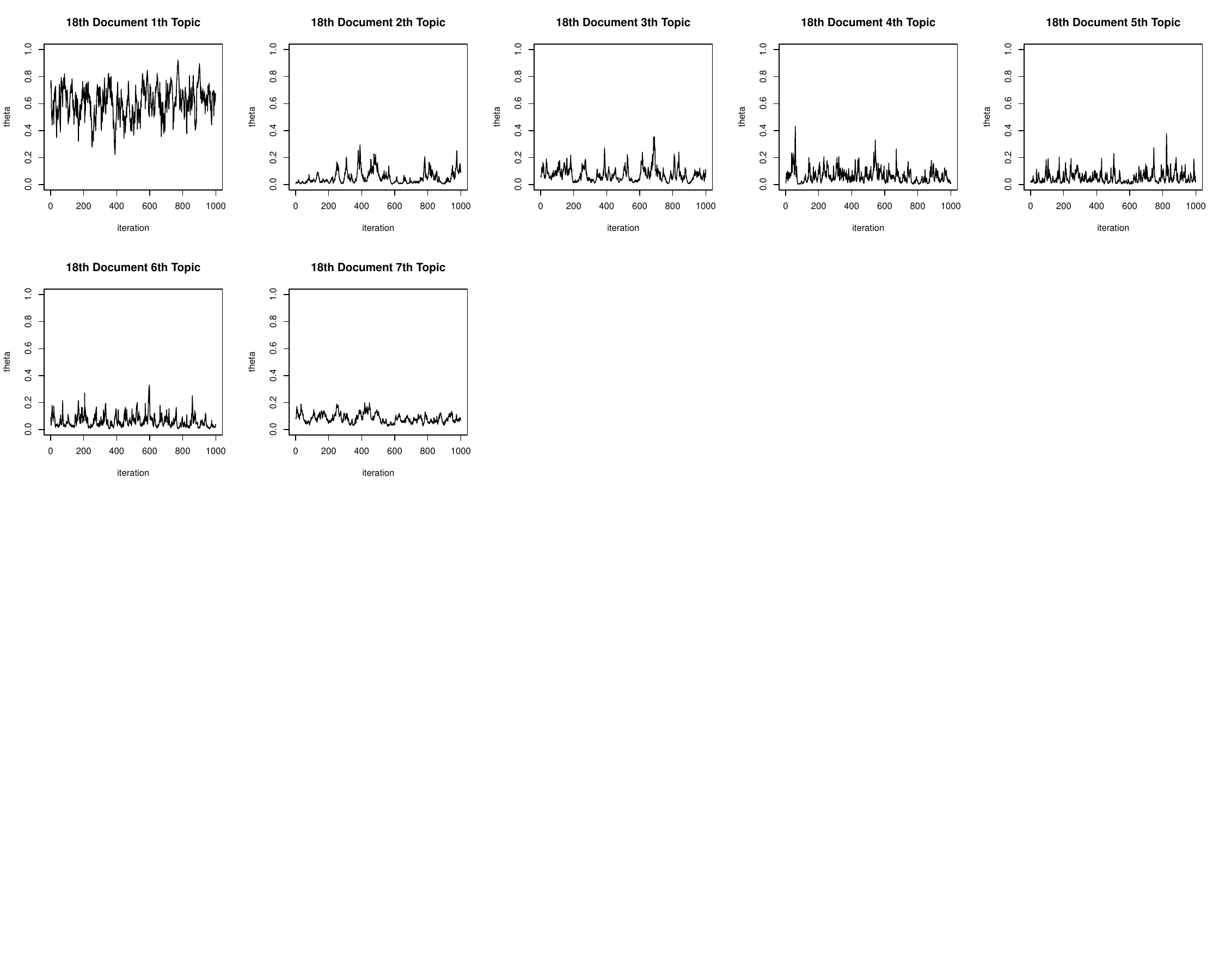}
         \caption{MCMC convergence of $\pmb\theta$ parameters for the 18th document in the subset of Privacy issue area. $\pmb\theta$ values are obtained by transforming the posterior samples of $\pmb\eta$ of the corresponding document. Horizontal red line indicates the true values of $\pmb\theta$ for the 18th document for each topic. We do not display the MCMC convergence for other documents, but all documents show similar level of convergence to the true value of $\pmb\theta$.}
	 \label{privacy_theta_mcmc}
\end{figure}

\begin{figure}[t!]
     \centering
         \includegraphics[width=\textwidth]{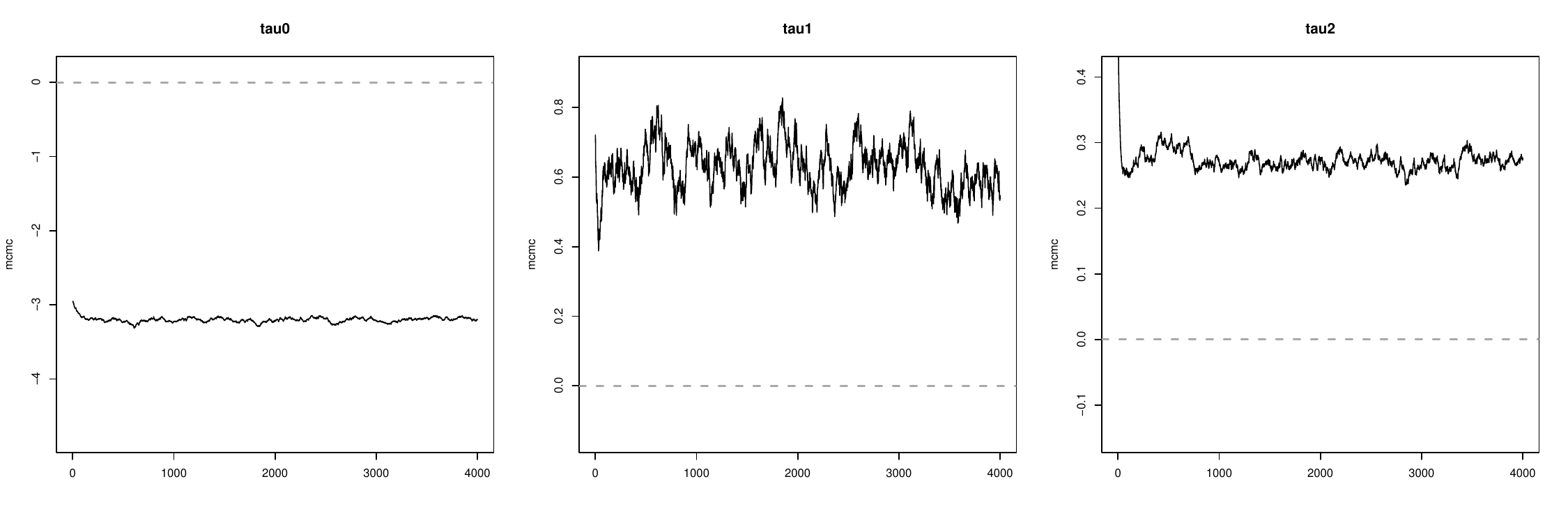}
         \caption{MCMC convergence of $\pmb\tau$ posterior samples for the SCOTUS application on Voting Rights issue area. Horizontal red line indicates the true values of $\pmb\tau$.}
	 \label{voting_tau_mcmc}
\end{figure}

\begin{figure}[t!]
     \centering
         \includegraphics[width=\textwidth]{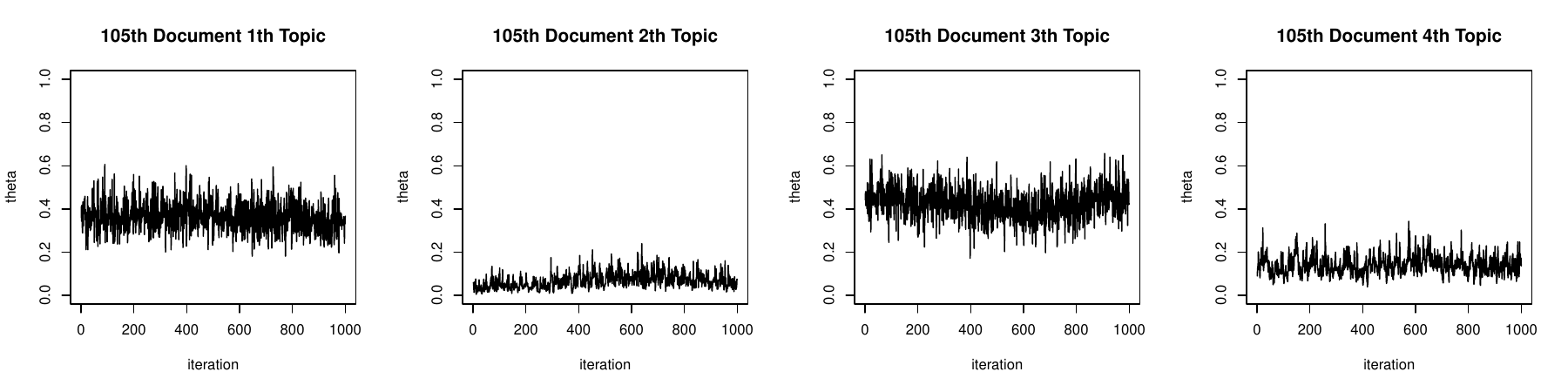}
         \caption{MCMC convergence of $\pmb\theta$ parameters for the 105th document in the subset of Voting Rights issue area. $\pmb\theta$ values are obtained by transforming the posterior samples of $\pmb\eta$ of the corresponding document. Horizontal red line indicates the true values of $\pmb\theta$ for the 105th document for each topic. We do not display the MCMC convergence for other documents, but all documents show similar level of convergence to the true value of $\pmb\theta$.}
	 \label{voting_theta_mcmc}
\end{figure}

\clearpage

\section{Comparison of the Predictive Performance against Existing Methods}
\label{sec:predict_prob}

In this section, we compare the predictive performance of the PCTM against two alternative models for document networks: the RTM and the LDA combined with Logistic Regression (LDA + Logistics).
In both alternative models, citations arise as a function of topic similarity at the word level. 
We use documents in the Privacy subset for this exercise. 
We choose paragraphs in \texttt{Gonzales v. Carhart} as our test set because \texttt{Gonzales v. Carhart} contains a sufficiently large number of citations and words to demonstrate how they contribute to the predictive performance.\footnote{\texttt{Gonzales v. Carhart} contains 12 citations, which is about 94 percentile of the distribution of the number of citations per document. It is the 9th latest document in our corpus.} 
We discard documents temporally later than \texttt{Gonzales v. Carhart}. 

Our exercise is essentially a leave-one-out cross-validation for each paragraph in \texttt{Gonzales v. Carhart}. 
Specifically, we take a paragraph in \texttt{Gonzales v. Carhart} as test data, and all other paragraphs in \texttt{Gonzales v. Carhart} and documents prior to it are assigned to the training data. 
Then we compute the predictive probability on the paragraph in the test set given our parameters fit on the training data. 
Note that due to the structure of this exercise, \texttt{Gonzales v. Carhart} will appear in both the training set and the test set. 
The above exercise is repeated for all 88 paragraphs in \texttt{Gonzales v. Carhart}. 

One challenge in this exercise is that the PCTM assigns topics to each paragraph while the RTM and the LDA assign topics to each word. 
That is, the RTM and the LDA do not recognize paragraphs in the data. 
Therefore, we treat the paragraph in the test data as if it is a new version of \texttt{Gonzales v. Carhart} when we compute the predictive probability in the RTM and the LDA.
In other words, we estimate topics of words in the test data from the topic probability for \texttt{Gonzales v. Carhart}. 

A formal description of the prediction exercise is as follows.
$\mathbf{W}_{iq}$ and $\mathbf{D}_{iq}$ are the data in a paragraph $q$ of a document $i$.
This corresponds to the test paragraph.
$\mathbf{W}^{train}, \mathbf{D}^{train}$ are the data in the training set.
This includes all paragraphs other than the paragraph $q$ of the document $i$ as well as all documents prior to the document $i$.
The parameters with $\hat{\cdot}$ symbol indicate that they are estimates based on the training data.
The following gives the posterior predictive probability for the PCTM.

\begin{equation}
\begin{split}
  &p(\mathbf{W}_{iq}, \mathbf{D}_{iq} \vert \mathbf{W}^{train}, \mathbf{D}^{train}) \\
  &= \sum_{k=1}^K \Big\{ p(\mathbf{W}_{iq} \vert z_{iq} = k, \hat{\boldsymbol \Psi})\\
  &\quad \times \prod_{j=1}^{i-1} \mathbb{P}(D_{iqj}^* > 0 \vert \hat{\boldsymbol \tau}, \hat{\boldsymbol \eta}, z_{iq} = k)^{\mathbb{I}\{D_{iqj}=1\}}\mathbb{P}(D_{iqj}^* < 0 \vert \hat{\boldsymbol \tau}, \hat{\boldsymbol \eta}, z_{iq} = k)^{\mathbb{I}\{D_{iqj}=0\}} \\
  &\quad \times p(z_{iq} = k \vert \hat{\boldsymbol \eta_i}) \Big\}
\end{split}
\end{equation}

By contrast, the following gives the posterior predictive probability for the RTM and LDA. 
We follow \cite{chang2009relational} for the notation of parameters.
$\boldsymbol\theta$ is a $N \times K$ document-topic matrix.
$\boldsymbol\eta$ is a $K$-length vector of coefficient and $\nu$ is intercept in the regression of citation on the topic.

\begin{equation}
\begin{split}
  &p(\mathbf{W}_{iq}, \mathbf{D}_{iq} \vert \mathbf{W}^{train}, \mathbf{D}^{train}) \\
  &= \sum_{\mathbf{z}} \Big\{ p(\mathbf{W}_{i} \vert \mathbf{Z}_{iq} = \mathbf{z}, \hat{\boldsymbol \Psi})\\
  &\quad \times \prod_{j=1}^{i-1}
  \left[\psi\left(\hat{\boldsymbol\eta}(\bar{\mathbf{Z}_{iq}} \circ \bar{\hat{\mathbf{Z}_j}}) + \hat{\nu}\right)\right]^{\mathbb{I}\{D_{iqj} = 1\}}
  \left[1-\psi\left(\hat{\boldsymbol\eta}(\bar{\mathbf{Z}_{iq}} \circ \bar{\hat{\mathbf{Z}_j}}) + \hat{\nu}\right)\right]^{\mathbb{I}\{D_{iqj} = 0\}} \\
  &\quad \times p(\mathbf{Z_{iq}} = \mathbf{z} \vert \hat{\boldsymbol \theta_i}) \Big\}
\end{split}
\end{equation}

Note that $\mathbf{Z}_{iq}$ is a vector with its length equal to the number of words in the test paragraph.
Since it is infeasible to compute all possible values of $\mathbf{Z}_{iq}$, we use Monte Carlo simulation to approximate its distribution. 
For LDA+Logistic model, the parameters are estimated by fitting LDA on the training data and then regressing the citation on the topics.

\begin{figure}
  \includegraphics[width=\textwidth]{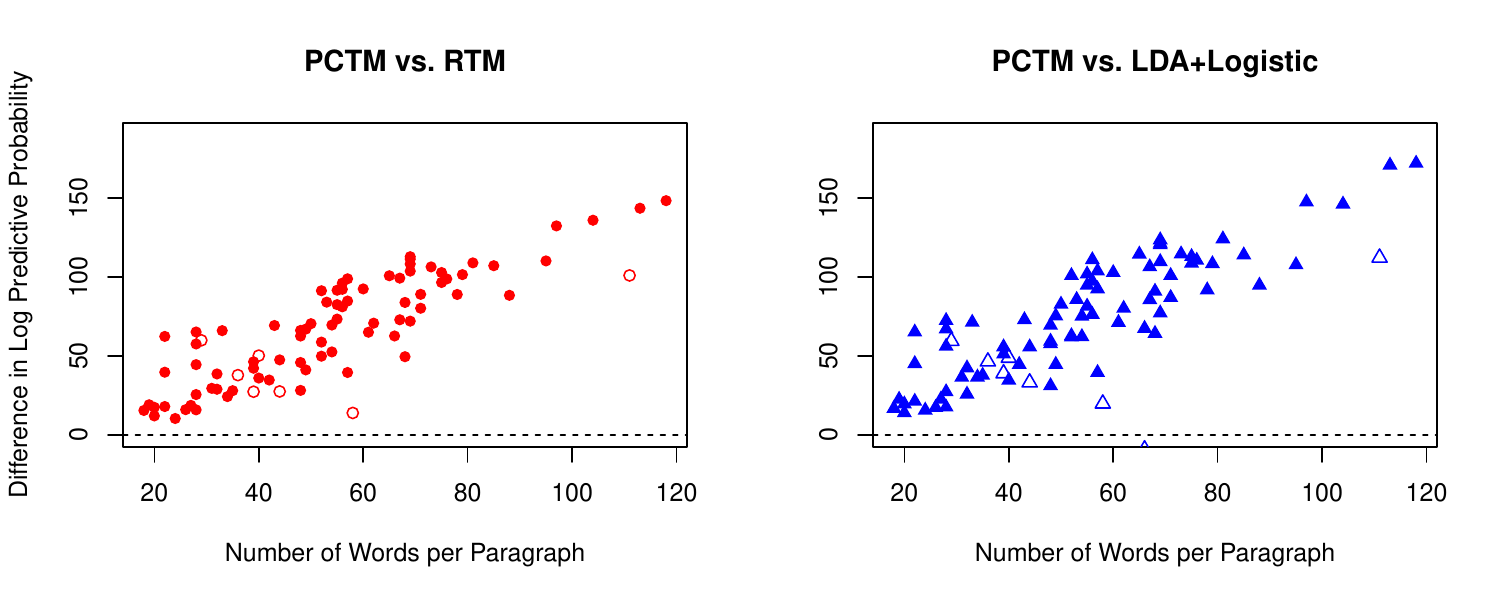}
  \caption{Difference in Predicted Probability with PCTM, RTM, and LDA + Logistic Regression. 
  The x-axis is the number of words per paragraph.
  The y-axis is the difference in the log posterior probability between PCTM and other models.
  The compared models are RTM for the left panel and LDA + Logistic regression for the right panel.
  Each symbol represents the difference in the log posterior probability between models for each paragraph. 
  Solid symbols are paragraphs without citations and hollow symbols are with citations.
  The prediction was performed by first fitting the models on a subset of the corpus temporally prior to the test paragraph, and then computing the predictive probability of the test paragraph as if the test paragraph is a new paragraph in the last document of the training corpus.  
  R package \texttt{lda} was used to fit the RTM and the LDA.
  Overall, the PCTM achieves higher posterior predictive probability compared to the RTM and the LDA + Logistic Regression models, particularly when a paragraph contains many words. 
  }
  \label{pred_privacy_realdata} 
\end{figure}

The results are displayed in Figure~\ref{pred_privacy_realdata}.
Each symbol represents the difference in the log posterior probability between models for each paragraph. 
The left panel compares the PCTM with the RTM and the right panel compares it with the LDA+Logistic regression.
Solid symbols denote the differences in the predictive probabilities for paragraphs without citations and hollow symbols are for ones with citations.
The main takeaway is that the PCTM almost always outperforms the other two models.
In particular, the improvement in predictive probability becomes greater when the prediction is made on paragraphs with more words. 
One explanation for this is that the PCTM suffers less from overfitting than the RTM or the LDA does with respect to predictions.
Since the RTM and the LDA assign topic parameters to each word, the model complexity for both models increases exponentially as more words are included in the document.
For the PCTM, on the other hand, increasing the number of words in paragraphs does not significantly impact the model complexity because the topic parameter is for paragraphs, not words. 

\clearpage

\section{Posterior Predictive Probability}
\label{sec:predictive}

The posterior probability of words and citations in a paragraph $p$ in a document $i$ can be computed by the following formula.
\begin{equation}
\begin{split}
  &p(\mathbf{W}_{ip}, \mathbf{D}_{ip} \vert \mathbf{W}^{train}, \mathbf{D}^{train}) \\
  &\propto \int_{\boldeta, \bPsi, \btau} \sum_{\mathbf{Z}} p(\mathbf{W}_{ip}, \mathbf{D}_{ip} \vert \mathbf{Z}, \boldeta, \bPsi, \btau) 
  \times p(\mathbf{Z}, \boldeta, \bPsi, \btau, \vert \mathbf{W}^{train}, \mathbf{D}^{train}) d\boldeta d\bPsi d\btau \\
  &\propto \int_{\boldeta, \bPsi, \btau} \sum_{\mathbf{Z}} p(\mathbf{W}_{ip}, \mathbf{D}_{ip} \vert \mathbf{Z}, \boldeta, \bPsi, \btau) 
  \times p(\mathbf{Z} \vert \boldeta, \bPsi, \btau, \mathbf{W}^{train}, \mathbf{D}^{train}) p( \boldeta, \bPsi, \btau \vert \mathbf{W}^{train}, \mathbf{D}^{train}) d\boldeta d\bPsi d\btau \\
  &\approx \sum_{k=1}^K \Big\{ p(\mathbf{W}_{ip}, \mathbf{D}_{ip} \vert z_{ip}^k = 1, \hat{\mathbf{Z}}^{train}, \hat{\boldeta}, \hat{\bPsi}, \hat{\btau}) \times \mathbb{P}(z_{ip}^k = 1 \vert \hat{\boldeta}) \Big\} \\
  &= \sum_{k=1}^K \Big\{ p(\mathbf{W}_{ip} \vert z_{ip}^k = 1, \hat{\bPsi})
  \times \prod_{j=1}^{i-1} p(D_{ipj} \vert \hat{\btau}, \hat{\boldeta}, z_{ip}^k = 1)
  \times \mathbb{P}(z_{ip}^k = 1 \vert \hat{\boldeta}) \Big\} \\
  &= \sum_{k=1}^K \Big\{ p(\mathbf{W}_{ip} \vert z_{ip}^k = 1, \hat{\bPsi})
  \times \prod_{j=1}^{i-1} p(D_{ipj}^* > 0 \vert \hat{\btau}, \hat{\boldeta}, z_{ip}^k = 1)^{\mathbb{I}\{D_{ipj}=1\}} p(D_{ipj}^* < 0 \vert \hat{\btau}, \hat{\boldeta}, z_{ip}^k = 1)^{\mathbb{I}\{D_{ipj}=0\}} \\
  &\quad \times \mathbb{P}(z_{ip}^k = 1 \vert \hat{\boldeta}) \Big\} \\
  &\propto \sum_{k=1}^K \Bigg\{ \prod_{v=1}^V \bPsi_{vk}^{W_{ipv}} 
  \times \prod_{j=1}^{i-1} \Big[\int_{t=0}^{\infty} p(D_{ipj}^* = t | \btau_0 + \btau_1 \kappa_j^{(i)} + \btau_2\boldeta_{jk}) dt\Big]^{\mathbb{I}\{D_{ipj}=1\}} \\
  &\quad \times \Big[\int_{t=-\infty}^{0} p(D_{ipj}^* = t | \btau_0 + \btau_1 \kappa_j^{(i)} + \btau_2\boldeta_{jk}) dt \Big]^{\mathbb{I}\{D_{ipj}=0\}} 
  \times \frac{\exp(\boldeta_{ik})}{\sum_{k'=1}^K \exp(\boldeta_{ik'})} \Bigg\} \\
\end{split}
\end{equation}
In the third line, we approximate the integral over $\boldeta$, $\bPsi$, and $\btau$ as well as the summation over $\mathbf{Z}$ in the training data.
We draw samples of these parameters from the posterior of the model fit on the training data for $\boldeta$, $\btau$, and $\mathbf{Z}$ in the training data, and we use an MLE estimate for $\bPsi$ (see Appendix~\ref{subsec:Psi}). 
The integrals in the last line can be easily computed because $D_{ipj}^*$ follows normal distributions with unit variance. 
We can also see that the posterior probability of a paragraph $p$ in a document $i$ being topic $k$ is proportional to the components inside the summation over $k$.

\clearpage 

\singlespacing
\pdfbookmark[1]{References}{References}
\bibliography{nettext}

\begin{thebibliography}{}

\bibitem[Ansolabehere and Snyder, 2008]{ansolabehere2008end}
Ansolabehere, S. and Snyder, J.~M. (2008).
\newblock {\em The end of inequality: One person, one vote and the
  transformation of American politics}.
\newblock WW Norton \& Company Incorporated.

\bibitem[Asuncion et~al., 2012]{asuncion2012smoothing}
Asuncion, A., Welling, M., Smyth, P., and Teh, Y.~W. (2012).
\newblock On smoothing and inference for topic models.
\newblock {\em arXiv preprint arXiv:1205.2662}.

\bibitem[Bai et~al., 2018]{bai2018neural}
Bai, H., Chen, Z., Lyu, M.~R., King, I., and Xu, Z. (2018).
\newblock Neural relational topic models for scientific article analysis.
\newblock In {\em Proceedings of the 27th ACM International Conference on
  Information and Knowledge Management}, pages 27--36.

\bibitem[Bailey and Maltzman, 2008]{bailey2008does}
Bailey, M.~A. and Maltzman, F. (2008).
\newblock Does legal doctrine matter? unpacking law and policy preferences on
  the us supreme court.
\newblock {\em American Political Science Review}, 102(3):369--384.

\bibitem[Benoit et~al., 2018]{quanteda}
Benoit, K., Watanabe, K., Wang, H., Nulty, P., Obeng, A., M{\"u}ller, S., and
  Matsuo, A. (2018).
\newblock quanteda: An r package for the quantitative analysis of textual data.
\newblock {\em Journal of Open Source Software}, 3(30):774.

\bibitem[Black and Spriggs, 2013]{black2013citation}
Black, R.~C. and Spriggs, J.~F. (2013).
\newblock The citation and depreciation of us supreme court precedent.
\newblock {\em Journal of Empirical Legal Studies}, 10(2):325--358.

\bibitem[Blei and Lafferty, 2007]{blei2007correlated}
Blei, D.~M. and Lafferty, J.~D. (2007).
\newblock A correlated topic model of science.
\newblock {\em The annals of applied statistics}, 1(1):17--35.

\bibitem[Blei et~al., 2003]{blei2003lda}
Blei, D.~M., Ng, A.~Y., and Jordan, M.~I. (2003).
\newblock Latent dirichlet allocation.
\newblock {\em Journal of machine Learning research}, 3(Jan):993--1022.

\bibitem[Broughman and Widiss, 2017]{broughman2017after}
Broughman, B.~J. and Widiss, D.~A. (2017).
\newblock After the override: An empirical analysis of shadow precedent.
\newblock {\em The Journal of Legal Studies}, 46(1):51--92.

\bibitem[Caselaw{\ }Access{\ }Project, 2024]{caselaw}
Caselaw{\ }Access{\ }Project (2024).
\newblock Caselaw access project.

\bibitem[Chang and Blei, 2009]{chang2009relational}
Chang, J. and Blei, D. (2009).
\newblock Relational topic models for document networks.
\newblock In {\em Artificial intelligence and statistics}, pages 81--88. PMLR.

\bibitem[Chang and Blei, 2010]{chang2010hrtm}
Chang, J. and Blei, D.~M. (2010).
\newblock {Hierarchical relational models for document networks}.
\newblock {\em The Annals of Applied Statistics}, 4(1):124 -- 150.

\bibitem[Chen et~al., 2013]{chen2013scalable}
Chen, J., Zhu, J., Wang, Z., Zheng, X., and Zhang, B. (2013).
\newblock Scalable inference for logistic-normal topic models.
\newblock {\em Advances in neural information processing systems}, 26.

\bibitem[Chen et~al., 2019]{chen2019snowboot}
Chen, Y., Gel, Y.~R., Lyubchich, V., and Nezafati, K. (2019).
\newblock Snowboot: bootstrap methods for network inference.
\newblock {\em arXiv preprint arXiv:1902.09029}.

\bibitem[Chen et~al., 2018]{chen2018use}
Chen, Y.-C., Wang, Y.~S., and Erosheva, E.~A. (2018).
\newblock On the use of bootstrap with variational inference: Theory,
  interpretation, and a two-sample test example.
\newblock {\em The Annals of Applied Statistics}, 12(2):846--876.

\bibitem[Clark and Lauderdale, 2010]{clark2010locating}
Clark, T.~S. and Lauderdale, B. (2010).
\newblock Locating supreme court opinions in doctrine space.
\newblock {\em American Journal of Political Science}, 54(4):871--890.

\bibitem[Clark and Lauderdale, 2012]{clark2012genealogy}
Clark, T.~S. and Lauderdale, B.~E. (2012).
\newblock The genealogy of law.
\newblock {\em Political Analysis}, 20(3):329--350.

\bibitem[Denny and Spirling, 2018]{denny2018text}
Denny, M.~J. and Spirling, A. (2018).
\newblock Text preprocessing for unsupervised learning: Why it matters, when it
  misleads, and what to do about it.
\newblock {\em Political Analysis}, 26(2):168--189.

\bibitem[Eom and Fortunato, 2011]{eom2011characterizing}
Eom, Y.-H. and Fortunato, S. (2011).
\newblock Characterizing and modeling citation dynamics.
\newblock {\em PloS one}, 6(9):e24926.

\bibitem[Fowler and Jeon, 2005]{fowler2005authority}
Fowler, J.~H. and Jeon, S. (2005).
\newblock The authority of supreme court precedent: a network analysis.
\newblock {\em Preprint as of June}, 29:2005.

\bibitem[Fowler and Jeon, 2008]{fowler2008authority}
Fowler, J.~H. and Jeon, S. (2008).
\newblock The authority of supreme court precedent.
\newblock {\em Social networks}, 30(1):16--30.

\bibitem[Fowler et~al., 2007]{fowler2007network}
Fowler, J.~H., Johnson, T.~R., Spriggs, J.~F., Jeon, S., and Wahlbeck, P.~J.
  (2007).
\newblock Network analysis and the law: Measuring the legal importance of
  precedents at the us supreme court.
\newblock {\em Political Analysis}, 15(3):324--346.

\bibitem[Hanmer and Ozan~Kalkan, 2013]{hanmer2013behind}
Hanmer, M.~J. and Ozan~Kalkan, K. (2013).
\newblock Behind the curve: Clarifying the best approach to calculating
  predicted probabilities and marginal effects from limited dependent variable
  models.
\newblock {\em American Journal of Political Science}, 57(1):263--277.

\bibitem[Hansford and Spriggs, 2006]{hansford2006politics}
Hansford, T.~G. and Spriggs, J.~F. (2006).
\newblock {\em The politics of precedent on the US Supreme Court}.
\newblock Princeton University Press.

\bibitem[Held and Holmes, 2006]{held2006bayesian}
Held, L. and Holmes, C.~C. (2006).
\newblock Bayesian auxiliary variable models for binary and multinomial
  regression.
\newblock {\em Bayesian analysis}, 1(1):145--168.

\bibitem[Imai et~al., 2016]{imai2016fast}
Imai, K., Lo, J., and Olmsted, J. (2016).
\newblock Fast estimation of ideal points with massive data.
\newblock {\em American Political Science Review}, 110(4):631--656.

\bibitem[Larsson et~al., 2017]{larsson2017speaking}
Larsson, O., Naurin, D., Derl{\'e}n, M., and Lindholm, J. (2017).
\newblock Speaking law to power: the strategic use of precedent of the court of
  justice of the european union.
\newblock {\em Comparative Political Studies}, 50(7):879--907.

\bibitem[Le and Lauw, 2014]{le2014probabilistic}
Le, T.~M. and Lauw, H.~W. (2014).
\newblock Probabilistic latent document network embedding.
\newblock In {\em 2014 IEEE International Conference on Data Mining}, pages
  270--279. IEEE.

\bibitem[Levin and Levina, 2019]{levin2019bootstrapping}
Levin, K. and Levina, E. (2019).
\newblock Bootstrapping networks with latent space structure.
\newblock {\em arXiv preprint arXiv:1907.10821}.

\bibitem[Linderman et~al., 2015]{linderman2015dependent}
Linderman, S., Johnson, M.~J., and Adams, R.~P. (2015).
\newblock Dependent multinomial models made easy: Stick-breaking with the
  p{\'o}lya-gamma augmentation.
\newblock {\em Advances in Neural Information Processing Systems}, 28.

\bibitem[Liu et~al., 2009]{liu2009topic}
Liu, Y., Niculescu-Mizil, A., and Gryc, W. (2009).
\newblock Topic-link lda: joint models of topic and author community.
\newblock In {\em proceedings of the 26th annual international conference on
  machine learning}, pages 665--672.

\bibitem[Lupu and Fowler, 2013]{lupu2013strategic}
Lupu, Y. and Fowler, J.~H. (2013).
\newblock Strategic citations to precedent on the us supreme court.
\newblock {\em The Journal of Legal Studies}, 42(1):151--186.

\bibitem[Lupu and Voeten, 2012]{lupu2012precedent}
Lupu, Y. and Voeten, E. (2012).
\newblock Precedent in international courts: A network analysis of case
  citations by the european court of human rights.
\newblock {\em British Journal of Political Science}, 42(2):413--439.

\bibitem[Nallapati et~al., 2008]{nallapati2008joint}
Nallapati, R.~M., Ahmed, A., Xing, E.~P., and Cohen, W.~W. (2008).
\newblock Joint latent topic models for text and citations.
\newblock In {\em Proceedings of the 14th ACM SIGKDD international conference
  on Knowledge discovery and data mining}, pages 542--550. ACM.

\bibitem[Newman, 2001]{newman2001clustering}
Newman, M.~E. (2001).
\newblock Clustering and preferential attachment in growing networks.
\newblock {\em Physical review E}, 64(2):025102.

\bibitem[Olsen and K{\"u}{\c{c}}{\"u}ksu, 2017]{olsen2017finding}
Olsen, H.~P. and K{\"u}{\c{c}}{\"u}ksu, A. (2017).
\newblock Finding hidden patterns in ecthr's case law: On how citation network
  analysis can improve our knowledge of ecthr's article 14 practice.
\newblock {\em International Journal of Discrimination and the Law},
  17(1):4--22.

\bibitem[Pelc, 2014]{pelc2014politics}
Pelc, K.~J. (2014).
\newblock The politics of precedent in international law: A social network
  application.
\newblock {\em American Political Science Review}, 108(03):547--564.

\bibitem[Porteous et~al., 2008]{porteous2008fast}
Porteous, I., Newman, D., Ihler, A., Asuncion, A., Smyth, P., and Welling, M.
  (2008).
\newblock Fast collapsed gibbs sampling for latent dirichlet allocation.
\newblock In {\em Proceedings of the 14th ACM SIGKDD international conference
  on Knowledge discovery and data mining}, pages 569--577.

\bibitem[Raftery et~al., 2012]{raftery2012fast}
Raftery, A.~E., Niu, X., Hoff, P.~D., and Yeung, K.~Y. (2012).
\newblock Fast inference for the latent space network model using a
  case-control approximate likelihood.
\newblock {\em Journal of computational and graphical statistics},
  21(4):901--919.

\bibitem[Roberts et~al., 2014]{roberts2013structural}
Roberts, M.~E., Stewart, B.~M., Tingley, D., Lucas, C., Leder-Luis, J.,
  Gadarian, S.~K., Albertson, B., and Rand, D.~G. (2014).
\newblock Structural topic models for open-ended survey responses.
\newblock {\em American Journal of Political Science}, 58(4):1064--1082.

\bibitem[Spaeth et~al., 2020]{scdb}
Spaeth, H.~J., Epstein, L., Martin, A.~D., Segal, J.~A., Ruger, T.~J., and
  Benesh, S.~C. (2020).
\newblock Supreme court database, version 2021 release 01.

\bibitem[Wang et~al., 2008]{wang2008measuring}
Wang, M., Yu, G., and Yu, D. (2008).
\newblock Measuring the preferential attachment mechanism in citation networks.
\newblock {\em Physica A: Statistical Mechanics and its Applications},
  387(18):4692--4698.

\bibitem[Xiao and Stibor, 2010]{xiao2010efficient}
Xiao, H. and Stibor, T. (2010).
\newblock Efficient collapsed gibbs sampling for latent dirichlet allocation.
\newblock In {\em Proceedings of 2nd asian conference on machine learning},
  pages 63--78. JMLR Workshop and Conference Proceedings.

\bibitem[Zhang and Lauw, 2020]{zhang2020topic}
Zhang, C. and Lauw, H.~W. (2020).
\newblock Topic modeling on document networks with adjacent-encoder.
\newblock In {\em Proceedings of the AAAI Conference on Artificial
  Intelligence}, volume~34, pages 6737--6745.

\bibitem[Zhang and Lauw, 2022]{zhang2022dynamic}
Zhang, D.~C. and Lauw, H. (2022).
\newblock Dynamic topic models for temporal document networks.
\newblock In {\em International Conference on Machine Learning}, pages
  26281--26292. PMLR.

\end{thebibliography}

\end{document}